\def\1{\'{\i}}
\def\2{\c c}
\def\cao{\c c\~ao\ }

\def\coes{\c c\~oes\ }

\def\ca{\c ca }

\def\cas{\c cas }

\def\et{{\it et al.} }

\def\eq{\begin{equation}}     
\def\eeq{\end{equation}}      
\def\dm{\begin{displaymath}}  
\def\edm{\end{displaymath}}   

\documentstyle[12pt,portug,epsf]{tese}   

\begin{document}



\newlength{\stack}     
\newcommand{\push}[1]{\setlength{\stack}{#1}}
\newcommand{\pop}[1]{\setlength{#1}{\stack}}


\parindent 30pt                 
\parskip 12pt plus 1pt          

\newlength{\oldbaselineskip}                    
\newlength{\newbaselineskip}                    
\setlength{\oldbaselineskip}{\baselineskip}     
\setlength{\newbaselineskip}{1.2 \baselineskip} 

\setlength{\baselineskip}{\newbaselineskip}     

%
%

\newenvironment{myitems}
{\begin{list}{$\bullet$}{\setlength{\rightmargin}{\leftmargin}}}{\end{list}}
\pagestyle{plain}


 
%
%

\begin{titlepage}
  \vspace*{2cm}
  \large
  \begin{center}
     {\bf GRADIENTES DE COR E O CEN\'ARIO DE EVOLU\c C\~AO SECULAR
      EM GAL\'AXIAS ESPIRAIS DE TIPO TARDIO}
  \end{center}
  \vspace{7cm}
  \normalsize
  \begin{center}
     {\bf DIMITRI ALEXEI GADOTTI} \end{center}
  \vspace{0.2cm}
  \begin{center}
     {Orientadora: Profa. Dra. Sandra dos Anjos} \end{center}
  \vspace{4cm}
  \begin{center}
  \parbox{11cm}
     {Disserta\c c\~ao apresentada ao Instituto Astron\^omico
     e Geof\1sico da Universidade de S\~ao Paulo
     para a obten\c c\~ao do t\1tulo de Mestre em Astronomia.}
  \end{center}
  \vspace{1.5cm}
  \begin{center} SETEMBRO DE 1999 \end{center}
\end{titlepage} 

%
%

\pagestyle{empty}
\vspace*{19cm}
\hspace*{11cm} {Aos meus pais.\\}
\hspace*{12.05cm} {E \`a Ina\^e.}
\newpage
\pagestyle{plain}

%
%

\pagestyle{empty}
\vspace*{2cm}
   {\huge \bf Agradecimentos}
\vspace*{1.5cm}

Agrade\c co ao NExGal -- ProNEx e ao CNPq pelo apoio financeiro durante a realiza\cao deste trabalho.

Agrade\c co \`a Secretaria do Departamento de Astronomia, com a qual sempre pude contar nas tarefas
de cunho administrativo, e aos professores com quem travei um contato mais estreito e que sempre me
foram uma fonte de d\'uvidas, inquieta\c c\~oes, \'arduos trabalhos, esclarecimentos 
e deslumbramentos. Agrade\c co
tamb\'em ao pessoal do OPD/LNA -- CNPq pela ajuda dedicada nas noites frias e escuras.

Em especial, agrade\c co a minha orientadora, Sandra dos Anjos, pelo papel fundamental que desempenhou
no aspecto profissional de minha vida, pela infinidade de li\coes que me deu a oportunidade de aprender, 
e pela presen\c ca sempre prestativa, dedicada e companheira.

Agrade\c co a Ronaldo E. de Souza pelas prazerosas e intrigantes discuss\~oes, e pelas respostas 
claras que tanto contribu\1ram para o engrandecimento deste trabalho. Outros pesquisadores que
contribu\1ram de forma fundamental a este trabalho foram Rob Kennicutt e Tim Beers, aos quais
deixo aqui meus agradecimentos.

Agrade\c co a todos os meus amigos e companheiros do Clube Alpino Paulista, que t\^em de me ouvir 
sempre extasiado nas noites frias e estreladas das montanhas. Muitas raz\~oes tenho eu para estimar
os colegas que ganhei no Departamento, e as amizades que aqui se solidificaram, em especial, N\'elson, 
Ricardo Schiavon, Daniel, J\'ulio, Jocel, Eraldo, Jairo, Ednilson, Jan, Alex Ign\'acio, Armando, 
Ronaldo (Monstro!!!), Alexandre, Rodrigo, Iran, Jaqueline, Bruno, Jorge, Lucimara e Adriano, j\'a
que em todos eles pude buscar conforto ao notar que a luz no fim do t\'unel era a luz de um outro trem! Deixei
para falar com distin\cao da Am\'elia, com quem tive o prazer de dividir um tranq\"uilo deslumbramento
com a ci\^encia em tardes de m\'usica nem t\~ao tranq\"uila. Outra distin\cao quero fazer para o 
Grupo de Astronomia Extragal\'actica do Departamento, com quem sempre aprendo muito a partir de
discuss\~oes acaloradas e bem-humoradas.

Um inesperado e delicioso agradecimento eu deixo aqui para Aline, que de forma repentina e decisiva
faz parte agora de minha vida, que mais do que eu pr\'oprio desejou o fechamento deste estudo, e por quem, 
por muitos, muitos motivos, eu tenho imenso orgulho e carinho.

Finalmente, quero expressar minha eterna gratid\~ao a minhas irm\~as, Tain\'a, T\'abata e Ina\^e, que
sempre compreenderam a minha aus\^encia e que, ainda assim, tanto amor me d\~ao; a minha m\~ae, por
me fazer sentir t\~ao orgulhoso de mim mesmo, e pelo apoio com o qual sempre posso contar; a meu
pai, inextingu\1vel fonte de incentivo e inspira\c c\~ao, com suas perguntas desconcertantes.
\newpage
\pagestyle{plain}

%
%

\pagestyle{empty}
\vspace*{3cm}
\begin{verse}
\hspace{0.5cm}``If the conquest of a great peak brings moments of exultation and bliss, which in the monotonous, 
materialistic existence
of modern times nothing else can approach, it also presents great dangers. It is not the goal of grand 
alpinism to face peril,
but it is one of the tests one must undergo to deserve the joy of rising for an instant above the 
state of crawling grubs.
On this proud and beautiful mountain, we have lived hours of fraternal, warm and exalting nobility.
Here, for a few days, we have ceased to be slaves and have really been men.''
\end{verse}
\hspace*{10cm}
{Lionel Terray}

\vspace*{3cm}
\begin{verse}
\hspace{0.5cm}``I worked very hard into the night, sitting at a small table in the kitchen, next to a window.
It was getting later and later -- about 2:00 or 3:00 {\sc a.m.} I'm working hard, getting all these
calculations packed solid with things that fit, and I am thinking, and I am concentrating,
and it's dark, and it's quiet \dots''
\end{verse}
\hspace*{10cm}
{Dick Feynman}

\newpage
\pagestyle{plain}


\pagenumbering{roman}          
\tableofcontents               
\listoffigures                 
\listoftables                  
\newpage

%
%

\addcontentsline{toc}{chapter}{RESUMO}
\vspace*{2cm}
   {\huge \bf Resumo}
\vspace{1.5cm}

N\'os realizamos um estudo estat\1stico do comportamento de perfis de cor em bandas largas (UBV) 
para 257 gal\'axias espirais do tipo Sbc, ordin\'arias e barradas, 
utilizando dados obtidos atrav\'es de fotometria fotoel\'etrica de abertura, 
dispon\1veis na literatura (Longo \& de Vaucouleurs 1983,1985). 
N\'os determinamos os gradientes de cor (B\,-V) e (U-B) para as gal\'axias da amostra total,
bem como os \1ndices de cor 
(B\,-V) e (U-B) de bojos e discos separadamente, utilizando m\'etodos estat\1sticos robustos. 
Utilizamos uma t\'ecnica de decomposi\cao 
bi--dimensional para modelar os perfis de brilho de bojos e discos em imagens dos arquivos 
do ``Digitised Sky Survey'' (DSS), obtendo par\^ametros 
estruturais caracter\1sticos para 39 gal\'axias. A aquisi\cao de imagens de 14 gal\'axias no
Laborat\'orio Nacional de Astrof\1sica permitiu-nos realizar um estudo fotom\'etrico comparativo, e 
atestar a validade dos resultados obtidos neste estudo.

Entre os principais resultados obtidos, destacam-se: {\bf (i)} -- 65\% das gal\'axias possuem 
gradientes de cor negativos (mais vermelhos no centro), 25\% possuem gradientes nulos, e 10\% apresentam 
gradientes positivos; {\bf (ii)} -- gal\'axias que apresentam gradientes de cor nulos tendem a ser
barradas; {\bf (iii)} -- os \1ndices de cor ao longo das gal\'axias com gradientes nulos s\~ao similares 
aos \1ndices de cor dos discos das gal\'axias com gradientes negativos;
{\bf (iv)} -- confirmamos a correla\cao entre os \1ndices de cor de bojos e discos, j\'a obtida
por outros autores; 
{\bf (v)} -- a aus\^encia de correla\cao entre os gradientes de cor e de metalicidade
sugere que o excesso de gal\'axias barradas com gradientes de cor nulos ou positivos reflete uma 
diferen\ca no comportamento da idade m\'edia da popula\cao estelar ao longo de gal\'axias
barradas e ordin\'arias; 
{\bf (vi)} -- gal\'axias
com gradientes de cor nulos ou positivos t\^em uma leve tend\^encia a apresentar bojos maiores e
com maior concentra\cao central de luz; e {\bf (vii)} -- confirmamos a correla\cao entre as 
escalas de comprimento de bojos e discos, j\'a obtida por outros autores.

Estes resultados s\~ao compat\1veis e favor\'aveis ao cen\'ario de evolu\cao secular, no qual barras
produzem fluxos radiais de massa para as regi\~oes centrais de gal\'axias, n\~ao somente
homogeneizando as popula\coes estelares ao longo de gal\'axias, produzindo discos e bojos com \1ndices 
de cor semelhantes, mas tamb\'em contribuindo para a forma\cao e/ou constru\cao de bojos.
\newpage

%
%

\addcontentsline{toc}{chapter}{ABSTRACT}
\vspace*{2cm}
   {\huge \bf Abstract}
\vspace{1.5cm}

We have done a statistical study of the behaviour of the broadband color profiles (UBV) for
257 Sbc galaxies, barred and unbarred, collecting data obtained through photoeletric aperture  
photometry, available in the literature (Longo \& de Vaucouleurs 1983,1985).
We have determined (B\,-V) and (U-B) color gradients for the total sample of galaxies, 
as well as (B\,-V) and (U-B) color indices of
bulges and disks separately, using robust statistical methods.
Applying a bi--dimensional decomposition technique to model the brightness profiles of bulges
and disks in images from the Digitised Sky Survey (DSS), we obtained characteristic structural parameters for 39
galaxies. The acquisition of images for 14 galaxies in the Laborat\'orio Nacional de Astrof\1sica
(Astrophysics National Laboratory) allowed us to do a comparative photometric study, and verify the 
validity of the results obtained in this work.

Among the main results obtained, we point out: {\bf (i)} -- 65\% of the galaxies have
negative color gradients (reddish inward), 25\% have zero gradients, and 10\% show positive gradients; 
{\bf (ii)} -- galaxies that show zero color gradients tend to be barred; {\bf (iii)} -- the color
indices along the galaxies with zero color gradients are similar to the color indices of the disks of the
galaxies with negative color gradients; {\bf (iv)} -- we confirm the correlation between the
color indices of bulges and disks, already found by other authors; 
{\bf (v)} -- the absence of correlation between color and metallicity gradients suggests that the
excess of barred galaxies with zero or positive color gradients reflects a difference in the 
behaviour of the mean age of the stellar population along barred and unbarred galaxies; 
{\bf (vi)} -- galaxies with zero
or positive color gradients show a slight tendency of having larger bulges, with a greater
central concentration of light; and {\bf (vii)} -- we confirm the correlation between the scale lenghts
of bulges and disks, already found by other authors.

These results are compatible and favourable to the secular evolutionary scenario, in which stellar 
bars induce radial mass fluxes to the central regions of galaxies, not only turning homogeneous
the stellar populations along the galaxies, producing disks and bulges with similar color indices, 
but also contributing to the formation and/or building of galactic bulges.
\newpage
\pagenumbering{arabic}          
   \newpage \chapter{Introdu\cao}

\hskip 30pt Uma das grandes realiza\coes de Edwin Hubble no per\1odo de 1926 a 1936 foi 
organizar e sintetizar, em um diagrama, padr\~oes de gal\'axias representativos das morfologias 
observadas dos objetos que se encontram pr\'oximos e brilhantes. Este diagrama, conhecido como Sistema 
de Classifica\cao de Hubble (SCH), com modifica\coes que foram sendo introduzidas durante algumas
d\'ecadas, \'e o que vem sendo amplamente utilizado at\'e hoje na classifica\cao morfol\'ogica de 
gal\'axias. Nesses trabalhos pioneiros, Hubble verificou
que as gal\'axias podem ser naturalmente divididas em duas classes distintas, de acordo com suas
caracter\1sticas morfol\'ogicas, com o padr\~ao comum de
apresentarem simetria rotacional em torno de um n\'ucleo central: as el\1pticas (E) e as espirais (S).
A classe das el\1pticas \'e ca-racterizada basicamente por apresentar uma \'unica componente,
com morfologia esferoidal ou elipsoidal, cujo grau de elipticidade permite definir o indicador 

\eq
n = {10{\left( 1 - {{b}\over a}\right) }},
\eeq

\noindent onde $a$ e $b$ s\~ao, respectivamente, os semi-eixos maior e menor da gal\'axia.
Dessa forma, $(1-b/a)$ denota a elipticidade do objeto, representando, portanto, diferentes est\'agios 
de elipticidade, desde as mais esf\'ericas, at\'e as mais achatadas, conforme as respectivas nota\coes 
(E0, E1, E$n$, \dots, E7). A classe das espirais foi definida basicamente como sendo aquela que 
possui pelo menos duas componentes, uma delas associada \`a regi\~ao central, 
aproximadamente esf\'erica, denominada 
bojo, e a outra relativa \`a distribui\cao de luz em um plano achatado, definida como disco fino.
Hubble percebeu que as espirais podiam ser subdivididas em duas fam\1lias distintas, definidas como 
ordin\'arias (S) e barradas (SB). As gal\'axias espirais ordin\'arias cumprem os requisitos da defini\cao 
b\'asica da 
classe, e portanto se caracterizam por terem somente duas componentes, o bojo e o disco, enquanto que 
nas barradas observa-se uma componente adicional, denominada barra. As espirais tamb\'em caracterizam-se
por terem diferentes est\'agios, ou seja, as fam\1lias podem ser divididas nos est\'agios a, b e c, 
conforme os crit\'erios a seguir: (1) a abertura 
e defini\cao dos bra\c cos 
espirais, (2) a raz\~ao entre as luminosidades de bojo e disco e 
(3) a resolu\cao do disco e dos bra\c cos em estrelas e regi\~oes H{\sc ii}.
Assim sendo, uma gal\'axia do tipo Sa (ou SBa), por exemplo, normalmente apresenta 
uma raz\~ao bojo/disco maior do que
aquela apresentada por uma gal\'axia do tipo Sc (ou SBc), bem como bra\c cos menos definidos e com menor 
n\'umero de regi\~oes H{\sc ii} e estrelas resolvidas.

Observando que as gal\'axias ao fim da seq\"u\^encia das el\1pticas (E7) t\^em caracter\1sticas morfol\'ogicas
relativamente similares \`aquelas no in\1cio da seq\"u\^encia das espirais (Sa), Hubble sugeriu
que as duas seq\"u\^encias fossem cont\1guas, e que as espirais barradas formassem uma seq\"u\^encia
paralela \`a de espirais ordin\'arias, sintetizando, desta forma, o seu sistema de classifica\cao de gal\'axias, 
como ilustrado em seu famoso diagrama, reproduzido na Figura 1.1. Mais ainda, definiu que
as gal\'axias situadas \`a esquerda em seu diagrama s\~ao genericamente denominadas 
do tipo ``early'' (jovem) em rela\cao \`aquelas situadas \`a direita, denominadas do tipo
``late'' (tardio). Atualmente, os termos ``jovem'' e ``tardio'' s\~ao utilizados somente para apontar a posi\cao
de uma gal\'axia no diagrama de classifica\c c\~ao, sem qualquer conota\cao ao significado evolutivo. 
Portanto, neste sistema de classifica\c c\~ao, uma gal\'axia Sc \'e mais tardia do que uma Sa, 
bem como uma Sa \'e mais jovem do que uma Sb, por exemplo.

\begin{figure}
\epsfxsize=15cm
\centerline{\epsfbox{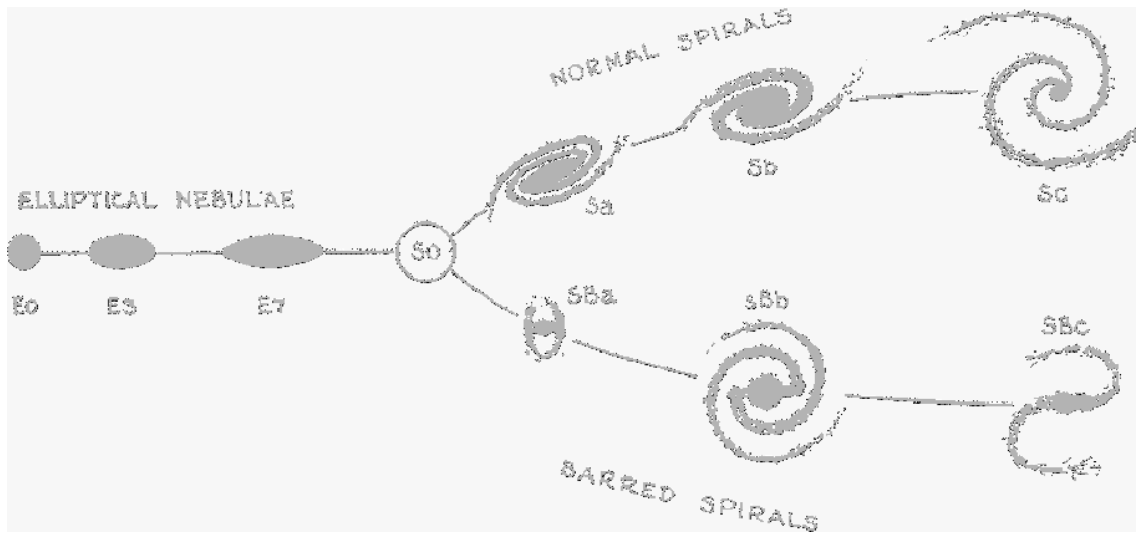}}
\caption{O sistema de classifica\cao morfol\'ogica de gal\'axias proposto por Hubble em 1936. 
Note que a classe das lenticulares ainda era de natureza um tanto hipot\'etica (Extra\1do de Hubble 1936).} 
\label{hubble}
\end{figure}

No entanto, ao perceber que as diferen\c cas entre propriedades de el\1pticas e espirais s\~ao
demasiado abruptas, Hubble introduziu uma nova classe de gal\'axias, 
a princ\1pio hipot\'etica, que \'e a classe das gal\'axias lenticulares, que teriam propriedades 
gerais intermedi\'arias entre as el\1pticas e as espirais. As lenticulares seriam ent\~ao mais tardias
do que as E7, por\'em n\~ao chegariam a apresentar estrutura espiral. Posteriormente, verificou-se a
exist\^encia real dessa classe de gal\'axias, tornando \'util e natural sua introdu\cao no sistema de 
classifica\c c\~ao, apesar de haver ainda hoje grande controv\'ersia em rela\cao \`a homogeneidade
das propriedades gerais desta classe. As gal\'axias lenticulares tamb\'em possuem duas grandes fam\1lias, 
a das ordin\'arias (S0) e a das barradas (SB0). Aquelas
gal\'axias cuja morfologia n\~ao apresenta simetria rotacional, e portanto n\~ao t\^em lugar no
diagrama de Hubble, s\~ao denominadas irregulares (Irr). Na \'epoca da elabora\cao do diagrama de Hubble, 
somente cerca de 3\% das gal\'axias eram classificadas como Irr (Hubble 1926,1936).

Ao longo dos anos, algumas etapas de refinamento foram aplicadas ao SCH, motivadas 
principalmente pelo aumento do n\'umero de cat\'alogos que surgiram a partir da inspe\cao de 
gal\'axias em placas fotogr\'aficas de grande campo. Na revis\~ao do
sistema de Hubble, por G. de Vaucouleurs (1963), as altera\c c\~oes\footnote{Parte 
destas altera\coes j\'a havia sido vislumbrada em Hubble (1936).} mais importantes introduzidas foram:

\paragraph{Refinamento das subclasses.} S\~ao introduzidos est\'agios intermedi\'arios  
na classe das lenticulares e na classe das espirais, como Sab e Sbc, por exemplo.
Al\'em disso, essa seq\"u\^encia foi
estendida para as subclasses cd, d, dm e m, onde m denota as espirais do tipo magel\^anicas. A
classe das irregulares \'e substitu\1da pelas subclasses Im e I, ordin\'arias ou barradas.

\paragraph{Refinamento da classifica\cao de barras.} S\~ao introduzidas as nota\coes SA para 
gal\'axias ordin\'arias e SAB para os casos transit\'orios (ou incertos) entre SA e SB.

\paragraph{Introdu\cao das variedades r e s.} S\~ao adicionados na nota\cao os s\1mbolos (r), para
aquelas gal\'axias em que os bra\c cos se iniciam tangentes a um anel, no qual a barra termina (no caso
de a gal\'axia ser barrada), e
(s), para as gal\'axias em que os bra\c cos partem das extremidades da barra (ou do bojo, no caso de a
gal\'axia ser ordin\'aria). Tamb\'em h\'a a nota\cao (rs) para os casos transit\'orios. Embora essa nota\cao
seja mais pr\'opria para gal\'axias barradas, tamb\'em pode ser utilizada para gal\'axias ordin\'arias.

Com o desenvolvimento de detectores, telesc\'opios e t\'ecnicas observacionais, e o explosivo aumento
do n\'umero de gal\'axias estudadas, novos cat\'alogos foram publicados e
pequenas altera\coes ao SCH foram sendo realizadas para 
representar com melhor precis\~ao diferentes sutilezas morfol\'ogicas, de tal forma que a classifica\cao
morfol\'ogica de gal\'axias revisada, e amplamente utilizada, foi sintetizada como pode ser visto no 
``Third Reference Catalog of Bright Galaxies'' (de Vaucouleurs \et 1991 -- doravante RC3). 
Uma dessas modifica\coes consistiu na introdu\cao do par\^ametro T (\1ndice do est\'agio de Hubble), que \'e 
definido por uma escala num\'erica, na qual gal\'axias do tipo Sa t\^em T = 1, gal\'axias Sab t\^em
T = 2, Sb t\^em T = 3, e assim por diante. Por outro lado, gal\'axias S0/a t\^em T = 0, e as de tipos mais 
jovens possuem T's negativos. Para uma revis\~ao sobre classifica\cao
de gal\'axias veja Sandage (1975); ou tamb\'em van den Bergh (1997).

Outros sistemas de classifica\cao de gal\'axias foram propostos com crit\'erios e par\^ame-tros
novos e distintos. Entretanto, os sistemas de classifica\cao posteriores ao de Hubble se baseiam, 
ainda que parcialmente, no SCH, e podem ser considerados complementares. Morgan (1958), por exemplo, 
utiliza a concentra\cao central de luz como crit\'erio de
classifica\cao e encontra correla\cao entre este par\^ametro, o tipo espectral dominante na
gal\'axia e o tipo de Hubble. Essa correla\cao \'e esperada, j\'a que a raz\~ao
bojo/disco diminui ao longo da
seq\"u\^encia de Hubble, quando se parte dos tipos jovens para os tipos tardios, e o tipo espectral
dominante nestas componentes \'e notavelmente distinto. Este sistema de classifica\c c\~ao, pouco 
utilizado nas d\'ecadas passadas, tem recentemente sido resgatado, devido \`a dificuldade em se 
utilizar o SCH em gal\'axias a grandes profundidades.
Abraham \et (1996), por exemplo, utilizam, para a classifica\cao de gal\'axias 
a altos ``redshifts'' no ``Hubble Deep Field'', os par\^ametros de concentra\cao central 
de luz e de assimetria, al\'em do tipo de Hubble, o que permite uma classifica\cao
mais quantitativa e objetiva.
A concentra\cao central de luz tem sido usada tamb\'em em algoritmos de classifica\cao autom\'atica 
(Abraham \et 1994). Aplica\coes recentes deste sistema de 
classifica\c c\~ao, utilizando a concentra\cao central de luz em gal\'axias no aglomerado de Virgo, podem
ser vistas em Koopmann \& Kenney (1998). van den Bergh (1960a,b) adiciona ao sistema de
Hubble cinco classes de luminosidade, referentes \`a luminosidade absoluta das gal\'axias, e encontra
correla\cao entre estas classes e a morfologia dos bra\c cos espirais, no sentido de que as gal\'axias
mais luminosas apresentam bra\c cos bem definidos e desenvolvidos. Mais recentemente, Elmegreen \&
Elmegreen (1982a,b) desenvolveram um sistema de classifica\cao morfol\'ogica dos bra\c cos espirais,
que parte do tipo 1 (floculento) em que os bra\c cos t\^em uma apar\^encia n\~ao uniforme, ca\'otica
e pouco definida, at\'e o tipo 12 (``grand design'') no qual os bra\c cos s\~ao sim\'etricos,
bem definidos e desenvolvidos. Nesse trabalho, os autores concluem que existe uma forte tend\^encia 
de as gal\'axias
barradas, ou com companheiras pr\'oximas, apresentarem bra\c cos com o padr\~ao ``grand design''.

Um sistema de classifica\cao que se diferencia substancialmente dos acima citados \'e aquele 
proposto por Humason (1936) e Morgan \& Mayall (1957; ver tamb\'em Sodr\'e \& Cuevas 1994), 
e consiste na classifica\cao espectral de gal\'axias. Neste sistema, uma gal\'axia \'e classificada
segundo seu espectro integrado, o que tamb\'em permite uma classifica\cao quantitativa e objetiva. 
\'E importante
ressaltar que existe uma forte correla\cao entre os tipos espectrais de gal\'axias e os tipos de
Hubble (e.g., Sodr\'e \& Cuevas 1997).

Apesar de o sistema de classifica\cao de Hubble ser demasiado simples e subjetivo, seu sucesso reside
exatamente em ignorar a mir\1ade de detalhes na estrutura das gal\'axias, que, se
considerados, for\c cariam
a introdu\cao de um sem--n\'umero de classes e tornariam a classifica\cao extremamente penosa e 
complexa. Ao contr\'ario, este sistema se concentra em caracter\1sticas e padr\~oes globais.

Por outro lado, van den Bergh (1997) argumenta que o sistema de classifica\cao de Hubble incorpora
somente gal\'axias luminosas, pr\'oximas e em ambientes pouco densos, como no campo ou em
aglomerados pobres. Desta forma, torna-se dif\1cil utilizar o sistema de Hubble para classificar,
por exemplo, gal\'axias a altos ``redshifts'', gal\'axias an\~as e gal\'axias de baixo brilho
superficial. Al\'em disso, van den Bergh enfatiza que o SCH n\~ao incorpora a classe das 
lenticulares de maneira satisfat\'oria, j\'a que esta classe parece ser tipicamente menos luminosa 
do que as el\1pticas e tamb\'em menos luminosa do que as espirais do tipo Sa. van den Bergh 
continua, salientando que o SCH t\~ao pouco incorpora adequadamente as gal\'axias cD's, tipicamente
presentes nas regi\~oes centrais de aglomerados, que parecem resultar de fus\~oes entre gal\'axias. 
O mesmo parece ocorrer com as el\1pticas, que no SCH s\~ao posicionadas em fun\cao da elipticidade projetada, 
mas que recentemente t\^em sido apontadas como tendo tamb\'em 2 fam\1lias: a das ``boxy'' e a das ``disky''.
Uma tentativa de incorporar as diferentes fam\1lias de el\1pticas, bem como as irregulares ordin\'arias
e barradas, foi proposta por Kormendy \& Bender (1996). Uma tentativa de incorporar ao SCH todas as classes 
que n\~ao est\~ao devidamente representadas na sequ\^encia de Hubble, utilizando um diagrama tri--dimensional, 
foi feita por van den Bergh (1997).

Evidentemente, o valor de um sistema de classifica\cao depende de sua utilidade, ou de sua 
capacidade em exibir correla\coes entre suas classes e v\'arios par\^ametros f\1sicos relevantes, 
conectando, por exemplo, as propriedades morfol\'ogicas aparentes nos sistemas e as suas propriedades 
f\1sicas. Assim, um sistema de classifica\cao \'util deve levar
\`a formula\cao de hip\'oteses e previs\~oes que possam contribuir para o 
desenvolvimento de modelos a res-peito da forma\c c\~ao, estrutura e evolu\cao de gal\'axias.
Nesse aspecto, \'e not\'oria a utilidade da classifica\cao de Hubble. O \1ndice de cor integrado de
gal\'axias, por
exemplo, \'e uma propriedade f\1sica que se correlaciona com os tipos morfol\'ogicos de Hubble, no
sentido de que gal\'axias de tipo jovem s\~ao mais vermelhas que as de tipo tardio, indicando a
popula\cao estelar que domina a emiss\~ao de luz em cada classe morfol\'ogica.

Na seq\"u\^encia das espirais, algumas tend\^encias podem ser identificadas quando se parte dos tipos
jovens para os tardios: (1) o aumento da fra\cao da massa total na forma de g\'as e poeira no meio 
interestelar, (2) o aumento da luminosidade absoluta das estrelas e regi\~oes H{\sc ii} mais
brilhantes resolvidas nos bra\c cos espirais e (3) o aumento no tamanho e no n\'umero de regi\~oes
H{\sc ii} resolvidas nos bra\c cos. A maior fra\cao de g\'as nos tipos tardios indica que estes
sistemas converteram g\'as em estrelas com menor rapidez do que os tipos mais jovens
(Larson 1990). Portanto, as espirais de tipo jovem consumiram praticamente todo o seu g\'as e hoje
observamos uma maior fra\cao de g\'as nas espirais tardias. Conseq\"uentemente, a taxa de forma\cao
estelar nas espirais tardias resulta ser tamb\'em mais elevada atualmente.
Luminosidade absoluta, massa total e a abund\^ancia de certos elementos qu\1micos
como o Oxig\^enio e o Ferro s\~ao outros exemplos de par\^ametros f\1sicos fundamentais que se 
correlacionam com os tipos de Hubble (veja Roberts \& Haynes 1994 para uma revis\~ao).

Apesar de ser um dos crit\'erios de classifica\cao no sistema de Hubble, 
principalmente para as gal\'axias vistas de perfil, a correla\cao entre
a raz\~ao bojo/disco e os tipos morfol\'ogicos apresenta consider\'avel dispers\~ao, i.e., a
tend\^encia desta raz\~ao de diminuir em dire\cao aos tipos tardios reflete somente um comportamento
m\'edio. Assim, podemos encontrar sistemas de mesma subclasse, mesma luminosidade total, mas com 
raz\~oes entre as luminosidades de bojo e disco bastante diversas. Como exemplo, entre muitos 
outros, pode-se citar
o caso de NGC 4800 e NGC 1068, que podem ser examinadas em ``The Hubble Atlas of Galaxies''
(Sandage 1961).
Ambos os sistemas s\~ao classificados como Sb, mas NGC 4800 possui um bojo muito mais proeminente que
NGC 1068. Isto ocorre porque o crit\'erio fundamental utilizado na classifica\cao \'e o grau de
abertura e defini\cao dos bra\c cos, e n\~ao a raz\~ao bojo/disco.

Evidentemente, toda essa variedade nas propriedades e na estrutura de gal\'axias deve estar relacionada
aos processos de forma\cao e evolu\cao de gal\'axias.
Por exemplo, gal\'axias de mesmo tipo morfol\'ogico, por\'em com diferentes raz\~oes bojo/disco, 
podem apresentar essa diferen\c ca devido ao fato de terem sofrido diferentes processos durante
suas evolu\c c\~oes. Os processos que envolvem a forma\cao e a evolu\cao das componentes bojo e disco
devem ter ocorrido de maneira distinta entre essas gal\'axias.
Uma informa\cao importante que pode nos trazer pistas a respeito dos diferentes processos evolutivos
que ocorrem em gal\'axias \'e a an\'alise de \1ndices de cor. Estes \1ndices proporcionam 
informa\coes a respeito da popula\cao estelar e da hist\'oria de forma\cao estelar em gal\'axias.

O estudo de cores em gal\'axias, em particular, da cor integrada em bandas largas, 
tem sido realizado para obter informa\coes a 
respeito da popula\cao estelar (e.g., Searle, Sargent \& Bagnuolo 1973; Tinsley 1980; Frogel 1985; Peletier 1989; 
Silva \& Elston 1994), bem como da extin\cao causada pela poeira interestelar (e.g., Evans 1994; 
Peletier \et 1994). O uso de t\'ecnicas de fotometria 
superficial de cores em bandas largas, principalmente nos estudos que envolvem um enfoque estat\1stico, 
praticamente n\~ao foi realizado. Uma exce\cao \'e a tese de doutoramento de R.S. de Jong, 
no ``Kapteyn Astronomical Institute'' (de Jong \& van der Kruit 1994;
de Jong 1996a,b,c).

Uma possibilidade de avaliar o comportamento da cor em gal\'axias, e portanto da distribui\cao da 
popula\cao estelar, que tem sido pouco explorada na literatura, \'e utilizar a distribui\cao radial de 
cor, ou seja, o perfil de cor. Esta informa\cao pode oferecer tamb\'em indica\coes sobre 
cen\'arios de forma\cao de bojos.        

Nessa Disserta\c c\~ao, estudamos os \1ndices de
cor (B\,-V) e (U-B) ao longo das componentes bojo e disco de gal\'axias espirais do tipo Sbc, 
ordin\'arias e barradas. Neste estudo, damos especial \^enfase \`as
previs\~oes dos cen\'arios de forma\cao de bojos nesta classe de gal\'axias. 
Assim, no restante deste Cap\1tulo discorreremos brevemente acerca de alguns dos principais cen\'arios de 
forma\cao e evolu\c c\~ao, e suas previs\~oes. Finalizaremos o Cap\1tulo 1 introduzindo com maior 
profundidade os objetivos deste trabalho.

\section{Forma\cao de bojos}

\hskip 30 pt Evidentemente, a forma\cao da componente esferoidal em gal\'axias espirais est\'a intimamente 
vinculada
\`a forma\cao de gal\'axias como um todo, de modo que um cen\'ario de forma\cao de gal\'axias deve poder
explicar as propriedades observadas nas componentes bojo, disco e halo, simultaneamente. Existem, atualmente, tr\^es 
principais cen\'arios su-geridos para a forma\cao dos bojos em gal\'axias espirais, que buscam se adequar, 
principalmente, \`as 
rela\coes observadas entre as propriedades de bojos e discos, j\'a que ainda n\~ao existe um consenso
a respeito do bojo ser somente a parte mais central do halo (Silk \& Bouwens 1999; Renzini 1999), ou 
de bojo e halo serem duas entidades estruturais distintas (Wyse, Gilmore \& Franx 1997).
Al\'em disso, as propriedades dos halos de gal\'axias s\~ao intrinsecamente mais dif\1ceis de
serem observadas.
Como veremos, as diferen\cas entre os 3 cen\'arios podem ser essencialmente resumidas na \'epoca de 
forma\cao do bojo em rela\cao \`a forma\cao do disco. Bojos podem ter sido formados em uma \'epoca anterior
\`a forma\cao do disco, ou podem ter sido formados posteriormente. Uma terceira hip\'otese \'e a de que
bojos e discos se formem concomitantemente.
Antes de passar a uma breve descri\cao
destes cen\'arios, vejamos algumas das propriedades observadas; muitas delas tornaram-se conhecidas 
atrav\'es de estudos do bojo e do disco da Gal\'axia.

\paragraph{Idades.} Em princ\1pio, a determina\cao das idades de bojos e discos seria um teste definitivo
na escolha de um cen\'ario de forma\c c\~ao. No entanto, como foi deixado bem claro por Combes (1999), o
bojo n\~ao tem necessariamente a mesma idade das estrelas que cont\'em, j\'a que pode ter sido
formado recentemente, constitu\1do por estrelas velhas. De qualquer forma, tem sido observado que os
discos t\^em, em m\'edia, cores mais azuis que as dos bojos, o que pode indicar que as estrelas dos bojos
s\~ao mais velhas. Entretanto, os gradientes de cor apresentados por gal\'axias espirais mostram que as
regi\~oes mais externas dos discos s\~ao mais jovens do que suas regi\~oes mais internas (de Jong 1996c).
De fato, Peletier \& Balcells
(1996) mostram que as cores de bojos e das regi\~oes internas dos discos s\~ao bastante similares,
concluindo que a diferen\c ca nas idades m\'edias das estrelas nestas componentes \'e menor do que
30\%, apesar de existir uma incerteza devido \`a degeneresc\^encia idade--metalicidade.

\paragraph{Metalicidades.} Na Gal\'axia, a metalicidade (i.e., a abund\^ancia Fe/H) m\'edia das regi\~oes
externas do bojo \'e similar \`a do disco na vizinhan\c ca solar. 
Por\'em, a distribui\cao de metalicidades das estrelas do bojo \'e muito mais alargada.
A distribui\cao de metalicidades das estrelas do halo, na vizinhan\c ca solar, \'e tamb\'em muito
alargada, mas o valor da metalicidade m\'edia \'e substancialmente inferior (Wyse, Gilmore \& Franx 1997).

\paragraph{Din\^amica.} Apesar de um certo suporte rotacional, os bojos s\~ao mantidos, em geral, 
pela dispers\~ao anisotr\'opica de velocidades
das estrelas que cont\^em, enquanto que os discos s\~ao mantidos quase integralmente pelo movimento rotacional
das estrelas. No entanto,
as estrelas nos bojos de baixa luminosidade t\^em maior momento angular, o que os torna mais similares
aos discos, enquanto que os bojos de alta luminosidade s\~ao dinamicamente mais semelhantes \`as gal\'axias 
el\1pticas 
(Silk \& Bouwens 1999). Al\'em disso, os bojos seguem o mesmo Plano Fundamental das gal\'axias el\1pticas,
e a mesma rela\cao luminosidade--metalicidade (Combes 1999 e refer\^encias a\1 contidas). Isso sugere que
a forma\cao dos bojos de alta luminosidade seja semelhante \`a forma\cao das el\1pticas, enquanto que a
origem dos bojos de baixa luminosidade esteja ligada \`a origem dos discos.

Vejamos agora com mais detalhes cada um dos principais cen\'arios propostos na literatura para a formacao de 
bojos, e as previs\~oes com rela\cao \`as
propriedades acima citadas, al\'em de outras descobertas e propriedades mais espec\1ficas.

\subsection{O cen\'ario monol\1tico de forma\c c\~ao}

\hskip 30pt Um estudo cl\'assico sobre a cinem\'atica de 221 estrelas an\~as, cujas \'orbitas cruzam a 
vizinhan\c ca solar, foi publicado em um famoso artigo de Eggen, Lynden-Bell \& Sandage (1962). Esse 
estudo indicou que estrelas de baixa metalicidade movem-se invariavelmente em \'orbitas bastante
exc\^entricas, enquanto que as estrelas de alta metalicidade possuem \'orbitas quase circulares.
Tamb\'em foi encontrada uma correla\cao entre a metalicidade estelar e o momento angular da estrela:
estrelas de baixa metalicidade possuem pouco momento angular. Tamb\'em foi mostrado que as estrelas de
alta metalicidade se concentram no plano do disco da Gal\'axia, enquanto que aquelas de baixa metalicidade
podem ser encontradas n\~ao somente no disco, mas como tamb\'em em v\'arias alturas distintas em rela\cao
ao plano Gal\'actico.

A interpreta\cao destas correla\coes levou os autores a formular um cen\'ario para a forma\cao da Gal\'axia
que, posteriormente, foi estendido para as outras gal\'axias, e deno-minado por monol\1tico. Neste cen\'ario,
o bojo seria formado atrav\'es do colapso radial do g\'as protogal\'actico de abund\^ancia qu\1mica primordial.
Este colapso termina rapidamente, em escalas de tempo da ordem de $10^{8}$ anos, quando o aumento do
momento angular na regi\~ao central interrompe a queda radial do g\'as. Entretanto, o colapso do g\'as na dire\cao
paralela ao momento angular continua, dando origem ao disco.

Como a densidade do g\'as nas regi\~oes centrais cresce rapidamente, tamb\'em cresce assim a taxa de
forma\cao estelar. Portanto, a primeira gera\cao de estrelas \'e formada durante o colapso, com 
\'orbitas de alta excentricidade, metalicidade baixa, e vai constituir o bojo. Enquanto a primeira
gera\cao estelar evolui, o g\'as remanescente que est\'a dando origem ao disco se enriquece com os
elementos qu\1micos resultantes da nucleoss\1ntese estelar. Assim, as estrelas que constituem o disco
possuem \'orbitas quase circulares, alta metalicidade, e s\~ao mais jovens que as estrelas do bojo.
\'E evidente que, neste cen\'ario, a forma\cao do bojo ocorre em uma \'epoca anterior \`a do disco.

Apesar da advert\^encia apontada acima por Combes (1999), as evid\^encias de uma popula\cao estelar
mais velha no bojo da Gal\'axia, e no de outras gal\'axias, t\^em levado pesquisadores a dar suporte ao 
cen\'ario monol\1tico (Silk \& Bouwens 1999).

No entanto, \'e dif\1cil conciliar as previs\~oes das metalicidades estelares neste cen\'ario com a 
largura da distribui\cao das metalicidades das estrelas no bojo Gal\'actico e, em especial, com a
descoberta de estrelas super-ricas em metais nesta componente (McWilliam \& Rich 1994). Tentativas
de concilia\cao levaram ao modelo de ciclo fechado (Ibata \& Gilmore 1995 e refer\^encias a\1
contidas; veja tamb\'em Fran\c cois, Vangioni-Flam \& Audouze 1990; Larson 1990), onde o g\'as \'e 
processado localmente, resultando na popula\cao estelar em cada ponto do sistema.

Al\'em disso, a an\'alise de imagens obtidas pelo Telesc\'opio Espacial Hubble por Ca-rollo \et (1997)
mostra que os bojos em gal\'axias espirais exibem freq\"uentemente pontos brilhantes, que s\~ao, 
muito provavelmente, regi\~oes de forma\cao estelar. Evid\^encias de que o mesmo fen\^omeno pode estar
ocorrendo no bojo da Gal\'axia s\~ao apresentadas em Rich \& Terndrup (1997). Pfenniger (1993)  
aponta que a simples exist\^encia de gal\'axias espirais sem bojos (Sd, Sm), ou com bojos muito
pequenos (Sc) j\'a \'e um ind\1cio de que o cen\'ario monol\1tico n\~ao \'e a \'unica possibilidade
para a forma\cao de gal\'axias. Outros fortes v\1nculos que devem ser obedecidos pelos modelos de
forma\cao que seguem o cen\'ario monol\1tico podem ser encontrados em Avila-Reese \& Firmani (1998).

\subsection{O cen\'ario hier\'arquico de forma\c c\~ao}

\hskip 30pt Toomre \& Toomre (1972) mostraram, atrav\'es de simula\coes num\'ericas, que encontros
entre gal\'axias podem produzir for\c cas de mar\'e suficientemente fortes para dar origem a 
v\'arias caracter\1sticas morfol\'ogicas observadas em gal\'axias, tais como pontes e caudas.
Entre os exemplos que podem endossar as previs\~oes das simula\c c\~oes, pode-se destacar 
NGC 4038 + NGC 4039 (``as Antenas'') e M 51 + NGC 5195 (a gal\'axia
``Redemoinho''). Estas simula\coes monstraram que o encontro entre gal\'axias certamente tem um importante
papel na evolu\cao destes objetos, e que caracter\1sticas morfol\'ogicas observadas poderiam ser produzidas
por tais encontros. Estudos posteriores (ver Alladin \& Narasimhan 1982 para uma revis\~ao)
indicaram que colis\~oes mais fortes, resultando na posterior
fus\~ao, entre gal\'axias espirais, poderiam dar origem a uma gal\'axia el\1ptica. Schweizer (1982),
por exemplo, mostra que NGC 7252 (tamb\'em conhecida como gal\'axia ``\'Atomos pela Paz'')
apresenta v\'arias caracter\1sticas que sugerem ser um resultado de
uma recente fus\~ao entre duas gal\'axias espirais de massas similares. No entanto, esta gal\'axia
possui um \'unico n\'ucleo, que tem uma distribui\cao de luminosidade seguindo a lei $r^{1/4}$, t\1pica
do perfil de luminosidade de gal\'axias el\1pticas. 

De maneira similar, a acresc\^encia de uma gal\'axia companheira an\~a por uma gal\'axia espiral 
poderia contribuir para a forma\cao da componente esferoidal de gal\'axias espirais. Este processo
poderia se repetir v\'arias vezes, j\'a que o encontro entre gal\'axias certamente n\~ao \'e um
raro fen\^omeno, contribuindo para a constru\cao de um bojo cada vez mais proeminente, constituindo,
assim, o cen\'ario hier\'arquico de forma\c c\~ao. Enquanto gal\'axias el\1pticas podem ser o resultado
de uma fus\~ao entre duas gal\'axias espirais de tamanhos similares, os bojos seriam formados pela
acresc\^encia de gal\'axias sat\'elites an\~as. 

A exist\^encia de v\'arios exemplos de gal\'axias espirais gigantes com numerosas companheiras an\~as 
favorece este cen\'ario, que possui duas vers\~oes principais. Na primeira delas, a fus\~ao das 
gal\'axias destr\'oi o disco j\'a existente, que volta a se formar atrav\'es da queda do g\'as 
remanescente. Nesta vers\~ao, o disco \'e continuamente destru\1do e recons-tru\1do, enquanto o bojo
torna-se cada vez mais importante. Assim, bojos ser\~ao mais velhos que discos. Seguindo este racioc\1nio,
Baugh, Cole \& Frenk (1996) elaboraram um modelo simples, que explica de maneira natural diversas
propriedades observadas em gal\'axias, tais como a rela\cao densidade--morfologia (segrega\cao
morfol\'ogica) e o efeito Butcher--Oemler. 
No entanto, uma das previs\~oes deste modelo \'e a de que os bojos em gal\'axias espirais de tipo
tardio, que possuem uma baixa raz\~ao bojo/disco, devem ser mais velhos do que aqueles em espirais
de tipo jovem, j\'a que para ter um disco maior a gal\'axia deve ter permanecido sem ser perturbada
por companheiras e adquirindo g\'as por um longo tempo. Essa previs\~ao n\~ao parece ser compat\1vel
com a observa\cao de que gal\'axias com alta raz\~ao bojo/disco n\~ao mostram evolu\cao na fun\cao de 
luminosidade at\'e
``redshifts'' da ordem de 1 (Wyse, Gilmore \& Franx 1997 e refer\^encias a\1 contidas).
Por outro lado, pode-se
construir modelos (Bouwens, Cay\'on \& Silk 1998) em que a acresc\^encia de uma gal\'axia companheira 
an\~a contribui tanto para a forma\cao do bojo quanto para a forma\cao do disco, sem destruir um
eventual disco j\'a existente (ver tamb\'em Kauffmann \& White 1993). 
Nesta vers\~ao do cen\'ario hier\'arquico, bojos e discos t\^em idades similares.

O cen\'ario hier\'arquico pode se adequar muito bem \`as idades observadas de bojos e discos.
Entretanto, se os bojos s\~ao formados via acresc\^encia, ent\~ao as metalicidades observadas em
bojos, e em particular no bojo da Gal\'axia, imp\~oem fortes v\1nculos na metalicidade dos objetos
acrescidos. Estes devem possuir metalicidades relativamente elevadas, o que traz limites na fra\cao do
bojo que pode ter sido acrescida recentemente (Combes 1999 e refer\^encias a\1 contidas). 
Al\'em disso, dada a correla\cao entre metalicidade e luminosidade, ent\~ao os bojos devem ter sido
formados por gal\'axias an\~as de luminosidades semelhantes \`as dos bojos (Wyse, Gilmore \& Franx 1997
e refer\^encias a\1 contidas).

No caso da Gal\'axia, as diferen\cas entre as propriedades cinem\'aticas e qu\1micas das estrelas no
bojo, nas Nuvens de Magalh\~aes e na (recentemente descoberta) gal\'axia an\~a esferoidal Sagittarius, 
indicam que, se a acresc\^encia de companheiras teve um papel importante na forma\cao do bojo
Gal\'actico, ent\~ao, certamente, estas companheiras eram bastante distintas das que n\'os temos
atualmente. No entanto, a exist\^encia de um grande n\'umero de estrelas no halo Gal\'actico com
\'orbitas retr\'ogradas (Larson 1990) \'e um ind\1cio de que a acresc\~ao de pequenos sat\'elites pode
ser importante na origem dos halos gal\'acticos. Por outro lado, esse processo pode n\~ao ter um papel
fundamental na origem de discos e bojos em geral, j\'a que a observa\cao de \'orbitas retr\'ogradas nestas 
componentes \'e rara (Wyse, Gilmore \& Franx 1997).

\subsection{O cen\'ario de evolu\cao secular}

\hskip 30pt Muitos estudos num\'ericos t\^em demonstrado que instabilidades din\^amicas em discos, 
tais como bra\c cos espirais\footnote{Ver, por exemplo, Zhang (1996,1998).}, mas, principalmente barras, 
podem ser respons\'aveis pela forma\cao de
bojos em escalas de tempo maiores que a escala de tempo din\^amica de gal\'axias (i.e., $ > 10^{8}$
anos).

Evid\^encias observacionais mostram que mais da metade das gal\'axias brilhantes 
no universo local possuem barras. Muito trabalho a respeito desta estrutura gal\'actica foi 
realizado para que o nosso conhecimento sobre barras possa ser considerado hoje pouco mais do 
que qualitativo (veja Friedli 1999 para uma revis\~ao).

Sabemos que, devido a instabilidades din\^amicas, \'e muito prov\'avel o surgimento 
espont\^aneo de uma barra em discos gal\'acticos, e
que estas estruturas tamb\'em podem ser induzidas por gal\'axias companheiras interactuantes.
Uma vez presentes, as barras produzem uma s\'erie de fen\^omenos evolutivos na gal\'axia
hospedeira. Estudos te\'oricos (e.g., Friedli \& Benz 1993, 
1995 e refer\^encias a\1 contidas) mostram que, atrav\'es de choques e torques gravitacionais, 
uma barra estelar \'e capaz de coletar
g\'as das regi\~oes externas do disco para as regi\~oes internas. Assim, deve ocorrer uma mistura em
grande escala do g\'as ao longo da gal\'axia, que, em princ\1pio, deveria afetar o comportamento dos 
perfis de abund\^ancia de alguns elementos qu\1micos.
Em acordo com essa previs\~ao, 
Martin \& Roy (1994) e Zaritsky, Kennicutt \& Huchra (1994) concluem, a partir de um estudo de
abund\^ancia qu\1mica em gal\'axias espirais, que as barradas tendem a apresentar gradientes 
da abund\^ancia O/H menos acentuados do que gal\'axias n\~ao--barradas. 
Al\'em disso, Sakamoto {\it et al.} (1999) mostram que gal\'axias barradas apresentam uma
maior concentra\c c\~ao central de g\'as molecular (CO) do que gal\'axias ordin\'arias.
Estes autores argumentam que estes resultados indicam que o transporte de g\'as ao longo da barra para
as regi\~oes centrais das gal\'axias deve ter ocorrido.

Por outro lado, simula\c c\~oes N--Corpos (e.g., Combes \& Sanders 1981) mostram que uma
barra que se desenvolve em um disco plano n\~ao permanece fina,
i.e., n\~ao permanece no plano do disco, j\'a que resson\^ancias 
orbitais (e/ou as instabilidades ``bar--buckling'' ou ``fire--hose'') provocam o aquecimento 
vertical da barra, que se manifesta na forma\c c\~ao de uma
estrutura perpendicular ao plano do disco, ou seja, no espessamento da barra, em uma escala de tempo
de cerca de 1 Giga-ano ap\'os a forma\cao desta. 
Este espessamento \'e mais importante na regi\~ao central da barra, que acaba desenvolvendo um 
bojo com morfologia retangular ou em forma de amendoim (Sellwood 1993).
Em uma s\'erie de
trabalhos recentes (Kuijken \& Merrifield 1995; Merrifield \& Kuijken 1999; Bureau \& 
Athanassoula 1999; Athanassoula \& Bureau 1999; Bureau \& Freeman 1999; Bureau, Freeman \& 
Athanassoula 1999) a natureza destes bojos foi estudada, considerando evid\^encias observacionais 
em favor de mecanismos de acresc\^encia ou de instabilidades em barras. Os resultados indicam que, 
apesar dos mecanismos de acresc\^encia serem poss\1veis e prov\'aveis, eles n\~ao s\~ao os mecanismos 
prim\'arios, e que instabilidades ``bar--buckling'' devem ser res-pons\'aveis pela ocorr\^encia da 
maioria dos bojos com estas morfologias. 
Estes resultados refor\c cam, portanto, a hip\'otese de que bojos retangulares ou em forma de amendoim 
s\~ao, de fato, gal\'axias barradas vistas de perfil, conforme sugerido por de Souza \& dos Anjos (1987), 
e que o cen\'ario de evolu\cao secular em gal\'axias barradas deve gerar as morfologias observadas nestes 
tipos de bojos. 

Al\'em disso, o aumento da concentra\c c\~ao de massa nas regi\~oes centrais da gal\'axia, 
provocado pelo transporte de g\'as  ao longo da barra, d\'a origem a \'orbitas estelares 
irregulares que transportam estrelas do disco para o bojo (e.g., Berentzen {\it et al.} 1998). 
Assim, temos, por um lado, o transporte de g\'as para as regi\~oes centrais das gal\'axias que
pode produzir surtos de forma\cao estelar e o enriquecimento qu\1mico destas regi\~oes. E, por outro lado,
o transporte de estrelas do disco para o bojo via aquecimento vertical da barra e a indu\cao de \'orbitas 
estelares irregulares. Dessa forma, os processos evolutivos relacionados a barras estelares podem ser 
respons\'aveis pela constru\c c\~ao de bojos gal\'acticos. 

Norman, Sellwood \& Hasan (1996), entre outros trabalhos 
te\'oricos, mostram que a concentra\c c\~ao central de massa, induzida pela barra, faz com que as
\'orbitas que a sustentam (entre essas as \'orbitas do tipo $x_{1}$, que s\~ao \'orbitas 
est\'aveis, de alta excentricidade, ao longo do eixo maior da barra) desapare\c cam. 
Portanto, a barra pode ser
destru\1da devido aos processos que ela pr\'opria induz. Estes autores v\~ao mais al\'em, 
sugerindo que a forma\c c\~ao da barra, sua dissolu\c c\~ao e conseq\"uente constru\c c\~ao
do bojo, possa ser um processo recorrente.

O transporte de g\'as para as regi\~oes centrais de gal\'axias, induzido pela barra, tem sido
estudado tamb\'em como uma poss\1vel alternativa para a alimenta\c c\~ao de n\'ucleos ativos 
de gal\'axias.
Atrav\'es de processos din\^amicos (desacoplamento), uma barra secund\'aria pode surgir interna
\`a barra prim\'aria e conduzir o g\'as at\'e pequenas escalas de dist\^ancia, pr\'oximas ao n\'ucleo 
ativo (Shlosman, Frank \& Begelman 1989; Shlosman, Begelman \& Frank 1990). Erwin \& Sparke 
(1998) apresentam gal\'axias onde a barra secund\'aria pode ser identificada. Em NGC 2681,
Erwin \& Sparke (1999) encontram uma hierarquia de tr\^es barras. Embora ainda n\~ao haja um
consenso a respeito da rela\c c\~ao das barras com os n\'ucleos ativos (Ho, 
Filippenko \& Sargent 1997; Knapen 1998), a forma\c c\~ao e dissolu\c c\~ao de barras hier\'arquicas
favorece ainda mais o cen\'ario no qual a constru\c c\~ao de bojos est\'a intimamente ligada
aos processos din\^amicos nos discos gal\'acticos (Friedli \& Martinet 1993). Neste \'ultimo
trabalho, \'e sugerido que a cont\1nua constru\c c\~ao do bojo pode transformar a morfologia
de uma gal\'axia. Assim sendo, uma gal\'axia do tipo Sc pode, atrav\'es destes processos de 
evolu\c c\~ao secular, tornar-se uma gal\'axia do tipo Sb, fazendo com que pelo menos a 
seq\"u\^encia tardia de Hubble ganhe um significado evolutivo entre tipos cont\1guos.

Muitos outros estudos observacionais d\~ao suporte ao cen\'ario de forma\cao de bojos via evolu\cao secular.
A similaridade das cores em bandas largas de bojos e das regi\~oes internas de discos, encontrada
por Peletier \& Balcells (1996) para uma amostra de espirais de tipo jovem, \'e um exemplo, j\'a que 
indica que as idades e as metalicidades das estrelas nessas
regi\~oes s\~ao semelhantes, embora a degeneresc\^encia idade--metalicidade nas cores de popula\coes 
estelares possa trazer incertezas. Courteau, de Jong \& Broeils (1996 -- veja tamb\'em de Jong 1996b) 
mostram que existe 
uma correla\cao entre as escalas de comprimento de bojos e discos para uma amostra de espirais de tipos
jovem e tardio. Estes resultados podem estar indicando que as forma\coes de bojo e disco n\~ao podem
ser fen\^omenos t\~ao distintamente separados como no cen\'ario monol\1tico, mas que deve existir uma
conex\~ao evolutiva entre essas duas componentes. Al\'em disso, os surtos de forma\cao estelar observados
por Carollo \et (1997) em bojos extragal\'acticos s\~ao naturalmente explicados no cen\'ario de evolu\cao
secular.

Outras evid\^encias que d\~ao apoio a este cen\'ario s\~ao encontradas em observa\coes a respeito da
din\^amica e cinem\'atica de bojos. Kormendy (1982) mostra que bojos triaxiais, que s\~ao dinamicamente 
semelhantes a barras, t\^em uma velocidade m\'axima de rota\cao maior do que os bojos de gal\'axias
ordin\'arias e, portanto, s\~ao mais semelhantes a discos. Neste artigo, Kormendy conclui que
``{\it uma fra\cao significativa do bojo em muitas gal\'axias barradas deve consistir de g\'as do 
disco que foi transportado para o centro pela barra. Na medida em que este g\'as se acumula, forma 
estrelas e d\'a origem a uma distribui\cao estelar muito centralmente concentrada, a qual \'e
fotometricamente semelhante a um bojo, mas dinamicamente semelhante a um disco.}''. Kormendy \&
Illingworth (1983) mostram que bojos de gal\'axias barradas t\^em, em geral, uma dispers\~ao central
de velocidade menor do que os bojos de gal\'axias ordin\'arias de mesma luminosidade. Como os
discos t\^em uma dispers\~ao central de velocidade menor do que bojos, este resultado \'e consistente
com a hip\'otese de que os bojos em gal\'axias barradas foram acrescidos com material do disco, 
transportado pela barra. A semelhan\c ca entre a distribui\cao de elipticidades de bojos e discos
tamb\'em foi utilizada como um argumento de que as forma\coes destas duas componentes s\~ao eventos
conexos (Kormendy 1993). Outra evid\^encia de que os processos evolutivos em barras contribuem para
a forma\cao de bojos vem do fato de que estrelas ricas em metais no bojo Gal\'actico possuem 
caracter\1sticas cinem\'aticas de barras, enquanto que as estrelas pobres em metais nesta regi\~ao
n\~ao possuem esta propriedade (Rich \& Terndrup 1997).

Entretanto, Wyse, Gilmore \& Franx (1997) apontam para um problema potencial para a aplicabilidade
geral deste cen\'ario de forma\cao de bojos. Este problema vem do fato de que as barras em gal\'axias
espirais de tipo jovem possuem um perfil de luminosidade constante, enquanto que aquelas em espirais de 
tipo tardio t\^em perfis exponenciais, ou ainda mais abruptos. Por outro lado, a decomposi\cao de
perfis de luminosidade de gal\'axias espirais mostra que os discos e os bojos em espirais tardias
s\~ao melhor ajustados por um perfil exponencial, enquanto que os bojos em espirais de tipo jovem
t\^em um perfil mais pr\'oximo \`a lei $r^{1/4}$ (e.g., Courteau, de Jong \& Broeils 1996). Estes resultados
podem indicar que o cen\'ario de evolu\cao secular somente tem um importante papel na forma\cao de
bojos em espirais tardias.

Por outro lado, analisando a densidade no espa\c co de fase das estrelas no bojo e no disco da Gal\'axia (que
t\^em valores t\1picos para outras gal\'axias tamb\'em), Wyse (1998) mostra que bojos de gal\'axias
n\~ao podem se formar atrav\'es da instabilidade din\^amica de discos estelares puros; uma componente
dissipativa (g\'as) precisa ser invocada para o processo, o que imp\~oe v\1nculos nos modelos que
se baseiam no cen\'ario de evolu\cao secular.

Apesar de Courteau, de Jong \& Broeils (1996) afirmarem que a correla\cao entre os comprimentos de escala de bojos e discos
\'e melhor compreendida em um modelo no qual o disco se forma primeiro e d\'a origem ao bojo, cada vez
mais torna-se claro que os 3 cen\'arios que acabamos de descrever devem ocorrer, e que a forma\cao de bojos
pode ser uma combina\cao destes 3 cen\'arios. No caso da Gal\'axia, 
por exemplo, a pequena diferen\c ca nas idades de aglomerados globulares de mesma metalicidade, e a
tamb\'em pequena varia\cao nas idades destes em rela\cao \`a dist\^ancia Galactoc\^entrica, favorece o
cen\'ario monol\1tico, j\'a que, neste cen\'ario, os aglomerados globulares se formariam rapidamente durante
o colapso da protogal\'axia. No entanto, a exist\^encia de alguns aglomerados globulares Gal\'acticos que
t\^em um comportamento an\^omalo indica que houve, em algum momento, a captura de companheiras
an\~as pela Gal\'axia (Stetson, VandenBergh \& Bolte 1996). Al\'em disso, todas as evid\^encias 
observacionais e os resultados dos estudos num\'ericos descritos acima tornam pouco prov\'avel 
que os processos de 
evolu\cao secular n\~ao tenham ocorrido, ou n\~ao ocorram, na Gal\'axia, principalmente porque muitas 
evid\^encias
apontam para a exist\^encia de uma barra em nosso sistema estelar (e.g., refer\^encias em Binney \& 
Tremaine 1987).

A quest\~ao que permanece para ser respondida por estudos futuros \'e a de determinar a import\^ancia relativa
de cada um destes cen\'arios, e como o papel de cada um deles varia em diferentes condi\coes
f\1sicas, como, por exemplo, em ambientes de diferentes densidades gal\'acticas. Certamente, o
estudo de gal\'axias em altos ``redshifts'', ou seja, a observa\cao de gal\'axias no processo de
forma\c c\~ao, trar\'a progressos na tentativa de obter estas respostas. Por exemplo, Abraham \et 
(1994,1996) e van den Bergh \et (1996) fazem um estudo das propriedades morfol\'ogicas de gal\'axias
no ``Hubble Deep Field'', utilizando um sistema de classifica\cao baseado na concentra\cao central
de luz e na assimetria dos objetos. Estes autores concluem que: {\bf(i)} gal\'axias barradas s\~ao
raras no ``Hubble Deep Field''; {\bf(ii)} a fra\cao de espirais de tipo jovem (i.e., com alta 
raz\~ao bojo/disco) \'e similar \`a mesma fra\cao no Universo local, e {\bf(iii)} a fra\cao de
objetos peculiares, ou em intera\cao ou fus\~ao \'e significativamente maior do que entre as gal\'axias
pr\'oximas. Este resultado pode favorecer o cen\'ario hier\'arquico. No entanto, Marleau \& Simard 
(1998) obt\^em par\^ametros estruturais para as gal\'axias no ``Hubble Deep Field'', atrav\'es da
decomposi\cao do perfil de brilho superficial destas gal\'axias em um perfil exponencial para o disco
e um perfil de S\'ersic para o bojo. Os resultados obtidos por Marleau e Simard contradizem os
descritos acima, j\'a que estes autores encontram que somente 8\% dos objetos t\^em bojos dominantes.
Este resultado claramente favorece um cen\'ario no qual a forma\cao do bojo ocorre posterior \`a do
disco, tal como no cen\'ario de evolu\cao secular.

\section{Gradientes de cor e o cen\'ario de evolu\cao secular -- Motiva\cao do trabalho}

\hskip 30pt O objetivo deste estudo \'e o de verificar a exist\^encia de correla\coes que possam
indicar a ocorr\^encia dos efeitos de evolu\cao secular em gal\'axias barradas, apresentados na
se\cao 1.1.3. Para tanto, realizamos duas principais abordagens distintas, por\'em complementares.
A primeira delas consiste, 
essencialmente, na determina\cao e an\'alise dos gradientes de cor de uma amostra de gal\'axias 
barradas e ordin\'arias. A segunda abordagem consiste na determina\cao e na an\'alise dos par\^ametros 
caracter\1sticos dos perfis de luminosidade de bojos e discos de parte de nossa amostra principal 
de gal\'axias. Na primeira abordagem, para determinar os gradientes de cor, utilizamos dados obtidos 
na literatura, enquanto que, na segunda abordagem, a decomposi\cao bojo/disco foi realizada atrav\'es 
de imagens digitalizadas do DSS (``Digitized Sky Survey''), adquiridas em placas fotogr\'aficas.
Finalmente, 
realizamos um estudo comparativo, com dados adquiridos por n\'os, em CCD (``Charge--Coupled Device''), 
em miss\~oes observacionais realizadas
no Observat\'orio do Pico dos Dias (Laborat\'orio Nacional de Astrof\1sica -- OPD/LNA).

\paragraph{Gradientes de cor (Cap\1tulo 2).} Uma importante ferramenta que pode trazer pistas a 
respeito dos diferentes processos evolutivos que ocorrem em gal\'axias \'e a an\'alise de \1ndices de cor. 
Veremos na se\cao 2.1 como estes \1ndices nos proporcionam informa\coes a respeito da popula\cao 
estelar e da hist\'oria de forma\cao estelar em gal\'axias. Uma conseq\"u\^encia natural do cen\'ario
de evolu\cao secular \'e a homogeneiza\cao das popula\coes estelares ao longo de bojo e disco, que deve
se manifestar, de maneira an\'aloga, na homogeneiza\cao dos \1ndices de cor ao longo da gal\'axia.
Para verificar este efeito, determinamos os gradientes de cor (B\,-V) e (U-B) para uma amostra de 257
gal\'axias de tipos morfol\'ogicos Sb, Sbc e Sc, barradas e ordin\'arias. Os gradientes foram calculados atrav\'es de
m\'etodos estat\1sticos robustos, utilizando dados da literatura, obtidos atrav\'es de fotometria 
fotoel\'etrica de abertura.

\paragraph{Fotometria superficial (Cap\1tulo 3).} Neste cap\1tulo, o objetivo \'e verificar, 
atrav\'es de t\'ecnicas de fotometria superficial, correla\coes que possam contribuir para corroborar, ou n\~ao, 
o cen\'ario de evolu\cao secular em gal\'axias barradas. Este cen\'ario prov\^e mecanismos
para a forma\cao e/ou constru\cao de bojos em gal\'axias espirais\footnote{Deixemos clara a distin\cao
entre os termos ``forma\c c\~ao'' e ``constru\c c\~ao'' de bojos. A forma\cao do bojo pressup\~oe a anterior
inexist\^encia desta componente. Por sua vez, a constru\cao do bojo se refere ao seu crescimento e,
portanto, necessita que esta componente j\'a exista, independentemente de qual tenha sido o seu processo
de forma\c c\~ao.}. 
Uma conseq\"u\^encia natural que seria esperada para as gal\'axias que sofreram estes efeitos de evolu\cao 
secular \'e a de que a componente bojo cresceria. Al\'em disto, como conseq\"u\^encia destes processos, dever\1amos esperar 
fortes correla\coes entre as dimens\~oes de bojo e disco, 
j\'a que estas duas componentes teriam seus processos de evolu\cao compartilhados. 
Portanto, uma correla\cao compat\1vel com o cen\'ario de evolu\cao secular seria a de que aquelas 
gal\'axias cujas popula\coes estelares de bojo e disco
s\~ao similares seriam tamb\'em aquelas que apresentam bojos mais proeminentes. 
Para verificar esta correla\c c\~ao, realizamos decomposi\coes bojo/disco 
para 39 gal\'axias, que constituem uma sub-amostra representativa de nossa amostra inicial.
O algoritmo de decomposi\cao ajusta perfis de brilho distintos para as componentes
bojo e disco ao perfil de brilho total da gal\'axia, fornecendo-nos par\^ametros caracter\1sticos, 
tais como o raio efetivo do bojo, o brilho superficial efetivo do bojo e a raz\~ao entre as luminosidades de
bojo e disco. \'E importante enfatizar que o algoritmo de decomposi\cao utiliza a imagem das gal\'axias
e, portanto, realiza um ajuste bi-dimensional, raramente apresentado na literatura. As imagens utilizadas
nessa abordagem s\~ao do DSS.
Como os gradientes de cor foram determinados com dados 
obtidos atrav\'es de fotometria fotoel\'etrica de abertura, e a decomposi\cao bojo/disco foi realizada em
imagens digita-lizadas, adquiridas em placas fotogr\'aficas, \'e relevante verificar se os
resultados assim obtidos s\~ao compat\1veis com aqueles obtidos atrav\'es do imageamento em CCD, j\'a que
essa \'e uma t\'ecnica mais moderna e refinada. Portanto, realizamos o imageamento em CCD para 14 gal\'axias 
de nossa amostra no OPD/LNA. Com essas imagens, os gradientes de cor foram determinados, e a decomposi\cao 
bojo/disco foi realizada, da mesma maneira como foi efetuado para o restante de nossa amostra. Esses dados
nos permitiram realizar um estudo comparativo, apresentado na se\cao 3.5.

Finalizando esta Disserta\c c\~ao, o Cap\1tulo 4 apresenta nossas conclus\~oes finais e nossas perspectivas
futuras para este trabalho.
   \newpage \chapter{Gradientes de Cor}

\hskip 30pt Neste cap\1tulo, apresentamos um estudo estat\1stico do comportamento dos \1ndices de 
cor ao longo da dire\cao radial de gal\'axias espirais tardias (Sb's, Sbc's e Sc's). O objetivo principal \'e o de
verificar, atrav\'es da informa\cao da distribui\cao de cor, e portanto, em princ\1pio, 
da popula\cao estelar, como os gradientes de cor se comportam nas gal\'axias 
da amostra. Em particular, estamos interessados em verificar a presen\c ca de gradientes de cor nulos, 
ou seja, \1ndices de cor semelhantes em bojos e discos, que indicariam, portanto, a homogeneiza\cao 
das popula\coes estelares, como previsto por processos de evolu\cao secular em gal\'axias barradas (ver
se\cao 1.1.3).
Para este fim, selecionamos
uma amostra de 257 gal\'axias dos tipos Sb, Sbc e Sc, ordin\'arias e barradas, com medidas dos \1ndices
de cor (B\,-V) e (U-B) ao longo de cada uma delas, publicadas em ``A General Catalogue of Photoeletric
Magnitudes and Colors of Galaxies in the U, B, V System'' (Longo \& A. de Vaucouleurs 1983,1985 -- 
doravante LdV83,85).

No decorrer deste cap\1tulo, vamos mostrar que, na maioria das gal\'axias, os \1ndices de 
cor variam suavemente da regi\~ao central 
para a regi\~ao perif\'erica, caracterizando, assim, um gradiente de cor. Veremos tamb\'em
que a maior parte das gal\'axias em nossa amostra apresenta um acentuado gradiente de cor negativo, 
i.e., t\^em \1ndices mais vermelhos nas regi\~oes centrais (bojo), e \1ndices mais azuis nas regi\~oes
perif\'ericas (disco).

A diferen\c ca entre os \1ndices de cor dos bojos e discos indica que a popula\cao estelar
dominante varia ao longo das gal\'axias. Este fato est\'a em acordo com a previs\~ao do cen\'ario
monol\1tico de forma\cao de gal\'axias espirais (ver se\cao 1.1.1), que identifica essa diferen\ca nas
popula\coes estelares de bojos e discos como conseq\"u\^encia de que a forma\cao destas componentes se
d\'a de forma distinta e separada.

Por outro lado, uma das conseq\"u\^encias naturais do cen\'ario de evolu\cao
secular em gal\'axias barradas
(ver se\cao 1.1.3) \'e a homogeneiza\cao das popula\coes estelares ao longo do bojo e disco, j\'a que 
os processos de evolu\cao promovem uma conex\~ao entre essas duas componentes.
Desde que haja uma quantidade
suficiente de g\'as no disco e desde que a barra esteja presente durante um intervalo
de tempo suficiente para que esse
g\'as seja transportado para as regi\~oes centrais da gal\'axia, dando origem a\1 a surtos de
forma\cao estelar, as popula\coes do bojo e disco dever\~ao ser mais semelhantes entre si do que aquelas
em uma gal\'axia que n\~ao sofreu tais processos. Essa homogeneiza\cao \'e ainda mais acentuada na 
medida em que a barra induz o transporte de estrelas do disco para o bojo atrav\'es do esquentamento
vertical do disco. Evidentemente, a homogeneiza\cao das popula\coes estelares deve se refletir de modo
an\'alogo no comportamento dos \1ndices de cor ao longo da gal\'axia. Assim sendo, uma gal\'axia que sofreu tais
processos deve apresentar \1ndices de cor semelhantes no bojo e no disco, ou seja, deve apresentar um gradiente de cor
aproximadamente nulo ou menos acentuado. De fato, cerca de 25\% das gal\'axias em nossa amostra 
apresentam este comportamento. Veremos que h\'a, entre essas gal\'axias, um excesso estatisticamente
significativo de gal\'axias barradas. Tamb\'em veremos que os \1ndices de cor destas gal\'axias s\~ao
similares aos \1ndices dos discos das gal\'axias com o t\1pico gradiente negativo. Estes resultados
podem estar indicando que, pelo menos em algumas gal\'axias, a barra atua unificando as popula\coes 
estelares de bojos e discos, em acordo com o cen\'ario de evolu\cao secular.

Na pr\'oxima se\c c\~ao, introduzimos os conceitos b\'asicos que fundamentam este
estudo. As se\coes 2.2 e 2.3 descrevem, respectivamente, a sele\cao da amostra estudada
e a determina\cao dos
gradientes de cor, bem como dos \1ndices de cor caracter\1sticos do bojo e do disco de cada gal\'axia. 
Os resultados s\~ao apresentados em 2.4, e analisados e discutidos na se\cao 2.5, que tamb\'em
traz um resumo de nossas conclus\~oes.

\section{Magnitudes, cores e popula\coes estelares}

\hskip 30pt As 1022 estrelas catalogadas no ``Almagesto'' de C. Ptolomeu est\~ao divididas  em
6 grupos distintos, ou magnitudes, de acordo com o brilho aparente desses objetos, de modo que as
estrelas mais brilhantes s\~ao de magnitude 1 e as estrelas mais t\^enues que podem ser vistas pelo
olho desarmado s\~ao de magnitude 6. Esta obra foi elaborada na primeira metade
do s\'eculo II A.D. e \'e uma compila\cao de todo o conhecimento de Astronomia adquirido at\'e ent\~ao
na Gr\'ecia Antiga. A ess\^encia dessa classifica\cao do brilho das estrelas perdurou e \'e utilizada at\'e
hoje. Acreditava-se, na \'epoca, que essa divis\~ao em magnitudes correspondia a
diferen\cas equivalentes em luminosidade, mas como a resposta do olho humano \`a radia\cao \'e 
logar\1tmica, i.e., nossos
olhos reconhecem raz\~oes iguais de luminosidade, mas n\~ao incrementos iguais na luminosidade\footnote
{Esta rela\cao \'e conhecida como a lei de Fechner dos est\1mulos.}, a escala de
magnitudes resulta ser uma escala logar\1tmica. Estudos posteriores por N. Pogson em 1856 mostraram que, de fato,
uma estrela de sexta magnitude \'e cerca de 100 vezes menos brilhante que uma de magnitude 1, mas somente
2.5 vezes menos brilhante que uma estrela de magnitude 5. Levando
estes fatos em considera\c c\~ao, podemos escrever:

\eq
m_{1} - m_{2} = {{-2.5\log_{10}}{\left({f_{1}}\over f_{2}\right)}}
\eeq

\noindent ou

\eq
{f_{1}\over f_{2}} = {10^{-0.4(m_{1} - m_{2})}},
\eeq

\noindent onde $m_{1}$ e $m_{2}$ s\~ao as magnitudes de duas estrelas distintas com fluxos de 
energia irradiada 
iguais a $f_{1}$ e $f_{2}$, respectivamente. Pode-se verificar imediatamente atrav\'es da
equa\cao (2.1) que a escala de magnitudes \'e uma escala relativa, e portanto, deve
ser fixa, arbitrando-se a magnitude de estrelas definidas como padr\~oes.

No entanto, a atmosfera terrestre absorve parte dos f\'otons emitidos pela estrela, e outros f\'otons s\~ao
absorvidos pelo sistema detector utilizado na observa\c c\~ao. Assim,
sendo $f_{\nu}$ o fluxo aparente de uma estrela em um intervalo de freq\"u\^encias, 
podemos expressar o fluxo integrado aparente $f$ em fun\cao do fluxo intr\1nseco em um intervalo de
freq\"u\^encias $f_{\nu}^{0}$ por:

\eq
f = {\int \limits_{0}^{\infty}f_{\nu}{\displaystyle d}{\nu}} = {\int \limits_{0}^{\infty}f_{\nu}^{0}
T_{\nu}F_{\nu}R_{\nu}{\displaystyle d}{\nu}},
\eeq

\noindent onde $T_{\nu}$ \'e a fra\cao de f\'otons transmitida pela atmosfera, $F_{\nu}$ \'e a
transmiss\~ao do sistema de filtros utilizado, $R_{\nu}$ \'e a efici\^encia na detec\c c\~ao de 
f\'otons pelo sistema de detec\c c\~ao do telesc\'opio e ${\nu}$ \'e a freq\"u\^encia da radia\cao
emitida.

Evidentemente, a fra\cao de f\'otons absorvida pela atmosfera \'e menor quando a estrela encontra-se
no z\^enite e aumenta em
dire\cao ao horizonte. De fato, a atenua\cao da luz emitida por uma estrela ao atravessar a atmosfera
segue a seguinte rela\c c\~ao:

\eq
m_{z} - m_{0} = k_{\nu} \sec z,
\eeq

\noindent onde $m_{0}$ \'e a magnitude da estrela no z\^enite e $m_{z}$ \'e a sua magnitude na
dist\^ancia zenital $z$. A constante $k_{\nu}$ \'e determinada ao se observar a magnitude de uma estrela
em v\'arias dist\^ancias zenitais. Conhecendo esta constante, podemos extrapolar a equa\cao (2.4) para
obter o valor da magnitude da estrela no caso em que a observa\cao fosse realizada acima da atmosfera
(``$\sec z = 0$''). Como a extin\cao atmosf\'erica varia de uma noite para outra, o valor de
$k_{\nu}$ deve
ser determinado em cada noite de observa\c c\~ao. Al\'em disso, 
como $k_{\nu}$ varia com o comprimento de onda da luz
observada, deve ser determinado para cada filtro utilizado. De fato, o efeito da
atmosfera terrestre n\~ao \'e somente atenuar a luz das estrelas, mas tamb\'em torn\'a-la mais
avermelhada, j\'a que espalha com maior efici\^encia luz com comprimentos de onda mais curtos.
Ap\'os determinar $k_{\nu}$, podemos corrigir os efeitos da extin\cao atmosf\'erica para todos
os outros objetos observados durante a noite, desde que conhe\c camos tamb\'em as dist\^ancias
zenitais em que estes foram observados.

A corre\cao para os outros termos da equa\cao (2.3), $F_{\nu}$ e $R_{\nu}$, \'e realizada ao
mesmo tempo em que se transforma as magnitudes obtidas para a escala padr\~ao, i.e., quando se
calibra os dados obtidos utilizando-se estrelas--padr\~ao observadas durante a noite, de modo que as
magnitudes instrumentais obtidas para essas estrelas possam ser comparadas
 \`as que est\~ao publicadas na literatura.
Por exemplo, Graham (1982) apresenta estrelas--padr\~ao para observa\coes no hemisf\'erio sul.

O fato de o meio interestelar ser preenchido por g\'as e poeira em baixas
densidades, faz com que tenhamos que corrigir as magnitudes estelares por ainda mais um 
fator, que \'e a extin\cao interestelar. Os gr\~aos de poeira presentes entre as estrelas absorvem mais fortemente 
a luz de comprimentos de onda mais curtos (como a luz ultravioleta, por exemplo), transformando os f\'otons
absorvidos em luz de comprimentos de onda mais longos (infravermelho). Por outro lado, a luz de comprimentos
de onda mais longos atravessa as camadas de poeira sem sofrer muita absor\c c\~ao. Portanto, o efeito da extin\cao 
interestelar \'e o de avermelhar a luz emitida pelas estrelas.

Como este trabalho versa sobre gal\'axias, precisamos conhecer o efeito da distribui\cao da poeira
interestelar na Gal\'axia sobre a luz emitida por estes objetos. Este efeito pode ser parametrizado
pelo o que se conhece como excesso de cor. Este par\^ametro corrige os valores observados dos efeitos 
da extin\cao (ou avermelhamento) Gal\'actica e pode ser descrito como uma fun\cao das coordenadas
gal\'acticas (l,b) do objeto. Al\'em disso, como as gal\'axias que estudamos aqui 
tamb\'em possuem gr\~aos
de poeira no seu pr\'oprio meio interestelar, tamb\'em precisamos fazer uma corre\cao pelos efeitos 
desta extin\cao (ou 
avermelhamento) intr\1nseca, i.e., pelos efeitos da absor\cao de luz pela poeira presente em cada uma das 
gal\'axias. Trataremos do avermelhamento Gal\'actico e do avermelhamento intr\1nseco mais adiante, na
determina\cao dos gradientes de cor (se\cao 2.3) e dos \1ndices de cor carater\1sticos de bojos e discos
(se\cao 2.3.1).

O sistema fotom\'etrico utilizado neste trabalho \'e o sistema UBV desenvolvido por Johnson e Morgan
em 1953. Consiste em 3 filtros que restringem a detec\c c\~ao da radia\cao emitida pela estrela
a uma determinada faixa de freq\"u\^encias, e cujas curvas de resposta podem ser vistas na Figura 2.1.
O comprimento de onda efetivo e a largura FWHM (``full width at half maximum'' -- entre par\^enteses)
dos filtros U, B e V
s\~ao, respectivamente, 365 (66), 445 (94) e 551 (88), em nanometros.

\begin{figure}
\epsfxsize=15cm
\centerline{\epsfbox{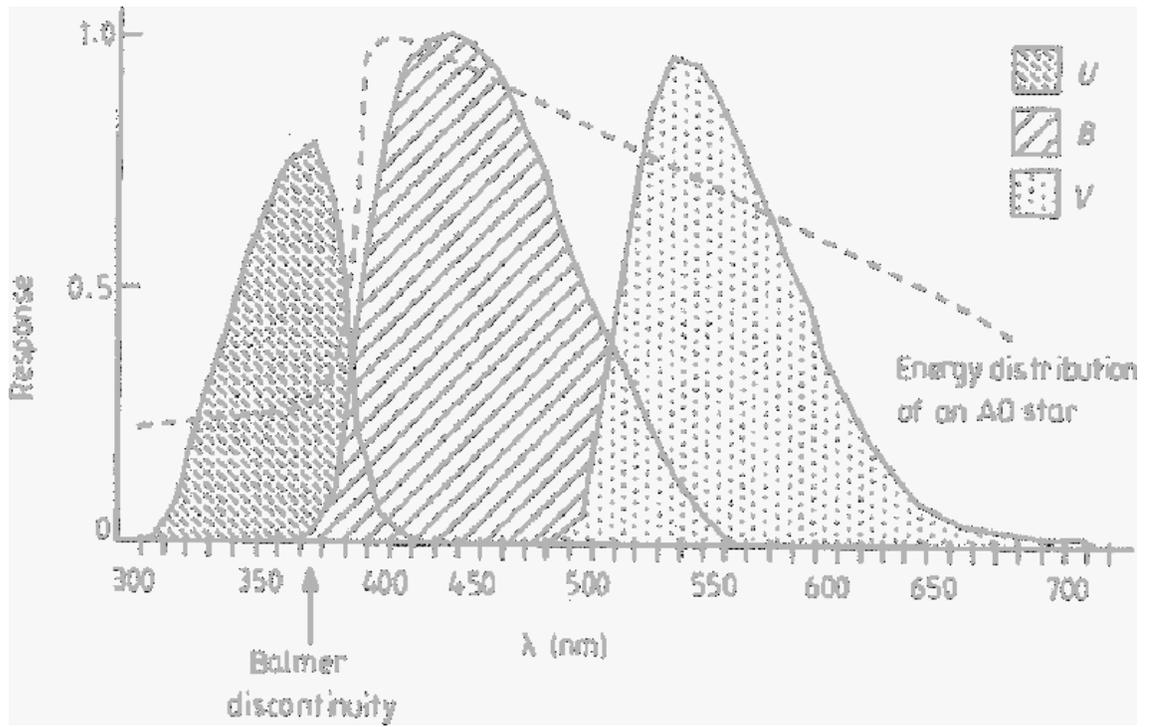}}
\caption{Curvas de resposta para os filtros U, B e V de Johnson e Morgan. No eixo das abscissas est\'a
o comprimento de onda em nanometros, e no eixo das ordenadas a resposta normalizada. A curva tracejada
indica uma distribui\cao espectral de energia t\1pica de uma estrela de tipo A0. A posi\cao da
descontinuidade de Balmer tamb\'em est\'a destacada. (Extra\1do de Kitchin 1998).} 
\label{ubv}
\end{figure}

Uma estrela vermelha \'e mais brilhante (possui magnitude menor) no filtro V do que no filtro B.
Como, inversamente,  uma estrela azul possui magnitude menor no filtro B do que no filtro V,
a diferen\ca entre essas duas magnitudes, denominada \1ndice de cor (B\,-V), \'e uma
medida da cor da estrela, pois a estrela vermelha ter\'a um \1ndice (B\,-V) maior do que a
azul. De fato, o \1ndice (B\,-V) de uma estrela azul, com tipo espectral mais jovem do que A0,  
ser\'a negativo, enquanto que o da vermelha ser\'a positivo. Assim, o \1ndice de cor (B\,-V) \'e
definido como

\eq
(B - V) \equiv m_{B} - m_{V} = c -2.5\log_{10} {{\int_{0}^{\infty} f_{\nu}T_{\nu}(B)F_{\nu}(B)R_{\nu}(B)
{\displaystyle d}{\nu}}\over{\int_{0}^{\infty} f_{\nu}T_{\nu}(V)F_{\nu}(V)R_{\nu}(V){\displaystyle d}{\nu}}}.
\eeq

De maneira an\'aloga se define o \1ndice de cor (U-B). A constante $c$ \'e definida de modo a se adequar \`a
escala padr\~ao de magnitudes, seguindo a conven\cao de que uma estrela de tipo espectral A0, 
como $\alpha$ Lyr (Vega), tenha a
mesma magnitude em todos os comprimentos de onda, e portanto, $(U-B) = (B-V) \equiv 0$ para
estrelas do tipo A0.

Assim sendo, o \1ndice de cor \'e uma medida da raz\~ao de fluxos em duas bandas fotom\'etricas
distintas e \'e, portanto, independente da dist\^ancia do objeto. Al\'em disso, como depende da forma
do espectro estelar, \'e uma medida aproximada da distribui\cao espectral da energia emitida
pela estrela. Como pode ser visto na Figura 2.2, o \1ndice de cor guarda realmente estreita rela\cao
com o tipo espectral.

\begin{figure}
\epsfxsize=15cm
\centerline{\epsfbox{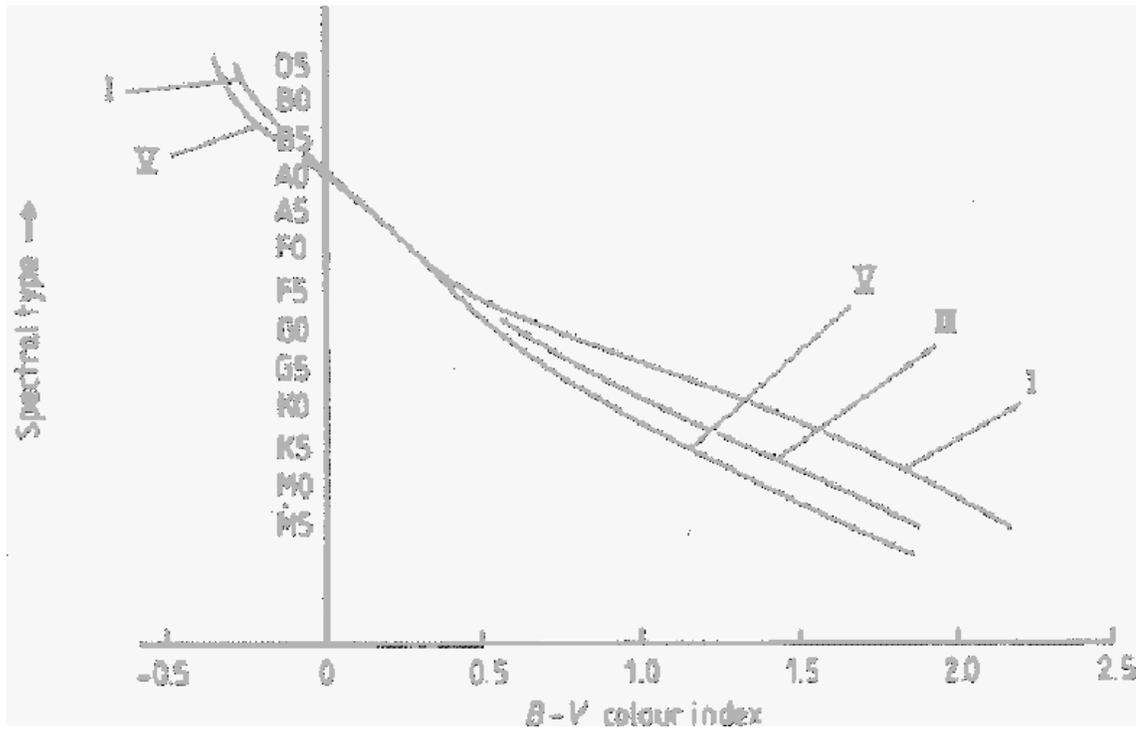}}
\caption{Rela\cao entre o tipo espectral e o \1ndice de cor (B\,-V) (Extra\1do de Kitchin 1998).} 
\label{tipespec}
\end{figure}

A distribui\cao espectral de energia emitida por uma estrela depende de alguns par\^ametros f\1sicos,
tais como temperatura, gravidade superficial e composi\cao qu\1mica. Como estrelas frias s\~ao vermelhas
e estrelas quentes s\~ao azuis, o \1ndice (B\,-V) \'e uma medida da temperatura da estrela. Al\'em
disso, como muitas linhas espectrais de absor\c c\~ao, relativas a elementos pesados, est\~ao presentes
na banda U, enquanto que a banda B \'e relativamente livre de linhas, o \1ndice (U-B) reflete, ainda
que grosseiramente, a composi\cao qu\1mica da estrela.
Estes mesmos \1ndices tamb\'em s\~ao afetados pela idade de uma estrela. Estrelas jovens tendem a ser 
mais azuladas, e se tornam mais vermelhas conforme evoluem no Diagrama HR.
Outra caracter\1stica do espectro estelar que
interefere no valor observado do \1ndice (U-B) \'e a descontinuidade de Balmer. Essa descontinuidade
consiste em uma queda abrupta na distribui\cao espectral de energia de uma estrela para comprimentos
de onda menores do que 364 nanometros. Essa queda \'e provocada pela absor\cao do cont\1nuo por
\'atomos de Hidrog\^enio no n\1vel n = 2. A descontinuidade de Balmer atinge um m\'aximo para
estrelas de tipo A0, na seq\"u\^encia principal, e para estrelas de tipo F0 para supergigantes.
Por outro lado, pode ser desprezada para estrelas mais quentes (tipos O e B) e mais frias (tipos 
G0 e os ainda mais frios). Na Figura 2.1 pode se observar a descontinuidade de Balmer em uma 
distribui\cao espectral de energia t\1pica de uma estrela A0.

A partir destas considera\coes podemos concluir que o \1ndice de cor na regi\~ao central de uma
gal\'axia, por exemplo, reflete o tipo espectral m\'edio das estrelas que dominam a emiss\~ao de luz nessa 
regi\~ao. Mais ainda, se existe uma varia\cao no \1ndice de cor ao longo de uma gal\'axia,
ent\~ao existe tamb\'em uma varia\cao na popula\cao estelar dominante ao longo desta gal\'axia, 
seja devido \`a idade m\'edia da popula\c c\~ao, \`a sua metalicidade, ou a ambos os efeitos.
Portanto, a maioria das gal\'axias, que apresenta gradientes de cor negativos, com os \1ndices
de cor diminuindo sistematicamente do centro para a periferia, possui popula\coes estelares bastante
distintas entre as regi\~oes centrais, dominadas pelo bojo, e as regi\~oes perif\'ericas, 
dominadas pelo disco. Por outro lado,
a homogeneiza\cao das popula\coes estelares ao longo de bojo e disco, promovida pelos processos de 
evolu\cao secular, pode provocar a homogeneiza\cao dos \1ndices de cor ao longo dessas duas componentes, 
de forma a atenuar os gradientes de cor negativos, tornando-os menos acentuados ou at\'e mesmo nulos.
Portanto, a observa\cao de que uma gal\'axia possui um gradiente de cor nulo pode ser um forte ind\1cio de 
que esta gal\'axia sofreu os processos de evolu\cao secular descritos na se\cao 1.1.3.

\section{Sele\cao da amostra}

\hskip 30pt Os dados que utilizamos para determinar os gradientes de cor foram extra\1dos do LdV83,85.
Na primeira etapa de sele\cao da amostra, somente utilizamos o LdV83, que se encontra dispon\1vel
eletronicamente no CDS (``Centre de Donn\'ees Astronomiques de Strasbourg'').
Este cat\'alogo \'e uma compila\cao de 16.680 observa\coes (fotometria fotoel\'etrica de abertura) 
de 3578 gal\'axias, coletadas em 150 fontes distintas, desde 1936 at\'e 1982.
S\~ao apresentados, para cada gal\'axia e
para diafragmas de diferentes aberturas, os \1ndices de cor (B\,-V) e (U-B), a magnitude V aparente,
o di\^ametro e o tipo morfol\'ogico da gal\'axia, bem como a fonte de observa\c c\~ao.

Selecionamos todas as gal\'axias no LdV83 com \1ndice do est\'agio de Hubble T = 3, 4 e 5, i.e.,
tipos morfol\'ogicos Sb, Sbc e Sc, ordin\'arias e barradas, com magnitudes menores do que 14 na banda B\footnote{
Este crit\'erio nos ajuda a garantir que a classifica\cao morfol\'ogica da gal\'axia \'e confi\'avel,
j\'a que sistemas mais t\^enues s\~ao, em geral, mais dif\1ceis
de serem classificados. De qualquer forma, a inclus\~ao de objetos com maiores magnitudes n\~ao aumentaria
substancialmente nossa amostra.},
segundo o RC3. Esta sele\cao resultou em 531 objetos com 2458 medidas atrav\'es de aberturas distintas. 
Posteriormente, inclu\1mos
os dados contidos no LdV85, seguindo os mesmos crit\'erios. Tamb\'em foram feitas as 
corre\coes apontadas neste suplemento, que dizem respeito a erros cometidos durante a confec\cao
do LdV83.

Em seguida, retiramos da amostra aquelas gal\'axias para as quais o \1ndice (B\,-V) foi medido em menos
do que 5 aberturas distintas. Assim, somente trabalhamos com aquelas gal\'axias para as quais um estudo
cuidadoso e completo do comportamento dos \1ndices de cor ao longo das componentes bojo e disco p\^ode
ser realizado. Al\'em disso, realizamos uma inspe\cao visual de toda a amostra, utilizando imagens do
DSS, eliminando objetos
peculiares, que apresentam perturba\c c\~oes, tais como faixas de poeira consp\1cuas e companheiras em forte
intera\c c\~ao, que poderiam prejudicar a an\'alise. Ap\'os este \'ultimo passo, a amostra final
consiste em 257 gal\'axias com 2906 medidas dos \1ndices de cor atrav\'es de aberturas distintas.

A escolha deste intervalo de classe morfol\'ogica est\'a vinculada ao fato de que estes sistemas s\~ao os mais
luminosos na banda B ao longo da seq\"u\^encia de Hubble (Roberts \& Haynes 1994; van den Bergh 1997), 
o que pode indicar, portanto, que s\~ao os objetos que apresentam as maiores taxas de forma\cao estelar na 
seq\"u\^encia das espirais. Al\'em disto, alguns estudos indicam que os objetos destas classes podem ser 
os que possuem maior 
instabilidade din\^amica, e portanto s\~ao mais vulner\'aveis \`a forma\cao de barras.
Mais ainda, outros estudos sugerem que os efeitos de evolu\cao secular somente s\~ao importantes
para gal\'axias de tipo tardio.
Vale notar que v\'arias gal\'axias foram distintamente classificadas no LdV83,85 e no RC3. Como a
incerteza no \1ndice do est\'agio de Hubble \'e da ordem de 2 (Lahav \et 1995), podemos considerar as
gal\'axias da presente amostra como pertencentes a uma \'unica sub-classe morfol\'ogica (T = 4 $\pm$ 1).

\section{C\'alculo dos gradientes de cor}

\hskip 30pt Os gradientes de cor (U-B) e (B\,-V) das gal\'axias da amostra foram determinados atrav\'es
dos \1ndices de cor (U-B) e (B\,-V) apresentados no LdV83,85 para v\'arias aberturas distintas ao longo de cada
gal\'axia.

Como o LdV83,85 \'e uma compila\cao de dados adquiridos por v\'arios observadores, utilizando
diferentes instrumentos e telesc\'opios, e em diferentes condi\coes atmosf\'ericas, \'e natural que 
algumas medidas pare\c cam inconsistentes entre si, j\'a que alguns dados podem estar mais perturbados
por erros internos. Por exemplo, autores diferentes podem fornecer valores bastante distintos do \1ndice
(U-B) para a mesma gal\'axia e no diafragma de mesma abertura. De fato, este \'e o caso de NGC 2377 
no diafragma de 
abertura de 2.6 minutos de arco, em que 3 fontes distintas atribuem ao \1ndice (U-B) os valores
0.11, 0.20 e 0.38! Tentar ajustar uma reta ao gradiente de cor em magnitudes, assumindo que os \1ndices 
de cor variam 
linearmente com o logaritmo da abertura, usando estes valores, atrav\'es da mais comumente utilizada e 
cl\'assica regress\~ao linear pelo m\'etodo dos M\1nimos Quadrados (MQ), resultar\'a em valores bastante incertos.

Assim sendo, torna-se muito importante o uso de t\'ecnicas estat\1sticas robustas para minimizar estas
incertezas. Escolhemos aplicar o m\'etodo da M\1nima Mediana dos Quadrados (MMQ), que foi apresentado pela 
primeira vez por Rousseeuw (1984). Este m\'etodo minimiza a {\em mediana} dos quadrados dos res\1duos, 
no lugar da soma dos quadrados dos res\1duos, como faz a cl\'assica regress\~ao linear pelo m\'etodo dos
m\1nimos quadrados. Os resultados obtidos podem resistir a at\'e quase 50\% de contamina\cao nos dados.
A Figura 2.3 ilustra a robustez do m\'etodo da m\1nima mediana (MMQ).

\begin{figure}[thb]
\epsfxsize=15cm
\centerline{\epsfbox{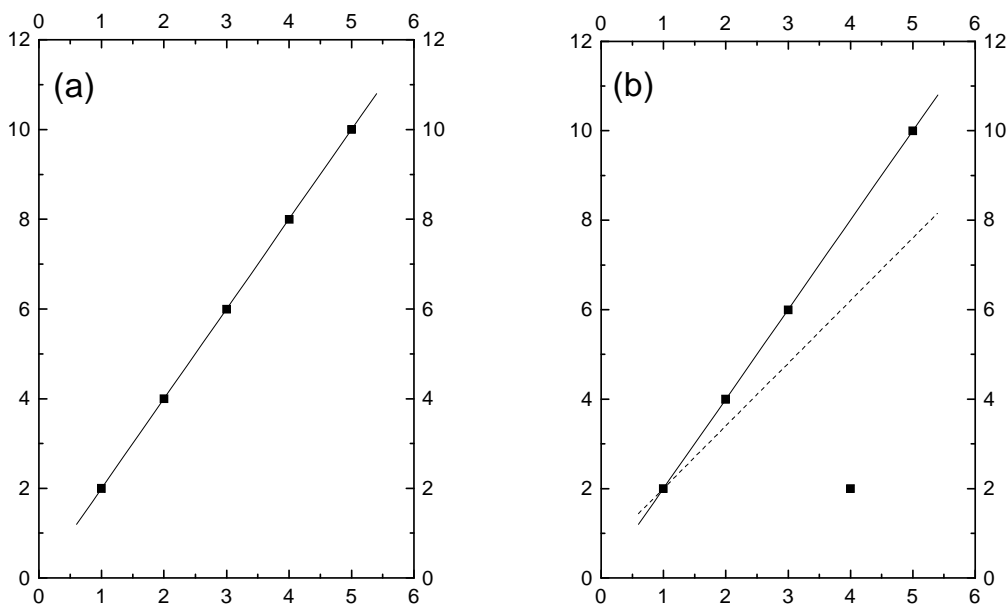}}
\caption{Um exemplo da robustez do m\'etodo de regress\~ao linear pela m\1nima mediana. Em (a), o resultado do
ajuste ao conjunto de pontos \'e indiferente do m\'etodo. Em (b), o m\'etodo cl\'assico (linha tracejada) \'e 
gravemente perturbado pela contamina\cao de um \'unico ponto, enquanto que o m\'etodo da m\1nima mediana 
(linha cheia)  
permanece atribuindo aos pontos o mesmo ajuste.} 
\label{lms}
\end{figure}

Para calcular os gradientes de cor das gal\'axias em nossa amostra, aplicamos o programa {\sc progress}, 
desenvolvido por Rousseeuw \& Leroy (1987). Este programa exe-cuta uma an\'alise robusta da regress\~ao
linear, e al\'em de usar o m\'etodo MMQ, fornece par\^ametros de regress\~ao
confi\'aveis, e permite identificar pontos que pare\c cam inconsistentes com o restante dos dados
(``outliers''). {\sc progress} primeiro determina os par\^ametros da regress\~ao atrav\'es do m\'etodo 
dos m\1nimos quadrados, e depois atrav\'es do m\'etodo da m\1nima mediana dos quadrados. Com esses resultados, o
algoritmo \'e capaz de separar pontos que estejam contaminando os dados e, atribuindo-lhes peso zero, 
recalcula o gradiente atrav\'es do m\'etodo dos m\1nimos quadrados. Com este algoritmo, os gradientes
obtidos t\^em, na maioria das vezes, os mesmos valores que aqueles obtidos quando se aplica {\em somente}
o m\'etodo da m\1nima mediana dos quadrados. No entanto, o m\'etodo MQ trabalha melhor
que o m\'etodo MMQ nos casos em que o n\'umero de pontos n\~ao \'e grande, caracter\1stica
importante no caso de nossa amostra. Assim, o algoritmo adotado em {\sc progress} fornece resultados 
mais confi\'aveis.

O c\'alculo de cada gradiente foi acompanhado cuidadosamente, pois, em alguns casos, o resultado da 
regress\~ao pelo m\'etodo cl\'assico \'e mais representativo do gradiente da gal\'axia do que aquele obtido 
atrav\'es do algoritmo de {\sc progress}.
Isto pode acontecer porque, ao tentar minimizar as incertezas, o m\'etodo da m\1nima mediana pode
caracterizar o gradiente global da gal\'axia utilizando somente alguns poucos pontos que se ajustam muito
bem a uma linha reta. Nos casos em que este ajuste \'e muito bom, o algoritmo de {\sc progress} n\~ao \'e 
capaz de recuperar o gradiente global da gal\'axia e descartar somente aqueles pontos que realmente s\~ao
``outliers''. Ao contr\'ario, descarta todos os pontos que n\~ao satisfazem aquele excelente ajuste
caracterizado por somente alguns poucos pontos. Portanto, em alguns casos, definimos o gradiente da gal\'axia
como sendo aquele obtido pelo m\'etodo cl\'assico.

Das 257 gal\'axias de nossa amostra, obtivemos 239 gradientes de cor (B\,-V) e 202 (U-B). O restante dos gradientes
foi desconsiderado, j\'a que: (1) o n\'umero de medidas dispon\1veis \'e insuficiente para uma boa determina\cao
do gradiente, e/ou (2) as medidas s\~ao demasiadamente inconsistentes, resultando em valores muito incertos para
os gradientes. Os gradientes que determinamos para as gal\'axias de nossa amostra, bem como os respectivos
erros, encontram-se na tabela A.1.

\begin{figure}
\epsfxsize=14cm
\centerline{\epsfbox{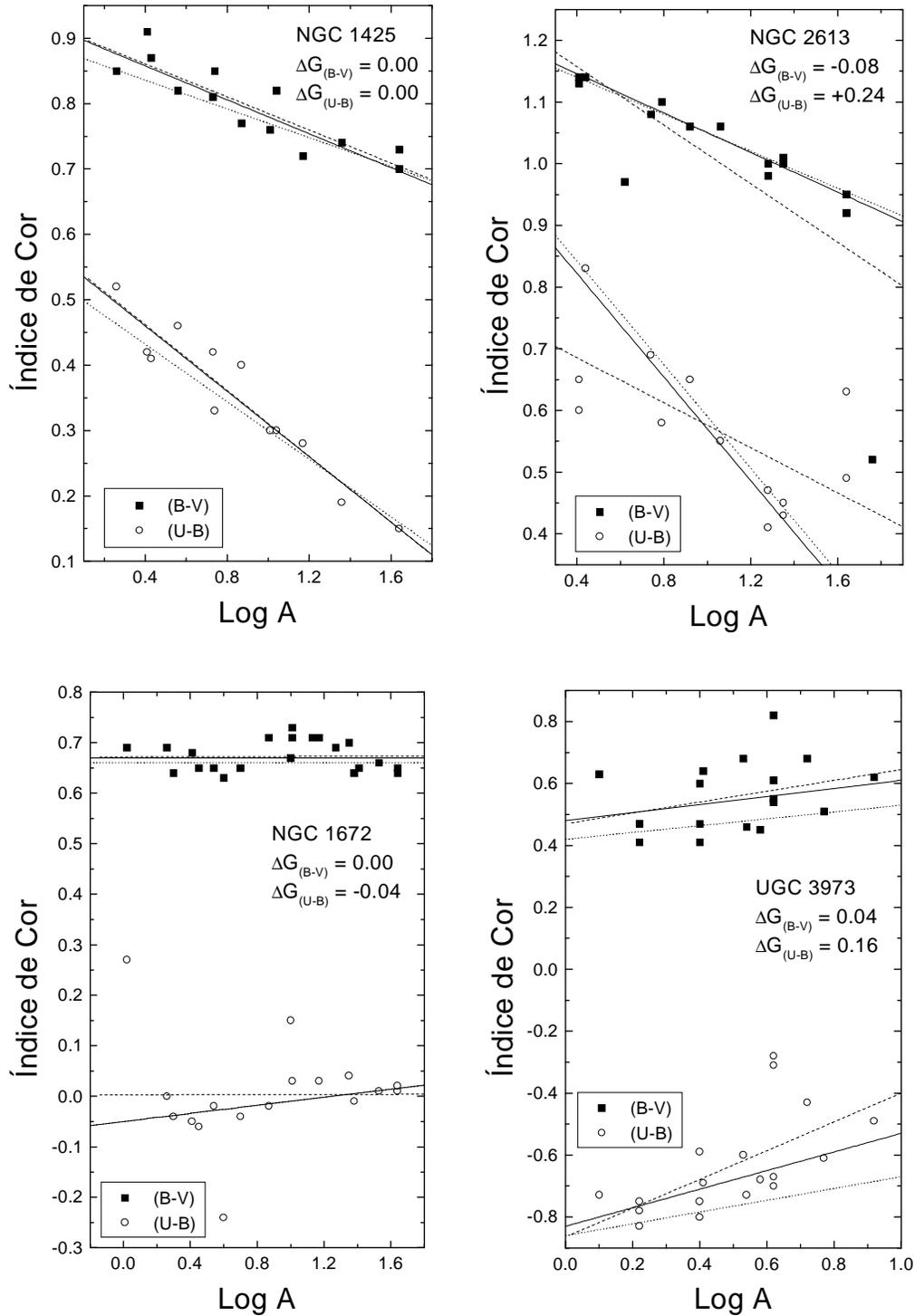}}
\caption{Exemplos de gradientes de cor para quatro gal\'axias em nossa amostra. O t\1pico gradiente negativo
de NGC 1425 e NGC 2613, o gradiente nulo de NGC 1672, e o raro gradiente positivo de UGC 3973 (confira os valores 
na Tabela A.1). Os \1ndices de cor em magnitudes est\~ao expressos em fun\cao do logaritmo da abertura do
diafragma em unidades de 0.1 minutos de arco. Os quadrados indicam o \1ndice (B\,-V), enquanto que os 
c\1rculos indicam o (U-B). As linhas tracejadas correspondem ao ajuste atrav\'es do m\'etodo dos m\1nimos
quadrados; linhas pontilhadas representam o ajuste pelo m\'etodo da m\1nima mediana e a linha cheia corresponde
ao ajuste determinado pelo algoritmo de {\sc progress}.} 
\label{exegrad}
\end{figure}

A Figura 2.4 exibe quatro exemplos do comportamento dos gradientes de cor para gal\'axias da presente
amostra. Nesta
figura pode-se avaliar os ajustes dos gradientes atrav\'es dos diferentes m\'etodos. As linhas tracejadas 
correspondem ao ajuste atrav\'es do m\'etodo dos m\1nimos quadrados; linhas pontilhadas representam o ajuste pelo
m\'etodo da m\1nima mediana e a linha cheia corresponde ao ajuste determinado pelo algoritmo de {\sc progress}.
Em cada gr\'afico est\'a indicado o valor da diferen\c ca entre o gradiente determinado pelo
m\'etodo dos m\1nimos quadrados e o determinado pelo algoritmo de {\sc progress} ($\Delta G \equiv G($MQ$) - 
G($PRO$)$). Nota-se que, para NGC 2613, o m\'etodo cl\'assico \'e perturbado pela \'ultima medida de (B\,-V). Para
esta gal\'axia, e para UGC 3973, os diferentes m\'etodos atribuem valores substancialmente diferentes para os
gradientes.

\begin{figure}
\epsfxsize=15cm
\centerline{\epsfbox{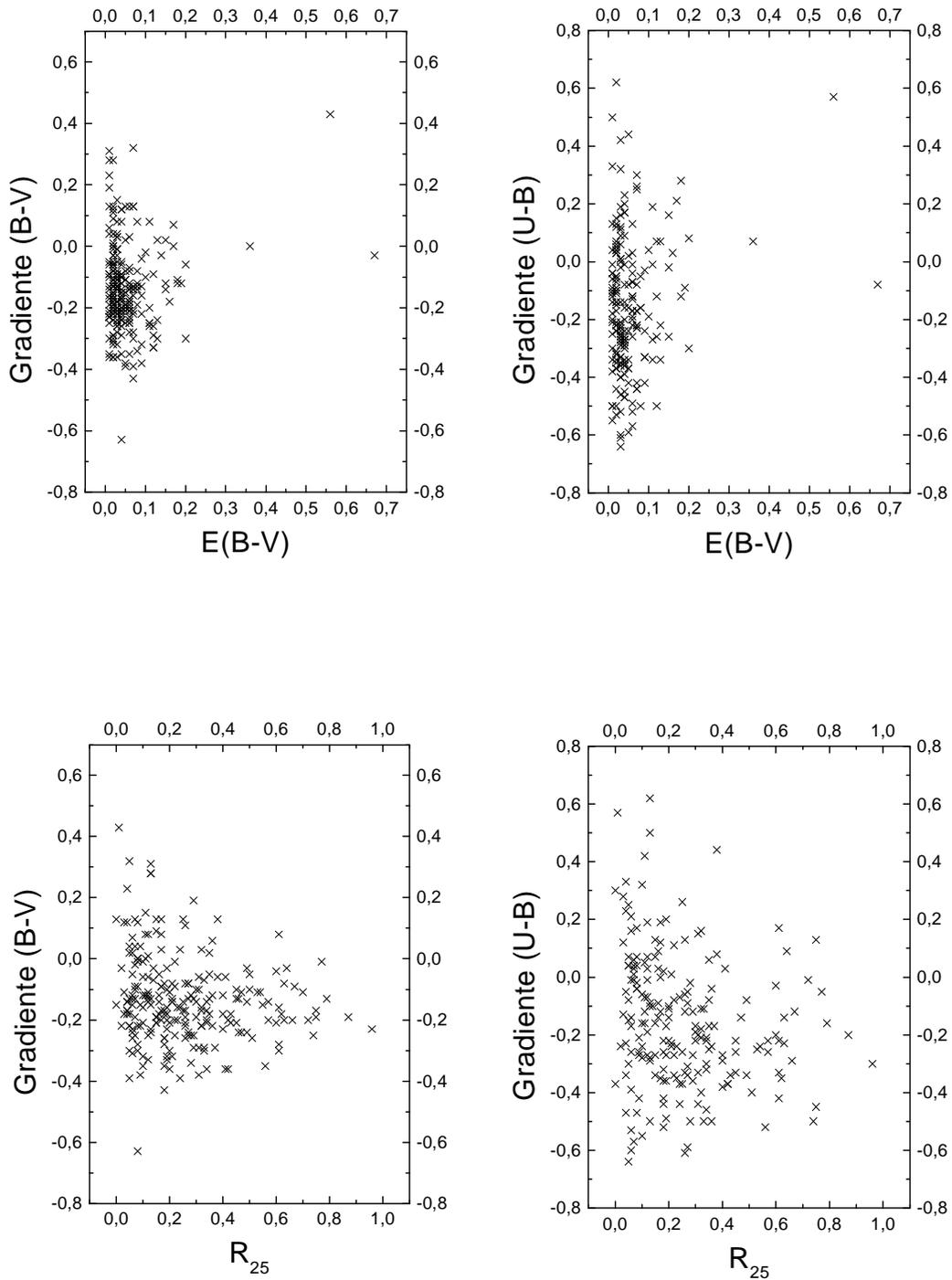}}
\caption{Gradientes de cor (B\,-V) e (U-B) para as gal\'axias de nossa amostra em fun\cao do excesso de cor
E(B\,-V) provocado pela extin\cao Gal\'actica (pain\'eis superiores), e em fun\cao da inclina\cao da gal\'axia, 
representada aqui pelo par\^ametro $R_{25}$ do RC3 (pain\'eis inferiores). A aus\^encia de correla\coes 
indica que as extin\coes Gal\'actica e intr\1nseca n\~ao interferem na determina\cao dos gradientes de cor.} 
\label{dust}
\end{figure}

Os dados em LdV83,85 {\em n\~ao} est\~ao corrigidos de fatores externos, tais como averme-lhamento Gal\'actico
e avermelhamento intr\1nseco, mas somente de fatores observacionais e da extin\cao atmosf\'erica. Na
determina\cao dos gradientes de cor, a corre\cao por averme-lhamento Gal\'actico n\~ao \'e necess\'aria,
j\'a que somente provocaria um deslocamento vertical do gradiente, n\~ao afetando sua inclina\c c\~ao, 
pois a mesma corre\cao \'e aplicada para toda a gal\'axia. Isto porque o excesso de cor produzido pela
extin\cao Gal\'actica pode ser expresso como fun\cao somente das coordenadas gal\'acticas do objeto\footnote{Na 
verdade, se a cor varia ao longo da
gal\'axia, o fato da extin\cao pela poeira interestelar variar com o comprimento de onda da luz faz com
que a corre\cao pelo avermelhamento seja distinta ao longo da gal\'axia. No entanto, estamos desprezando
este efeito, considerando que este somente produz perturba\coes muito pequenas em compara\cao com os 
erros fotom\'etricos e na determina\cao dos gradientes.}.
A corre\cao pelo avermelhamento intr\1nseco \'e
dif\1cil de prever corretamente, devido aos problemas de inclina\cao das gal\'axias e da quest\~ao 
da espessura \'optica
de gal\'axias espirais (e.g., Giovanelli \et 1994,1995; de Jong 1996c). De qualquer maneira, tal corre\cao
acarretaria modifica\coes sem import\^ancia nos gradientes de cor, tendo em vista as incertezas envolvidas.
De fato, a Figura 2.5 mostra que os avermelhamentos Gal\'actico e intr\1nseco n\~ao interferem na 
determina\cao dos gradientes de cor. Os pain\'eis superiores desta figura mostram que n\~ao h\'a qualquer
correla\cao entre os gradientes que determinamos e o avermelhamento Gal\'actico, representado pelo excesso
de cor $E(B-V)$. Por outro lado, como a extin\cao intr\1nseca varia com a inclina\cao do disco da gal\'axia
em rela\cao \`a nossa linha de visada, elaboramos os pain\'eis inferiores da figura, que exibem os gradientes
em fun\cao do par\^ametro $R_{25}$ do RC3, que representa a inclina\cao da gal\'axia.
Nestes pain\'eis, as gal\'axias cujos discos encontram-se mais inclinados em rela\cao \`a nossa linha de visada, 
portanto vistos de perfil, s\~ao as que possuem valores maiores de $R_{25}$.
A aus\^encia de 
correla\coes indica que o avermelhamento intr\1nseco n\~ao interfere na determina\cao dos gradientes. No 
entanto, essas extin\coes produzem efeitos bastante consider\'aveis nos {\em \1ndices} de cor.
Voltaremos a tratar deste assunto na pr\'oxima se\c c\~ao.

\subsection{Determina\cao dos \1ndices de cor caracter\1sticos dos bojos e discos}

\hskip 30pt Com a determina\cao dos gradientes de cor podemos avaliar o comportamento global dos \1ndices
de cor ao longo das gal\'axias, e comparar os valores dos gradientes com outras propriedades desses objetos,
tais como, a presen\c ca de barras, o brilho superficial, gradientes de abund\^ancia qu\1mica etc. Entretanto,
uma das propostas deste estudo \'e avaliar e comparar 
os \1ndices de cor das componentes disco e bojo separadamente, ou seja, \'e 
fazer um estudo que permita encontrar diferen\c cas entre as popula\coes 
estelares das componentes bojo e disco. Dessa forma, poderemos avaliar as diferen\c cas
entre as popula\coes estelares nos bojos e discos em cada uma das gal\'axias da nossa amostra.

Aplicaremos, inicialmente, uma primeira aproxima\cao na
determina\cao dos \1ndices de cor caracter\1sticos de bojos e discos, para poder verificar se existe uma 
correla\cao entre as cores destas componentes, conforme apresentado anteriormente 
por Peletier \& Balcells (1996). Estes autores encontram correla\cao
entre as cores de bojo (tomada como sendo aquela \`a metade do raio efetivo) e do disco (tomada como sendo aquela
a dois comprimentos de escala) para os \1ndices (U-R), (B-R), (R-K) e (J-K), para uma amostra de 30 gal\'axias 
espirais de tipo jovem.

Em nossa primeira metodologia, os \1ndices foram estimados, em primeira apro-xima\c c\~ao, atrav\'es da inspe\cao 
dos gradientes. O \1ndice do
bojo foi tomado como sendo aquele medido atrav\'es do diafragma de menor abertura dispon\1vel. O \1ndice do
disco foi tomado como sendo aquele medido na abertura dispon\1vel que mais se aproxima daquela equivalente \`a 
isofota de brilho superficial em B igual a 25 mag arcsec$^{-2}$, sendo este valor aquele apresentado no RC3.
Em alguns casos, quando os dados dispon\1veis n\~ao alcan\c cam as dimens\~oes requisitadas, foi feita uma
extrapola\cao linear do gradiente determinado. Por outro lado, naqueles casos que apresentam medidas com 
valores distintos para a mesma abertura, o valor m\'edio foi considerado. Nenhuma corre\cao foi aplicada
aos va-lores obtidos nessa metodologia. Os resultados da nossa estimativa ser\~ao apresentados na se\cao 2.4.6.
No entanto, podemos adiantar que, de fato, existe uma correla\cao entre os \1ndices de cor de bojos e discos
para as gal\'axias em nossa amostra.

No entanto, esta estimativa inicial apresenta o inconveniente de que, em muitos casos, os valores obtidos 
n\~ao s\~ao 
representativos da mesma regi\~ao em gal\'axias diferentes. O fato das gal\'axias apresentarem tamanhos
angulares distintos e de terem sido observadas com conjuntos de diafragmas de aberturas distintas, faz com que,
em muitos casos, os \1ndices caracter\1sticos sejam tomados em dist\^ancias galactoc\^entricas distintas de
gal\'axia para gal\'axia.

Para comparar \1ndices de cor caracter\1sticos de bojos e discos em regi\~oes semelhantes nas diferentes 
gal\'axias, adotamos a seguinte metodologia. Definimos o \1ndice de cor caracter\1stico do bojo
como sendo aquele a uma dist\^ancia galactoc\^entrica de 1/5 do raio efetivo da gal\'axia. O \1ndice 
caracter\1stico do disco foi definido como sendo aquele a uma dist\^ancia
de 2 raios efetivos. Desta forma, definimos o gradiente de cor, $G$, como sendo,

\eq
{G} \equiv {{\Delta (X-Y)} \over {\Delta \log A}},
\eeq

\noindent onde $(X-Y)$ representa o \1ndice de cor e $A$ \'e a abertura do diafragma em unidades de
0.1 arcmin. Indicando os valores inicial e final da sequ\^encia de aberturas medidas 
em cada gal\'axia com os subescritos $i$ e $f$, temos:

\eq
{G} = {{{(X-Y)}_{f} - {(X-Y)}_{i}} \over {\log {A}_{f} - \log {A}_{i}}}.
\eeq

Para determinar o \1ndice do bojo, fa\c camos as seguintes substitui\c c\~oes: ${(X-Y)}_{f} = {(X-Y)}_{ef}, 
{(X-Y)}_{i} = {(X-Y)}_{b}, {A}_{f} = {A}_{ef}$ e ${A}_{i} = {A}_{ef}/5$, onde ${(X-Y)}_{ef}$ \'e o 
\1ndice de cor efetivo (dado pelo RC3), ${(X-Y)}_{b}$ \'e o \1ndice caracter\1stico do bojo e ${A}_{ef}$ 
\'e a abertura efetiva, tamb\'em segundo o RC3. Desse modo, obtemos o \1ndice caracter\1stico do bojo como sendo:

\eq
{{(X-Y)}_{b}} = {{(X-Y)}_{ef} - 0.7 G}.
\eeq

Analogamente, podemos obter o \1ndice de cor do disco (a 2 raios efetivos) como sendo:

\eq
{{(X-Y)}_{d}} = {{(X-Y)}_{ef} + 0.3 G}.
\eeq

As equa\coes (2.8) e (2.9) foram utilizadas no programa {\sc progress} 
para determinar os \1ndices de cor caracter\1sticos de bojo 
e disco para cada gal\'axia. Note que, como estamos obtendo estes \1ndices em dist\^ancias 
galactoc\^entricas normalizadas pelo raio efetivo das gal\'axias, garantimos que as gal\'axias  
est\~ao sendo igualmente amostradas. Ou seja, garantimos que estes \1ndices representam as mesmas regi\~oes em 
cada gal\'axia, mesmo quando observamos gal\'axias a distintas dist\^ancias.

Por\'em, antes de avaliar os resultados obtidos, \'e preciso corrigir estes valores pelos
efeitos da extin\cao Gal\'actica e da extin\cao intr\1nseca. Ao contr\'ario dos gradientes, \'e \'obvio que
os efeitos da extin\cao sobre os \1ndices de cor n\~ao podem ser desprezados.

A natureza dessas extin\coes
\'e a mesma, qual seja, o fato da poeira presente no meio interestelar, principalmente no plano do disco
de uma gal\'axia espiral, absorver e espalhar a luz. A diferen\c ca reside em que a extin\cao Gal\'actica
se refere ao efeito da poeira na Gal\'axia sobre a luz emitida por objetos extragal\'acticos, enquanto que 
a extin\cao intr\1nseca se refere ao efeito da poeira presente nas gal\'axias sobre a luz emitida pela estrelas
que estas cont\^em. No entanto, como a extin\cao varia com o comprimento de onda da luz, esta n\~ao s\'o \'e
atenuada, como tamb\'em torna-se mais avermelhada. Assim, para podermos corrigir os \1ndices de cor, precisamos
conhecer o excesso de cor produzido pelas extin\c c\~oes. Este excesso de cor \'e definido como:

\eq
{E(X-Y)} = {(X-Y) - {(X-Y)}_{0}},
\eeq

\noindent onde o subescrito $0$ indica o valor intr\1nseco, i.e., corrigido, exatamente como a equa\cao (3.59) de
Binney \& Merrifield (1998). Os valores de $E(B-V)$ para a extin\cao Gal\'actica variam de acordo com as
coordenadas gal\'acticas das gal\'axias observadas, e foram determinados atrav\'es dos mapas recentemente 
obtidos por Schlegel, Finkbeiner \& Davis (1998). Para determinar $E(U-B)$, utilizamos a rela\cao

\eq
{{E(U-B)}\over {E(B-V)}} = {0.72\pm 0.03},
\eeq

\noindent que \'e a equa\cao (3.1.37) de Kitchin (1998).

O excesso de cor provocado pela extin\cao intr\1nseca \'e muito mais complicado de se determinar, j\'a
que o nosso conhecimento sobre a distribui\cao de poeira no meio interestelar das gal\'axias \'e ainda 
bastante prim\'ario (ver, e.g., Tully \et 1998). Existem v\'arios estudos (e.g., van Houten 1961; Bruzual, 
Magris \& Calvet 
1988) que investigam os efeitos de extin\cao sobre as cores integradas de gal\'axias, e que assumem 
aproxima\coes para a distribui\cao de poeira, considerando simetria esf\'erica ou plano--paralela. Entretanto, 
os efeitos radiais da extin\cao sobre a cor raramente s\~ao determinados. Perfis 
de cores produzidos por modelos de poeira foram constru\1dos por de Jong (1996c) para avaliar
at\'e que ponto a exist\^encia de gradientes de cor pode ser atribu\1da ao avermelhamento pela extin\cao por poeira. 
Os resultados indicam que o avermelhamento provocado por poeira n\~ao tem um papel muito importante  
nos perfis de cor.

Por outro lado, sabemos que a extin\cao intr\1nseca \'e
proporcional \`a inclina\cao do disco da gal\'axia em rela\cao \`a nossa linha de visada, j\'a que, no caso
de uma gal\'axia com alta inclina\c c\~ao, a luz emitida pelas estrelas nas regi\~oes mais afastadas desta
gal\'axia ter\'a de atravessar uma quantidade maior de poeira, e ser\'a proporcionalmente mais atenuada.
Estes efeitos provocados pela inclina\cao podem ser corrigidos. Desta forma, aplicamos uma corre\cao 
que simula a situa\cao em que todas as gal\'axias est\~ao sendo observadas de face, 
quando os efeitos da extin\cao intr\1nseca s\~ao minimizados.

Obtivemos o $E(B-V)$ provocado pela extin\cao intr\1nseca, em fun\cao da inclina\cao da gal\'axia, 
atrav\'es do seguinte racioc\1nio. Giovanelli \et 
(1994), utilizando imagens de gal\'axias do tipo Sc na banda I, determinaram uma rela\cao para o coeficiente
de extin\cao intr\1nseca em fun\cao da inclina\cao da gal\'axia, que \'e:

\eq
{{A}_{I}} = {1.12(\pm 0.05) \log {{a}\over {b}}},
\eeq

\noindent onde $a$ e $b$ s\~ao, respectivamente, os eixos maior e menor da gal\'axia. Por outro lado, 
Elmegreen (1998) mostra que os coeficientes de extin\cao nas bandas U, B e V s\~ao, respectivamente,
3.81, 3.17 e 2.38 vezes maior que o coeficiente de extin\cao na banda I, segundo observa\coes na Gal\'axia.
Como a Gal\'axia \'e do tipo Sbc, podemos utilizar estes valores como sendo caracter\1sticos das
gal\'axias em nossa amostra, o que implica em admitir que a poeira nas gal\'axias tem propriedades 
semelhantes na Gal\'axia. Assim, temos que:

\eq
{{A}_{U}} = {4.27 \log {{a}\over {b}}},
\eeq

\eq
{{A}_{B}} = {3.55 \log {{a}\over {b}}},
\eeq

\eq
{{A}_{V}} = {2.67 \log {{a}\over {b}}}.
\eeq

Como

\eq
{E(X-Y)} = {{A}_{X} - {A}_{Y}},
\eeq

\noindent segundo a equa\cao (3.59) de Binney \& Merrifield (1998), ent\~ao

\eq
{E(U-B)} = {0.68 \log {{a}\over {b}}},
\eeq

\eq
{E(B-V)} = {0.87 \log {{a}\over {b}}}.
\eeq

Para determinar as corre\coes a serem aplicadas atrav\'es das equa\coes (2.17) e (2.18), substitu\1mos
$\log a/b$ pelo par\^ametro $R_{25}$ do RC3, que pode ser considerado como equi-valente. \'E interessante
comparar as corre\coes que aplicamos com as que s\~ao
aplicadas no RC3. Este cat\'alogo utiliza as seguintes rela\coes para gal\'axias do tipo Sbc:

\eq
{E(U-B)} = {0.38 R_{25}},
\eeq

\eq
{E(B-V)} = {0.33 R_{25}}.
\eeq

Portanto, a corre\cao que estamos aplicando \'e $\sim$ 2--3 vezes maior do que a aplicada no RC3. De fato, quando
da confec\cao deste cat\'alogo, acreditava-se que os discos de gal\'axias espirais fossem quase transparentes.
Trabalhos recentes (Giovanelli \et 1994,1995 e refer\^encias a\1 contidas; Tully \et 1998) mostram que a
espessura \'optica dos discos \'e bem maior do que se acreditava. Assim, nossas corre\coes consideram que
os discos possuem alta espessura \'optica.

Embora nossa amostra se componha de gal\'axias pr\'oximas, aplicamos tamb\'em a corre\cao K para compensar 
os efeitos do desvio
para o vermelho provocados pela velocidade de recess\~ao das gal\'axias sobre os \1ndices de cor. Utilizamos as 
seguintes rela\c c\~oes, seguindo os crit\'erios do RC3:

\eq
{{(B-V)}_{0}} = {(B-V) - 6.5 \times {10}^{-6} cz},
\eeq

\eq
{{(U-B)}_{0}} = {(U-B) - 4.5 \times {10}^{-6} cz},
\eeq

\noindent onde $cz$ \'e a velocidade da gal\'axia ao longo da linha de visada. Para as gal\'axias em nossa
amostra, este par\^ametro cobre o seguinte intervalo de valores: -295 Km/s (NGC 224) $\leq cz \leq$ 
8720 Km/s (UGC 4013). Por\'em, o valor t\1pico \'e $cz \sim$ 2000 Km/s, resultando em corre\coes da 
ordem de mil\'esimos de magnitude.

\section{Resultados}

\hskip 30pt A seguir apresentamos os principais resultados obtidos, quais sejam, gradientes de cor (U-B) e (B\,-V)
para cada gal\'axia em nossa amostra, bem como os \1ndices de cor (U-B) e (B\,-V) caracter\1sticos para o
bojo e o disco de cada uma dessas gal\'axias. Como o objetivo principal consiste em verificar a 
exist\^encia de ind\1cios 
que possam corroborar, ou n\~ao, o cen\'ario de evolu\cao secular em gal\'axias barradas (se\cao 1.1.3), procuramos
identificar diferen\c cas entre as propriedades de gal\'axias barradas e ordin\'arias.
\'E importante ressaltar que inclu\1mos na classe de barradas as fam\1lias fortemente (SB) e fracamente (SAB) 
barradas, conforme classificadas no RC3.

Inicialmente, apresentamos os resultados relativos aos gradientes de cor das gal\'axias em nossa
amostra, tais como aqueles relativos \`a distribui\cao dos gradientes, bem como compara\coes com resultados 
obtidos por outros autores. S\'o ent\~ao os
resultados concernentes aos \1ndices de cor ser\~ao exibidos.

\subsection{Distribui\cao dos gradientes}

\hskip 30pt A avalia\cao da distribui\cao dos gradientes nas gal\'axias desta amostra requer um cuidado especial, 
j\'a que esta se comp\~oe de gal\'axias vistas em v\'arios \^angulos de inclina\c c\~ao. Desta forma, 
vamos avaliar a distribui\cao de gradientes de cor, separando a amostra inicial em 2 sub-amostras, 
uma contendo gal\'axias vistas de face, e a outra com gal\'axias vistas de perfil. O crit\'erio utilizado 
para a separa\cao destas sub-amostras consiste em definir um limite para o \^angulo de inclina\c c\~ao adequado 
para a separa\cao destas duas sub-amostras. Levando em conta que temos como objetivo comparar os nossos 
resultados com os de outros autores, utilizamos o mesmo valor de inclina\cao definido por de Jong \& van der 
Kruit (1994), ou seja, as gal\'axias com $R_{25} \leq 0.20$, que corresponde a $b/a \geq 0.625$, s\~ao 
consideradas de face. Aquelas que n\~ao s\~ao inclu\1das neste crit\'erio fazem parte ent\~ao da sub-classe 
das gal\'axias vistas de perfil.

A Figura 2.6 exibe a distribui\cao dos gradientes para: (a) gal\'axias
ordin\'arias vistas de face, (b) barradas vistas de face, (c) ordin\'arias vistas de perfil, e (d)  
barradas vistas de perfil. A Tabela 2.1 apresenta, para cada um desses diagramas, o n\'umero total de
gal\'axias, a m\'edia da distribui\cao 
(e seu respectivo erro padr\~ao) e o desvio padr\~ao na distribui\c c\~ao, calculados atrav\'es do
ajuste de uma Gaussiana. Nestes histogramas, cada barra corresponde a um intervalo de 0.05 magnitudes.
Como a mediana dos erros nos gradientes \'e igual a 0.03 mag para (B\,-V) e 0.04 mag para (U-B), este intervalo
\'e significativo.

\begin{figure}
\epsfysize=22cm
\centerline{\epsfbox{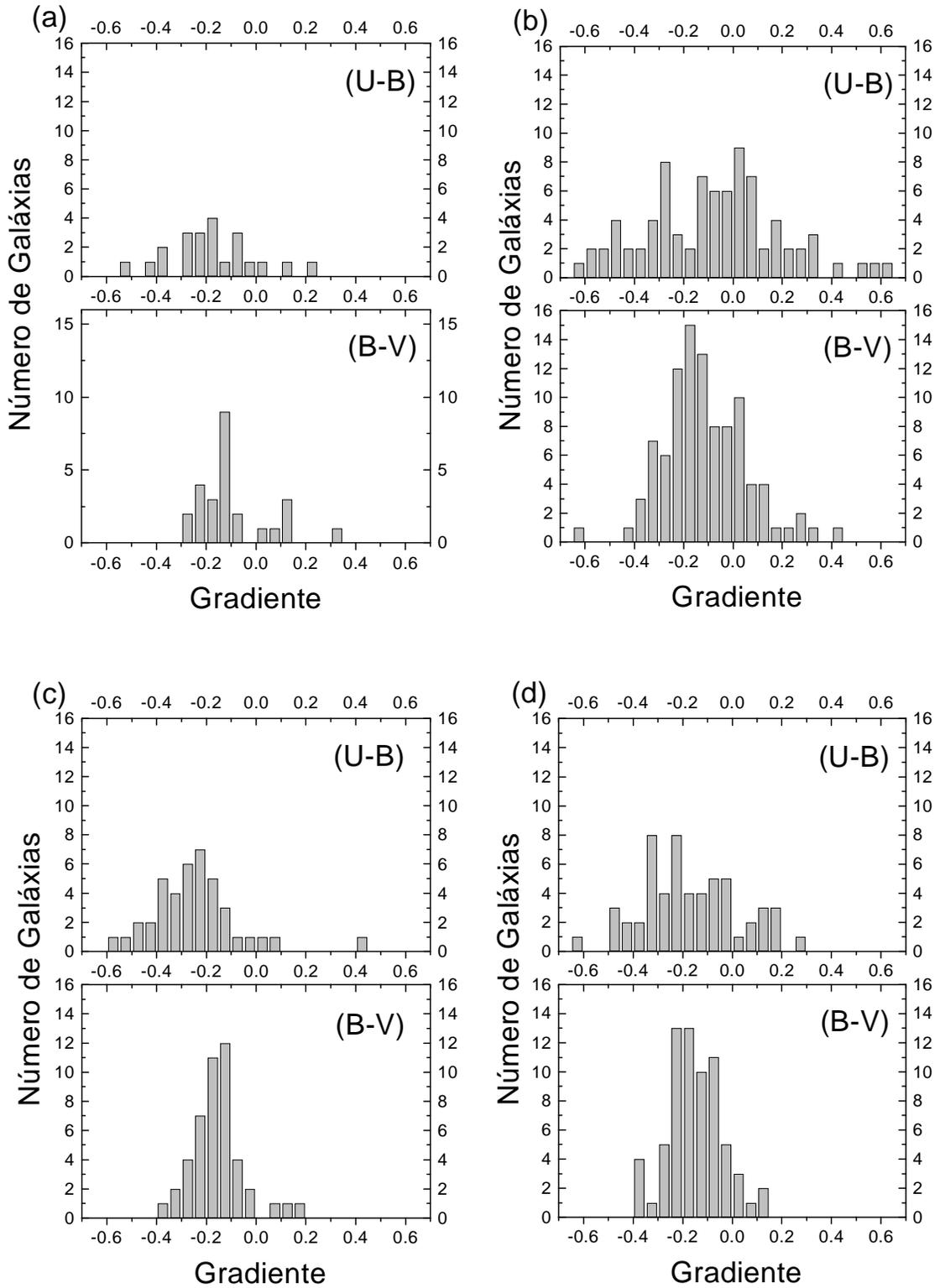}}
\caption{Distribui\cao dos gradientes de cor (U-B) e (B\,-V) para: (a) gal\'axias
ordin\'arias vistas de face, (b) barradas vistas de face, (c) ordin\'arias vistas de perfil, e (d)  
barradas vistas de perfil.} 
\label{dgrade-2}
\end{figure}

\begin{table}
\caption{Valores m\'edios (e respectivos erros) e desvios para as distribui\coes apresentadas nas
Figuras 2.6 e 2.7. N \'e o n\'umero total de gal\'axias em cada distribui\c c\~ao.}
\begin{center}
\begin{tabular}{||c|cccc|cccc||}
\hline
\hline
Diagrama	& N   	& m\'edia & erro    & desvio & N       & m\'edia & erro  & desvio   \\
(Figura 2.6)	&      	&	  & (B\,-V) &	     &	       &         & (U-B) & 	    \\
\hline
(a)		& 26   	& -0.14   & 0.01    & 0.06   & 22      & -0.19   & 0.03  & 0.14   \\
(b)		& 98  	& -0.14   & 0.02    & 0.15   & 82      & -0.08   & 0.03  & 0.27   \\
(c)		& 46   	& -0.16   & 0.01    & 0.07   & 41      & -0.26   & 0.02  & 0.12   \\
(d)		& 68   	& -0.16   & 0.01    & 0.10   & 56      & -0.19   & 0.03  & 0.20   \\
\hline
Diagrama	&     	&  	  &  	    &        &         &         &       &        \\
(Figura 2.7)	&     	&         &         &        &         &         &       &        \\
(a)		& 72   	& -0.16	  & 0.01    & 0.07   & 63      & -0.24   & 0.02  & 0.13   \\
(b)		& 166  	& -0.15	  & 0.01    & 0.13   & 138     & -0.13   & 0.02  & 0.25   \\
\hline
\hline
\end{tabular}
\end{center}
{\footnotesize{Diagramas -- Figura 2.6: (a): ordin\'arias vistas de face; (b): barradas vistas de face; 
(c):  
ordin\'arias vistas de perfil; (d): barradas vistas de perfil. Figura 2.7: (a): todas as gal\'axias ordin\'arias; 
(b): todas as barradas.}}
\end{table}

Algumas informa\coes b\'asicas podemos obter avaliando os resultados obtidos com as distribui\coes na 
Figura 2.6, bem como com os valores dos gradientes m\'edios listados na Tabela 2.1.
A primeira \'e a de que as distribui\coes dos gradientes em (U-B) s\~ao sistematicamente mais alargadas
do que em (B\,-V), tanto para as gal\'axias de face como para as de perfil. Este alargamento 
se reflete nas estimativas dos desvios, que s\~ao maiores para esta cor. Existem v\'arias causas 
prov\'aveis para esta diferen\c ca. 
Em primeiro lugar, os erros
fotom\'etricos na banda U s\~ao mais expressivos e, em segundo lugar, o n\'umero de medidas em (U-B)
para a determina\cao do gradiente \'e, em geral, menor do que em (B\,-V), o que certamente pode diminuir
a precis\~ao na estimativa do gradiente. H\'a ainda a quest\~ao da descontinuidade de Balmer (ver se\cao 2.1) e, evidentemente,
esta propriedade pode tamb\'em ser intr\1nseca das gal\'axias em nossa amostra.
A segunda informa\cao diz respeito \`as gal\'axias vistas de perfil. Nestas gal\'axias, existe uma tend\^encia
dos gradientes serem mais negativos (mais acentuados) quando comparados com as gal\'axias vistas de face, 
em particular em (U-B). Este comportamento pode estar ligado ao fato de a extin\cao intr\1nseca 
diferencial ser mais pronunciada nestas
gal\'axias, cujos discos apresentam maior inclina\cao com rela\cao \`a nossa linha de visada.
Tamb\'em pode-se notar que, em (U-B), a maior parte das gal\'axias barradas (ver pain\'eis (b) e (d)), se concentra
em valores de gradientes que s\~ao menos acentuados do que as gal\'axias ordin\'arias (ver pain\'eis (a) e (c)), 
conforme pode-se verificar atrav\'es dos valores m\'edios, que se encontram na Tabela 2.1. Esta 
caracter\1stica desaparece em (B\,-V).
A quarta informa\cao diz respeito ao alargamento das distribui\c c\~oes.
Nota-se que a largura das distribui\coes \'e maior para as barradas, tanto para as gal\'axias de face 
como para as de perfil. Isto significa que uma propor\cao maior de gal\'axias barradas apresenta
gradientes nulos ou positivos. Em (U-B), por exemplo, $55\% \pm 8\%$ das gal\'axias barradas vistas de face t\^em 
gradientes nulos ou positivos. Por outro lado, nas ordin\'arias vistas de face esta fra\cao \'e 
de $32\% \pm 12\%$. J\'a em (B\,-V), cerca de $41\% \pm 6\%$ das barradas vistas de face t\^em gradientes 
nulos ou positivos, enquanto que $31\% \pm 11\%$ das ordin\'arias vistas de face apresentam este efeito.

Com a Figura 2.6 foi feita uma avalia\cao das diferen\cas nas distribui\coes dos gradientes para
gal\'axias vistas de face e de perfil, separadamente. 
Na Figura 2.7, colocamos todas as gal\'axias da amostra, sem separar portanto as de face e as de perfil, 
para avaliar a distribui\cao dos gradientes de cor nas ordin\'arias (a) e nas barradas (b).
Verificamos,
mais uma vez, que as distribui\coes s\~ao similares em (B\,-V), mas, em (U-B), gal\'axias barradas
tendem a apresentar gradientes menos acentuados do que aqueles apresentados por gal\'axias ordin\'arias 
(veja os valores da Tabela 2.1). Mais uma vez tamb\'em, se observa que as distribui\coes s\~ao mais
alargadas para as gal\'axias barradas. Estas propriedades podem estar relacionadas com os efeitos provocados
pelas barras nas gal\'axias que as cont\^em, em particular, aos processos de evolu\cao secular.

\begin{figure}
\epsfysize=22cm
\centerline{\epsfbox{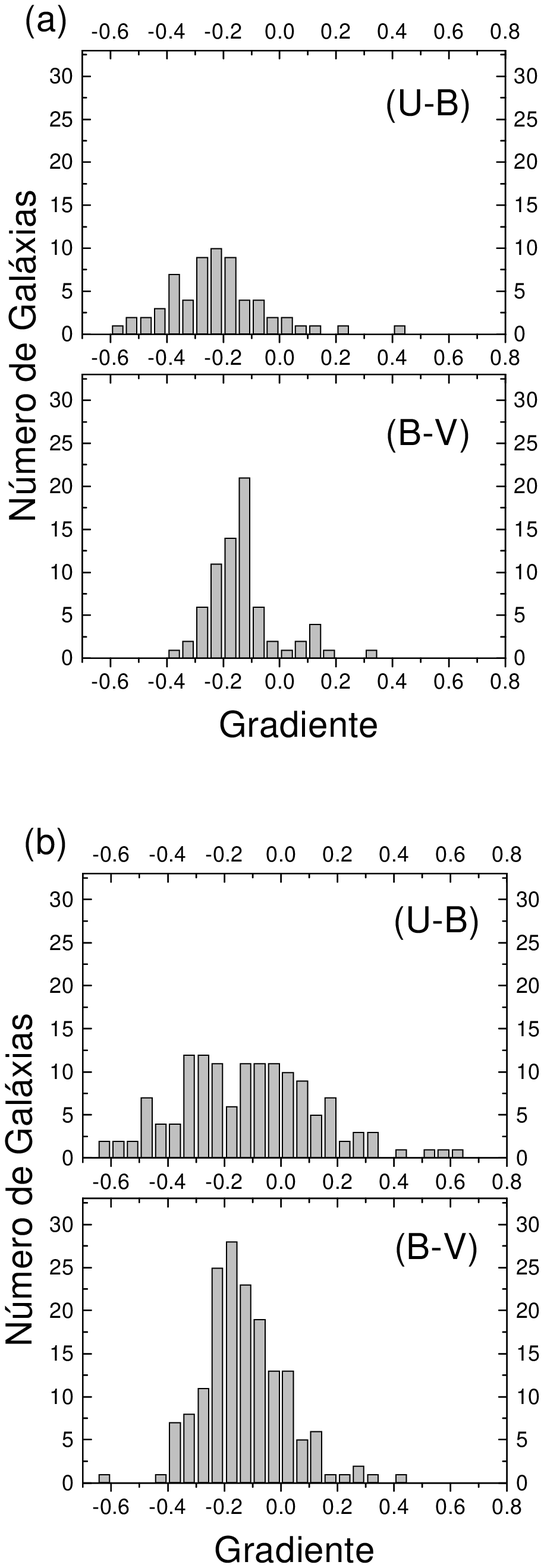}}
\caption{Distribui\cao dos gradientes de cor (U-B) e (B\,-V) para: (a) todas as gal\'axias ordin\'arias, e 
(b), todas as barradas. Como se v\^e, gal\'axias barradas tendem a apresentar gradientes em (U-B) 
menos acentuados do que gal\'axias ordin\'arias. No entanto, esta caracter\1stica desaparece em (B\,-V).} 
\label{dgrade2-1}
\end{figure}

\begin{figure}
\epsfysize=22cm
\centerline{\epsfbox{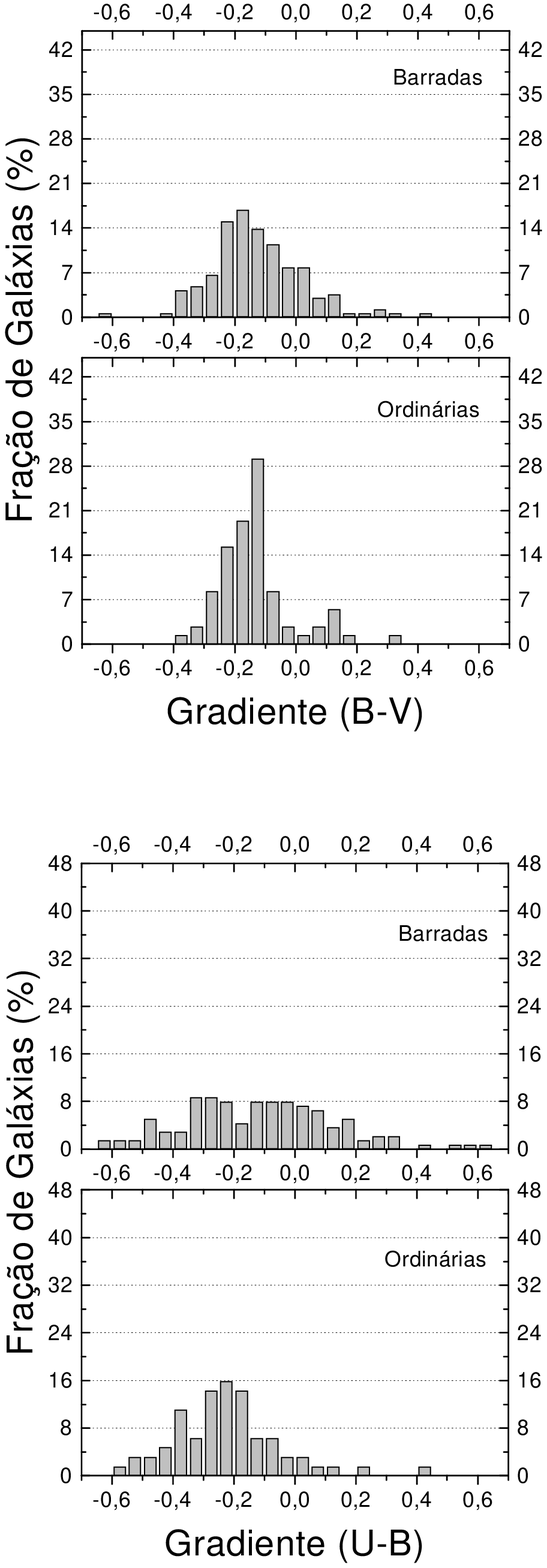}}
\caption{Compara\cao normalizada entre as distribui\coes dos gradientes de cor em gal\'axias ordin\'arias e
barradas. Gal\'axias barradas apresentam uma maior fra\cao de gal\'axias com gradientes pr\'oximos de zero, 
ou positivos, em ambas as cores.} 
\label{dgrade2-2}
\end{figure}

\begin{figure}
\epsfysize=22cm
\centerline{\epsfbox{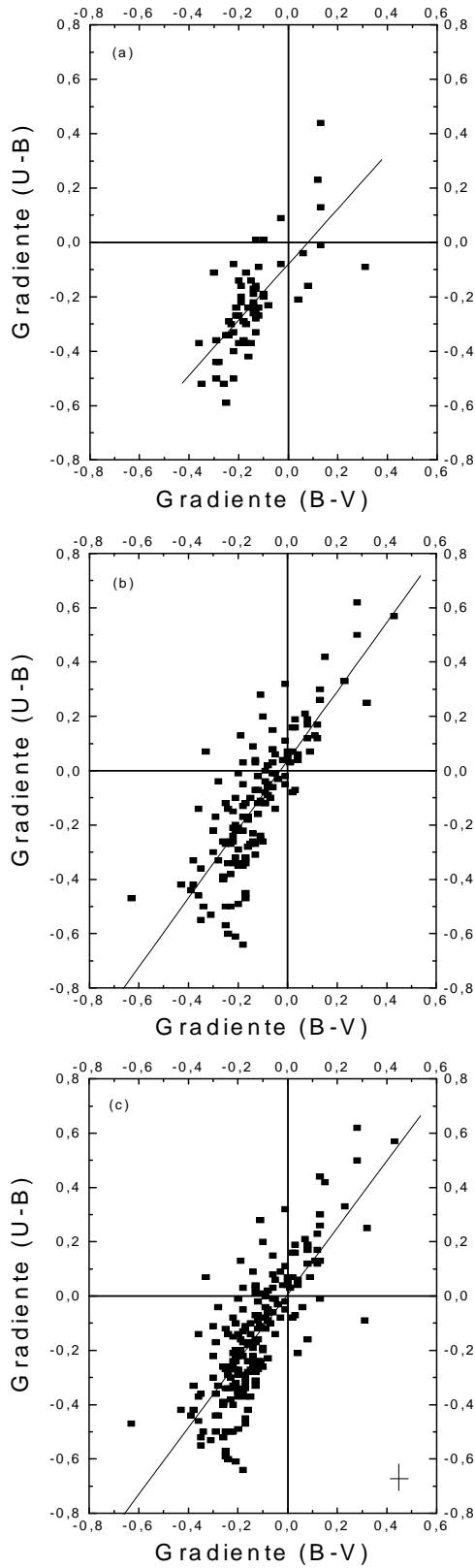}}
\caption{Gradientes de cor (U-B) em fun\cao dos gradientes (B\,-V) para: (a) gal\'axias ordin\'arias, (b) 
barradas e (c) toda a nossa amostra. A linha cheia corresponde a um ajuste linear para os pontos. 
Claramente, os gradientes s\~ao correlacionados.} 
\label{dgrade-1}
\end{figure}

A Figura 2.8 exibe as mesmas distribui\coes apresentadas na Figura 2.7. No entanto, nesta figura fazemos 
uma compara\cao direta entre a distribui\cao dos gradientes para barradas e ordin\'arias, considerando cada 
cor separadamente, normalizando o n\'umero de gal\'axias barradas, em cada intervalo de gradiente,
pelo n\'umero {\em total} de gal\'axias
barradas, e fazendo o mesmo com as ordin\'arias. Verifica-se que, em ambas as cores, a fra\cao de gal\'axias
barradas que apresentam gradientes nulos, ou positivos, \'e maior do que esta mesma fra\cao
de gal\'axias ordin\'arias. \'E interessante observar que a propor\cao de barradas com $G_{B-V} \geq 0$ 
\'e $37\% \pm 5\%$, enquanto que nas ordin\'arias temos 
$24\% \pm 6\%$. Portanto, dentro das incertezas, estes dois n\'umeros s\~ao semelhantes, apesar de que existe uma 
propor\cao ligeiramente maior de barradas com $G_{B-V} \geq 0$. J\'a no caso de (U-B), existem $47\% \pm 6\%$ de 
barradas com $G_{U-B} \geq 0$, enquanto que apenas $19\% \pm 5\%$ das ordin\'arias t\^em 
$G_{U-B} \geq 0$. Como o \1ndice (U-B) \'e mais sens\1vel \`a metalicidade, bem como \`a idade da 
popula\cao estelar, isto significa que uma propor\cao 
maior de barradas deveria apresentar gradientes nulos ou positivos, o que, de fato, ocorre. 
Esta diferen\c ca no comportamento de $G_{B-V}$ e $G_{U-B}$ {\em n\~ao} pode ser atribu\1da aos erros de 
avalia\cao dos gradientes. De fato, os erros s\~ao maiores no $G_{U-B}$, mas os mesmos erros 
devem afetar barradas e ordin\'arias.

A Figura 2.9 mostra os gradientes de cor (U-B) em fun\cao dos gradientes (B\,-V) para: (a) 
gal\'axias ordin\'arias, (b) barradas e (c) toda a nossa amostra. Como se pode ver, os gradientes 
s\~ao correlacionados. Al\'em disso, a correla\cao est\'a presente de forma equivalente para barradas e
para ordin\'arias. De fato, os coeficientes de correla\cao de Pearson R s\~ao: (a) 0.71, (b) 0.80 e 
(c) 0.78. O
valor ligeiramente inferior de R para as ordin\'arias est\'a provavelmente ligado ao fato de haver
um n\'umero menor de gal\'axias neste grupo. A mesma correla\cao foi verificada separando-se as 
gal\'axias vistas de face, sem diferen\c cas significativas. Estas correla\c c\~oes, na verdade, s\~ao esperadas, 
porque o gradiente representa, em \'ultima an\'alise, uma varia\cao na popula\cao estelar entre a regi\~ao 
interna e externa. Ocorre que estas varia\coes afetam tanto a cor (B\,-V), como a (U-B). Os modelos de 
Larson e Tinsley (1978), por exemplo, mostram que para uma popula\cao formada em um \'unico ``burst'' (com 
dura\cao de $2 \times 10^{7}$ anos), em uma gal\'axia espiral,
o \1ndice de cor (B\,-V) varia de $\Delta (B-V) \sim 1.1$, enquanto que o 
\1ndice de cor (U-B) varia de $\Delta (U-B) \sim 1.5$, entre duas popula\coes com cerca de 
10 Giga-anos de diferen\c ca de idade. Portanto, devemos esperar que 
$\Delta (U-B) / \Delta (B-V) \sim 1.4$. Por outro lado, como 
$G_{B-V} = \Delta (B-V) / \Delta \log A$ e $G_{U-B} = \Delta (U-B) / \Delta \log A$, e esperamos que os gradientes de cor
sejam causados por varia\coes na popula\cao estelar, usando estes modelos podemos prever que $G_{U-B} / G_{B-V} \sim 1.4$.
A correla\cao que observamos na Figura 
2.9 tem uma inclina\cao $G_{U-B} / G_{B-V} \sim 1.2$, bastante pr\'oxima do previsto por estes modelos.

\subsection{Tr\^es categorias para os gradientes: \hspace{-0.2cm} negativos, nulos e positivos}

\hskip 30pt O cen\'ario monol\1tico de forma\c c\~ao, que na d\'ecada de 60 parecia explicar definitivamente 
a quest\~ao da forma\cao de gal\'axias, \'e consistente com uma distribui\cao radial de popula\coes estelares, 
onde as popula\coes mais velhas, ou mais avermelhadas, se concentram na regi\~ao central da gal\'axia, e as 
estrelas jovens e mais azuis nas regi\~oes externas.
Este comportamento da distribui\cao radial de popula\coes estelares se traduz em um gradiente de cor negativo, 
i.e., cores mais vermelhas nas regi\~oes centrais e mais azuis nas regi\~oes perif\'ericas. 
As figuras 2.6 e 2.9 indicam que a grande maioria das gal\'axias ana-lisadas neste estudo, t\^em, de fato, 
gradientes negativos. Entretanto, com a descoberta de popula\coes estelares ricas em metais no bojo 
da Gal\'axia, a possibilidade de outros cen\'arios de forma\cao de gal\'axias tem sido explorada, 
conforme discutido na se\cao 1.1. Notamos em nossa amostra, que, 
al\'em das gal\'axias com gradientes negativos, 
existem gal\'axias com gradientes nulos, e gal\'axias com gradientes positivos. De fato, podemos separar
as gal\'axias em nossa amostra em tr\^es categorias distintas em rela\cao ao comportamento do gradiente
de cor. Definimos as gal\'axias com gradientes negativos como sendo aquelas que apresentam um gradiente
$\leq -0.10$. As gal\'axias com gradientes nulos s\~ao aquelas em que o gradiente se encontra na faixa
de valores $-0.10 < G < 0.10$. Por fim, as gal\'axias com gradientes positivos s\~ao aquelas em que
$G \geq 0.10$. A Tabela 2.2 exibe a distribui\cao das gal\'axias em nossa amostra, segundo estas 3
categorias, para ambas as cores, e para uma quarta categoria, na qual a defini\cao de gal\'axias com
gradientes nulos \'e mais restrita, seguindo o crit\'erio $-0.05 < G < 0.05$. A coluna (1) desta tabela
apresenta o n\'umero total de gal\'axias em cada categoria, enquanto que a coluna (2) d\'a a fra\cao
da amostra total que representa a categoria. Assim, temos aproximadamente 65\% de gal\'axias com o
t\1pico gradiente negativo, 25\% de gal\'axias com gradientes nulos e aproximadamente 10\% com gradientes positivos,
com pequenas varia\coes em cada cor. Por sua vez, as colunas (3), (4) e (5) indicam, respectivamente,
a fra\cao de gal\'axias ordin\'arias, levemente barradas e barradas em cada categoria. A coluna (6) indica a 
fra\cao total de barradas (SAB+SB) e, finalmente, a coluna (7) apresenta o n\'umero de gal\'axias com
AGN (``Active Galactic Nuclei''), bem como a fra\cao da categoria que estas gal\'axias representam. 
As gal\'axias com AGN foram 
identificadas atrav\'es do cat\'alogo de V\'eron-Cetty \& V\'eron (1998). A import\^ancia de se fazer uma
an\'alise da fra\cao de gal\'axias com AGN em cada categoria decorre da sugest\~ao apresentada por alguns
autores (e.g., Shlosman, Frank \& Begelman 1989; Shlosman, Begelman \& Frank 1990) 
de que as barras podem gerar mecanismos para abastecer n\'ucleos ativos,
atrav\'es de processos semelhantes \`aqueles que podem levar \`a evolu\cao secular.

{\small
\begin{table}
\caption{Distribui\cao das gal\'axias de nossa amostra em rela\cao \`as v\'arias categorias a serem
avaliadas.}
\begin{center}
\begin{tabular}{||ccccccccc||}
\hline
\hline
		     & Cor     & Total & Amostra    & SA   & SAB             & SB    & SAB+SB          & AGN\\
               	     &         & (1)   & (2)        & (3)  & (4)             & (5)   & \hskip0.4cm(6)  & (7)\\
\hline
G $\geq$ 0.1         & (B\,-V) & 18    & 8\%        & 33\% & 22\%            & 45\%  & \hskip0.4cm67\% & 6(33\%)\\
-0.1 $<$ G $<$ 0.1   &         & 55    & 23\%       & 13\% & 45\%            & 42\%  & \hskip0.4cm87\% & 11(20\%)\\ 
-0.05 $<$ G $<$ 0.05 &         & 25    & 10\%       & 12\% & 40\%            & 48\%  & \hskip0.4cm88\% & 6(24\%)\\ 
G $\leq$ -0.1        &         & 166   & 69\%       & 36\% & 40\%            & 24\%  & \hskip0.4cm64\% & 18(11\%)\\
\hline
G $\geq$ 0.1         & (U-B)   & 27    & 13\%       & 11\% & 41\%            & 48\%  & \hskip0.4cm89\% & 9(33\%)\\ 
-0.1 $<$ G $<$ 0.1   &         & 47    & 24\%       & 19\% & 38\%            & 43\%  & \hskip0.4cm81\% & 10(21\%)\\ 
-0.05 $<$ G $<$ 0.05 &         & 22    & 11\%       & 18\% & 32\%            & 50\%  & \hskip0.4cm82\% & 6(27\%)\\
G $\leq$ -0.1        &         & 128   & 63\%       & 41\% & 35\%            & 24\%  & \hskip0.4cm59\% & 13(10\%)\\
\hline
\hline
\end{tabular}
\end{center}
{\footnotesize{(1): n\'umero de gal\'axias na categoria; (2): fra\cao da amostra total na categoria; (3): 
fra\cao de gal\'axias ordin\'arias; (4): fra\cao de gal\'axias levemente barradas; 
(5): fra\cao de gal\'axias barradas; 
(6): fra\cao total de barradas; (7): gal\'axias com AGN.}}
\end{table}
}

A fra\cao de gal\'axias barradas (SAB+SB) em toda a amostra \'e de aproximadamente 70\%. Podemos ver na 
Tabela 2.2 que existe um excesso 
de gal\'axias barradas entre aquelas gal\'axias que apresentam gradientes nulos ou positivos. Para 
o \1ndice de cor (B\,-V), a fra\cao de 
gal\'axias barradas com o gradiente negativo \'e de 64\%, enquanto que esse valor sobe para 87\% entre as
gal\'axias com o gradiente nulo. Em (U-B), apenas 59\% das gal\'axias com o gradiente negativo s\~ao barradas,
enquanto que 81\% das gal\'axias com o gradiente nulo s\~ao barradas. Se considerarmos o crit\'erio mais 
restrito para a defini\cao de gal\'axias com gradientes nulos ($-0.05 < G < 0.05$), esta diferen\ca \'e
ligeiramente mais acentuada. Portanto, isto indica que as gal\'axias barradas est\~ao sendo super-representadas 
na categoria de objetos com gradiente nulo ou
positivo, indicando que a barra age definitivamente como um mecanismo de homogeneiza\cao do \1ndice de cor 
das gal\'axias.

Outra caracter\1stica interessante \'e que, em (U-B), a fra\cao de gal\'axias barradas com o gradiente
positivo \'e ainda maior. Isto se deve ao fato de que esta cor \'e mais sens\1vel \`a popula\cao e \`a metalicidade.
No entanto, este comportamento n\~ao aparece em (B\,-V). Al\'em disso, a 
fra\cao de gal\'axias com AGN aumenta progressivamente das gal\'axias com gradientes negativos para as com
gradientes nulos e, por fim, para as com gradientes positivos. Isto pode indicar que, pelo menos em parte, 
o fen\^omeno AGN est\'a correlacionado com a presen\c ca de barras.

Para, mais uma vez, fazer uma avalia\cao destas propriedades de maneira a minimizar os efeitos da 
extin\cao intr\1nseca, recalculamos todos os valores apresentados na Tabela 2.2 somente para gal\'axias
vistas de face. Estas perfazem um n\'umero de 124 em (B\,-V) e 104 em (U-B). A Tabela 2.3 \'e equivalente
\`a Tabela 2.2, mas apresenta os valores calculados para as gal\'axias vistas de face, separadamente.

{\small
\begin{table}
\caption{Distribui\cao das gal\'axias de nossa amostra (vistas de face) em rela\cao \`as v\'arias categorias 
a serem avaliadas.}
\begin{center}
\begin{tabular}{||ccccccccc||}
\hline
\hline
			  & Cor     & Total & Amostra & SA   & SAB  & SB   & SAB+SB             & AGN\\
         		  &         & (1)   & (2)     & (3)  & (4)  & (5)  & \hskip0.4cm(6)     & (7)\\
\hline
G $\geq$ 0.1              & (B\,-V) & 14    & 11\%    & 28\% & 29\% & 43\% & \hskip0.4cm72\%     & 5(36\%)\\
-0.1 $<$ G $<$ 0.1        &         & 32    & 26\%    &  9\% & 53\% & 38\% & \hskip0.4cm91\%     & 4(12\%)\\ 
-0.05 $<$ G $<$ 0.05      &         & 17    & 14\%    &  6\% & 47\% & 47\% & \hskip0.4cm94\%     & 4(24\%)\\ 
G $\leq$ -0.1             &         & 78    & 63\%    & 25\% & 47\% & 28\% & \hskip0.4cm75\%     & 8(10\%)\\
\hline
G $\geq$ 0.1         	  & (U-B)   & 19    & 18\%    & 10\% & 37\% & 53\% & \hskip0.4cm90\%     & 7(37\%)\\ 
-0.1 $<$ G $<$ 0.1        &	    & 30    & 29\%    & 17\% & 53\% & 30\% & \hskip0.4cm83\%     & 2(7\%)\\ 
-0.05 $<$ G $<$ 0.05      &         & 16    & 15\%    & 12\% & 38\% & 50\% & \hskip0.4cm88\%     & 5(31\%)\\
G $\leq$ -0.1             &	    & 55    & 53\%    & 27\% & 42\% & 31\% & \hskip0.4cm73\%     & 4(7\%)\\
\hline
\hline
\end{tabular}
\end{center}
{\footnotesize{(1): n\'umero de gal\'axias na categoria; (2): fra\cao da amostra total na categoria; (3): 
fra\cao de gal\'axias ordin\'arias; (4): fra\cao de gal\'axias levemente barradas; 
(5): fra\cao de gal\'axias barradas; 
(6): fra\cao total de barradas; (7): gal\'axias com AGN.}}
\end{table}
}

A fra\cao de gal\'axias barradas na sub-amostra de gal\'axias vistas de face \'e de 79\%. Este aumento pode estar
vinculado ao fato de que \'e mais f\'acil identificar a barra em uma gal\'axia quando esta \'e vista de face
do que quando vista de perfil. Atrav\'es da Tabela 2.3, verifica-se que as propriedades da amostra total,
apresentadas na Tabela 2.2 e nos par\'agrafos acima, permanecem aproximadamente iguais na sub-amostra de gal\'axias vistas
de face. A \'unica diferen\ca \'e a de que, nessa sub-amostra, o excesso de gal\'axias barradas entre as
gal\'axias com gradientes nulos \'e ligeiramente menos significativo, o que ocorre em ambas as cores.

\subsection{Compara\cao com os gradientes de Prugniel \& H\'eraudeau (1998)}

\hskip 30pt Prugniel \& H\'eraudeau (1998 -- doravante PH98) apresentam, em um cat\'alogo, 
estimativas de gradientes de cor (U-B) e (B\,-V)
para a maior parte das gal\'axias em nossa amostra, calculados atrav\'es de dados mais recentes que os 
apresentados em LdV83,85, incluindo fotometria com
observa\coes em CCD. Para evitar os efeitos da contamina\cao dos dados com
observa\coes imperfeitas, estes autores optaram por atribuir pesos estat\1sticos distintos para cada
fonte de observa\c c\~ao.

Evidentemente, \'e interessante comparar os valores dos gradientes obtidos por n\'os com os obtidos por
Prugniel \& H\'eraudeau para as gal\'axias em comum nos dois trabalhos. Na Figura 2.10, o painel 
superior exibe a correla\cao entre os gradientes apresentados em PH98. Este diagrama \'e an\'alogo \`a
Figura 2.9(c), que apresenta a correla\cao para os nossos gradientes. Nota-se que a correla\cao para os
gradientes de PH98 \'e ligeiramente menor. De fato, o valor do coeficiente de correla\cao de Pearson R 
\'e igual 0.68 para PH98, e igual a 0.78 para os nossos valores. Os outros dois pain\'eis da Figura 2.10
apresentam os nossos gradientes em fun\cao dos de PH98. A linha tracejada indica o caso em que a 
correla\cao seria perfeita. A linha cheia mostra o ajuste linear para os pontos do gr\'afico. Como era
de se esperar, a correla\cao \'e bastante boa, com R = 0.85 para (B\,-V) e R = 0.81 para (U-B).

\begin{figure}
\epsfysize=20cm
\centerline{\epsfbox{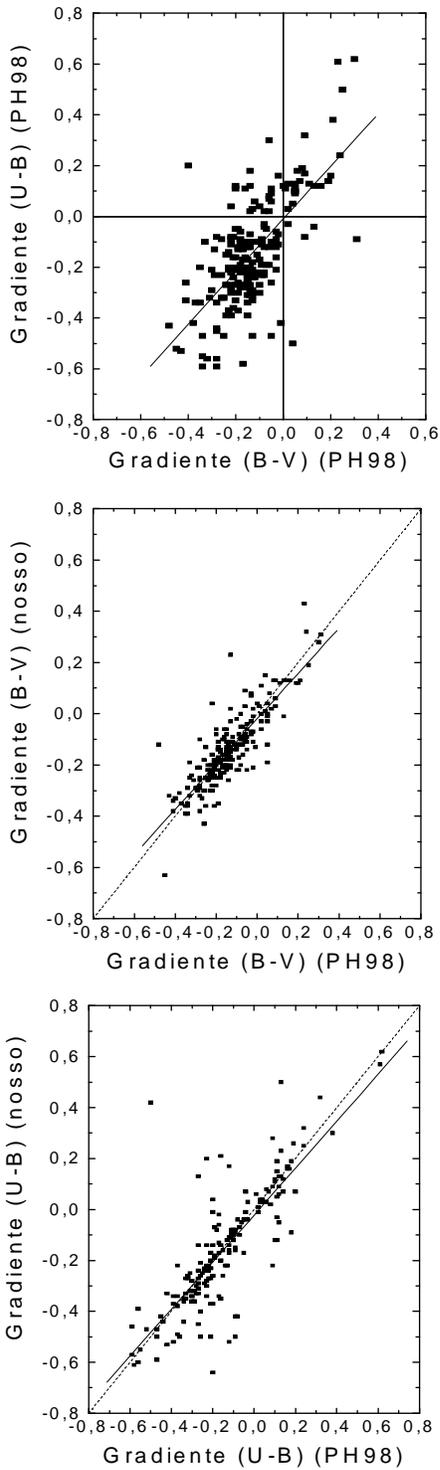}}
\caption{Compara\cao com os gradientes obtidos por PH98. O diagrama superior mostra que a correla\cao
entre os gradientes (U-B) e (B\,-V) em PH98 \'e ligeiramente menor do que a apresentada pelos valores obtidos
por n\'os (ver Figura 2.9(c)). Os outros dois diagramas apresentam os nossos gradientes em fun\cao dos de
PH98. A linha tracejada indica uma correla\cao perfeita. A linha cheia mostra o ajuste 
linear para os pontos do gr\'afico.} 
\label{prugniel-1}
\end{figure}

\begin{figure}
\epsfysize=22cm
\centerline{\epsfbox{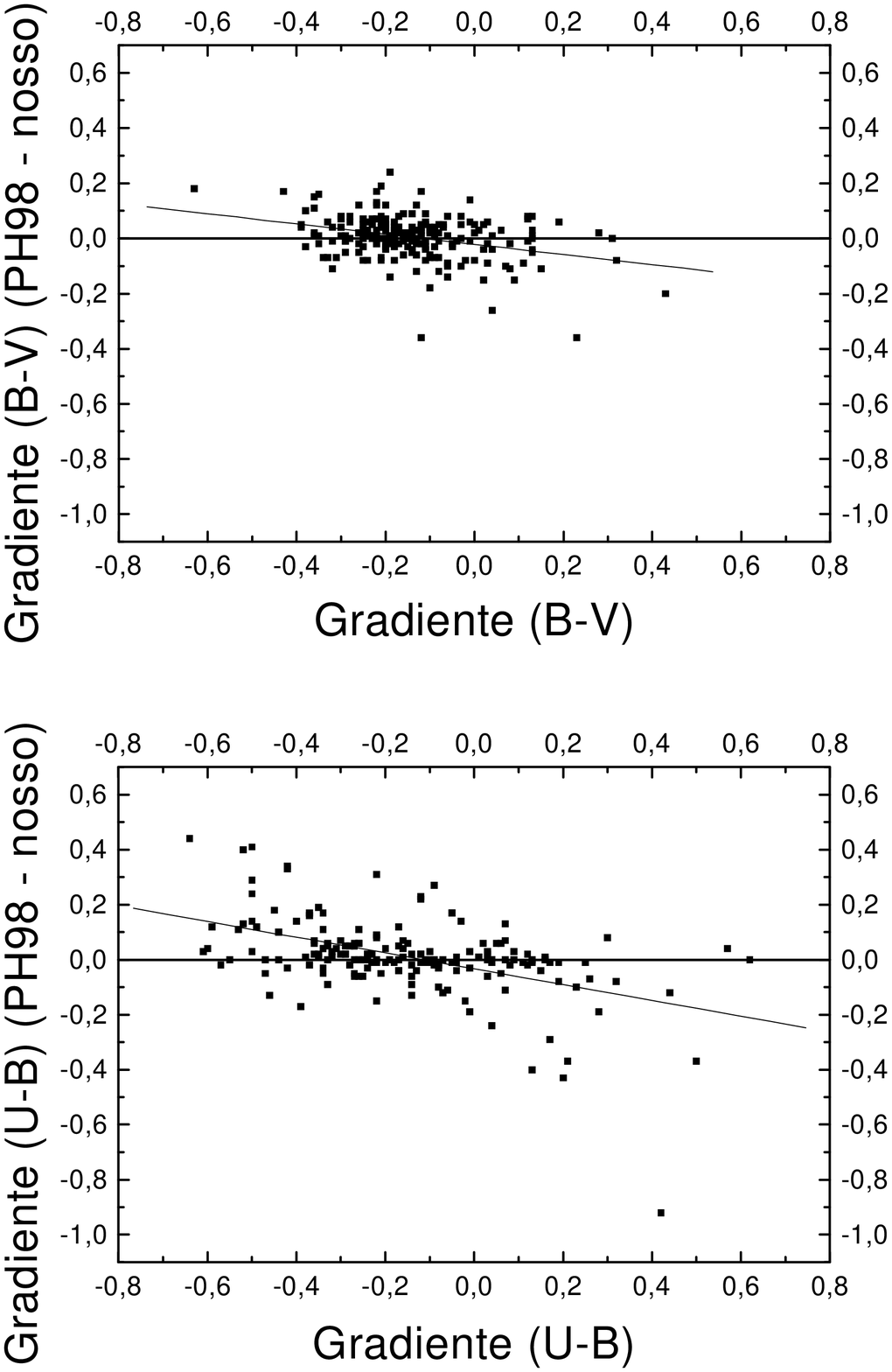}}
\caption{A diferen\ca entre os valores dos gradientes obtidos aqui e os de PH98 em fun\cao dos gradientes
por n\'os determinados. A linha cheia indica o ajuste linear para os pontos.} 
\label{prugniel-3}
\end{figure}

\begin{figure}
\epsfysize=21cm
\centerline{\epsfbox{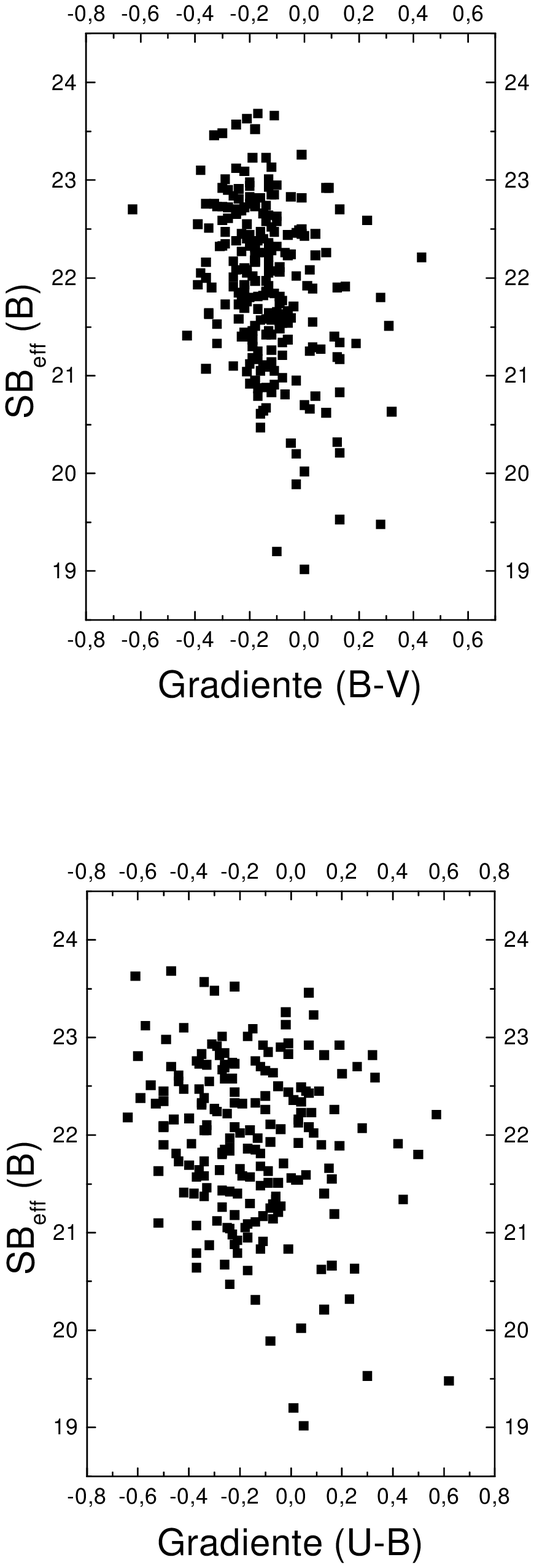}}
\caption{O brilho superficial efetivo na banda B, obtido por PH98 para as gal\'axias em comum com a
nossa amostra, em fun\cao dos gradientes de cor determinados por n\'os. Evidentemente, n\~ao h\'a uma forte 
correla\cao entre esses par\^ametros. Pode-se notar, entretanto, uma leve tend\^encia de as gal\'axias com
gradientes nulos ou positivos apresentarem uma maior concentra\cao central de luz.} 
\label{prugniel-2}
\end{figure}

A Figura 2.11 apresenta a diferen\ca entre os valores obtidos aqui e os de PH98 em fun\cao dos nossos
gradientes. Verifica-se que os gradientes que s\~ao os mais acentuados, tanto negativa quanto positivamente,
s\~ao os que apresentam as maiores diferen\c cas, o que \'e natural de se esperar. \'E importante notar que
n\~ao h\'a nenhuma diferen\ca sistem\'atica entre os dois trabalhos. A linha cheia indica o ajuste linear
para os pontos. O valor m\'edio das diferen\cas \'e de 0.004 em (B\,-V) e de 0.011 em (U-B).

PH98 tamb\'em apresentam o brilho superficial efetivo (i.e., o brilho superficial interno ao raio efetivo
da gal\'axia) na banda B para grande parte das gal\'axias em nossa 
amostra. Este par\^ametro indica a concentra\cao central de luz da gal\'axia.
Para verificar se existe alguma correla\cao entre o brilho superficial efetivo e o gradiente de
cor em uma gal\'axia, elaboramos os gr\'aficos apresentados na Figura 2.12. Esta figura exibe os dados
apresentados por PH98 em fun\cao dos gradientes de cor (U-B) e (B\,-V) por n\'os determinados. Como se v\^e,
n\~ao h\'a correla\cao forte. O que se pode notar, entretanto, \'e uma leve tend\^encia de as gal\'axias com
gradientes nulos ou positivos apresentarem uma maior concentra\cao central de luz. Isto tamb\'em indica 
que os gradientes estimados no presente trabalho n\~ao est\~ao afetados por efeitos sistem\'aticos 
devido aos objetos de maior brilho superficial.

\subsection{A morfologia quantitativa de barras de Martin (1995)}

\begin{figure}
\epsfysize=20cm
\centerline{\epsfbox{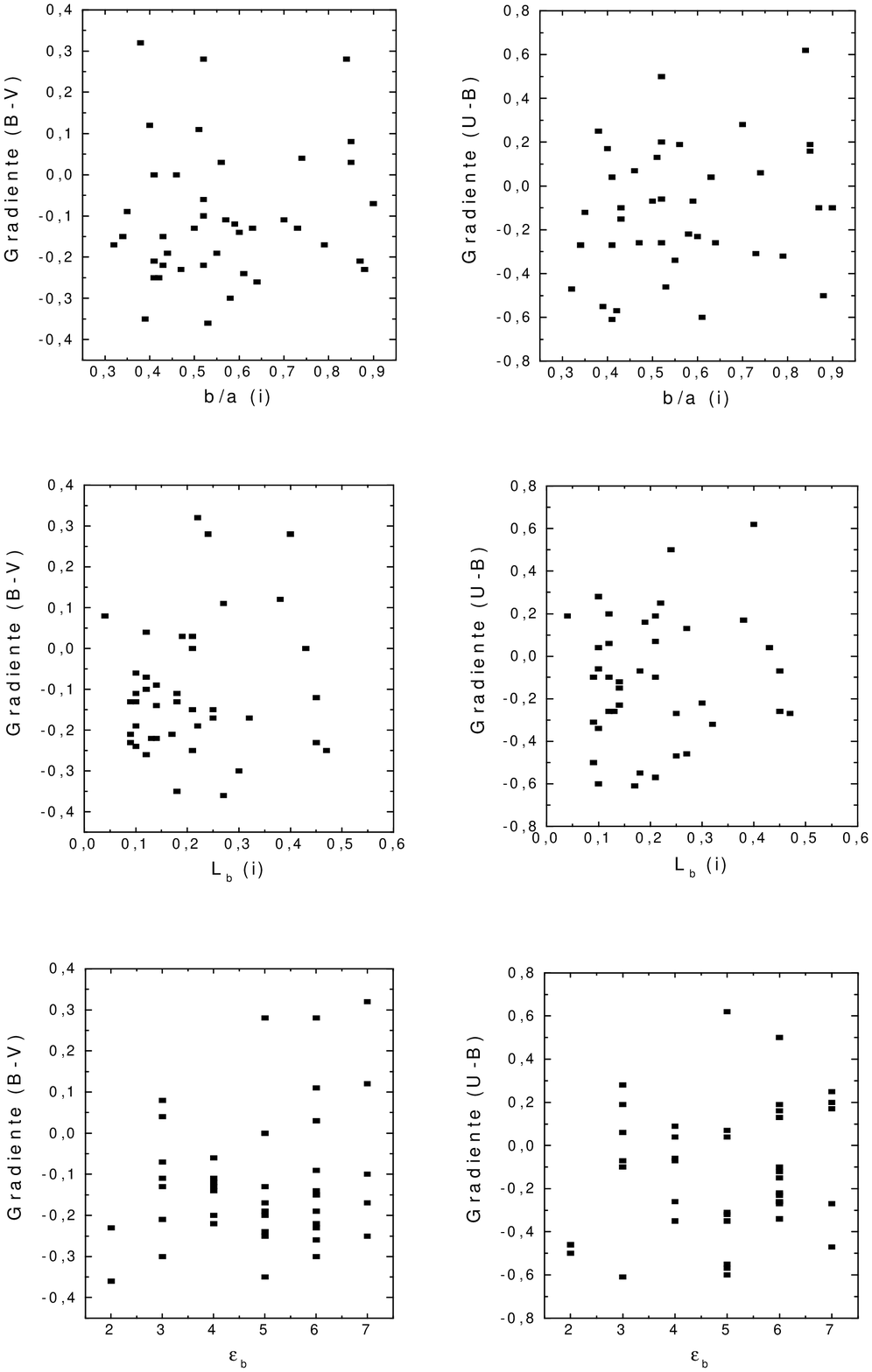}}
\caption{Gradientes de cor em fun\cao dos par\^ametros obtidos em M95 para as barras de 45 gal\'axias de
nossa amostra. Nos diagramas superiores os gradientes est\~ao em fun\cao da raz\~ao axial das barras. Os
diagramas centrais apresentam os gradientes em fun\cao do comprimento normalizado das barras, e nos diagramas
inferiores em fun\cao da elipticidade aparente das barras. \'E evidente que n\~ao existe correla\cao alguma.} 
\label{martin-1}
\end{figure}

\hskip 30pt Martin (1995 -- doravante M95) realiza visualmente medidas da raz\~ao axial, comprimento 
do eixo maior
(normalizado pela isofota de 25 mag arcsec$^{-2}$), e elipticidade aparente de barras em gal\'axias espirais.
Os dois primeiros par\^ametros s\~ao corrigidos por efeitos de inclina\c c\~ao.
Neste trabalho, encontra-se uma rela\cao entre o comprimento da barra e o di\^ametro do bojo, no sentido
de que gal\'axias com bojos menores possuem barras menores. Al\'em disso, encontra-se aparentemente
uma correla\cao entre a presen\ca de forma\cao estelar nuclear intensa e a raz\~ao axial da barra,
no sentido de que barras fortes (i.e., com raz\~ao axial $<$ 0.6) est\~ao presentes em gal\'axias com
surtos de forma\cao estelar nuclear.

Felizmente, 45 gal\'axias de nossa amostra foram estudadas em M95, o que nos permite avaliar a exist\^encia
de correla\coes entre os gradientes de cor e os par\^ametros obtidos por Martin. A Figura 2.13 exibe os
diagramas que apresentam os gradientes de cor em fun\cao dos par\^ametros obtidos em M95 para as barras
destas gal\'axias. Como se v\^e, n\~ao h\'a correla\c c\~oes. Isto significa que, apesar da correla\cao 
entre a presen\c ca de barra e a exist\^encia de gradiente (de cor ou metalicidade) nulo ou positivo, 
o valor do gradiente de cor n\~ao depende da morfologia das pr\'oprias barras. Tanto barras de maior ou menor
extens\~ao, como de maior ou menor raz\~ao axial, residem em gal\'axias que 
apresentam o mesmo comportamento do gradiente de cor.

\subsection{Gradientes de abund\^ancia O/H}

\hskip 30pt Martin \& Roy (1994 -- doravante MR94) e Zaritsky, Kennicutt \& Huchra (1994 -- doravante ZKH94) 
apresentam o gradiente da 
abund\^ancia O/H para gal\'axias espirais, determinado atrav\'es de observa\coes de regi\~oes H{\sc ii}.
Martin \& Roy concluem que gal\'axias barradas tendem a apresentar gradientes menos acentuados do que
gal\'axias ordin\'arias. Tamb\'em concluem que os gradientes tornam-se menos acentuados \`a medida que
o comprimento normalizado ou a elipticidade aparente das barras aumenta (ver se\cao 2.4.4). Zaritsky, 
Kennicutt \& Huchra encontram evid\^encias de que a presen\ca de uma barra induz a atenua\cao do
gradiente.

\begin{figure}
\epsfysize=21cm
\centerline{\epsfbox{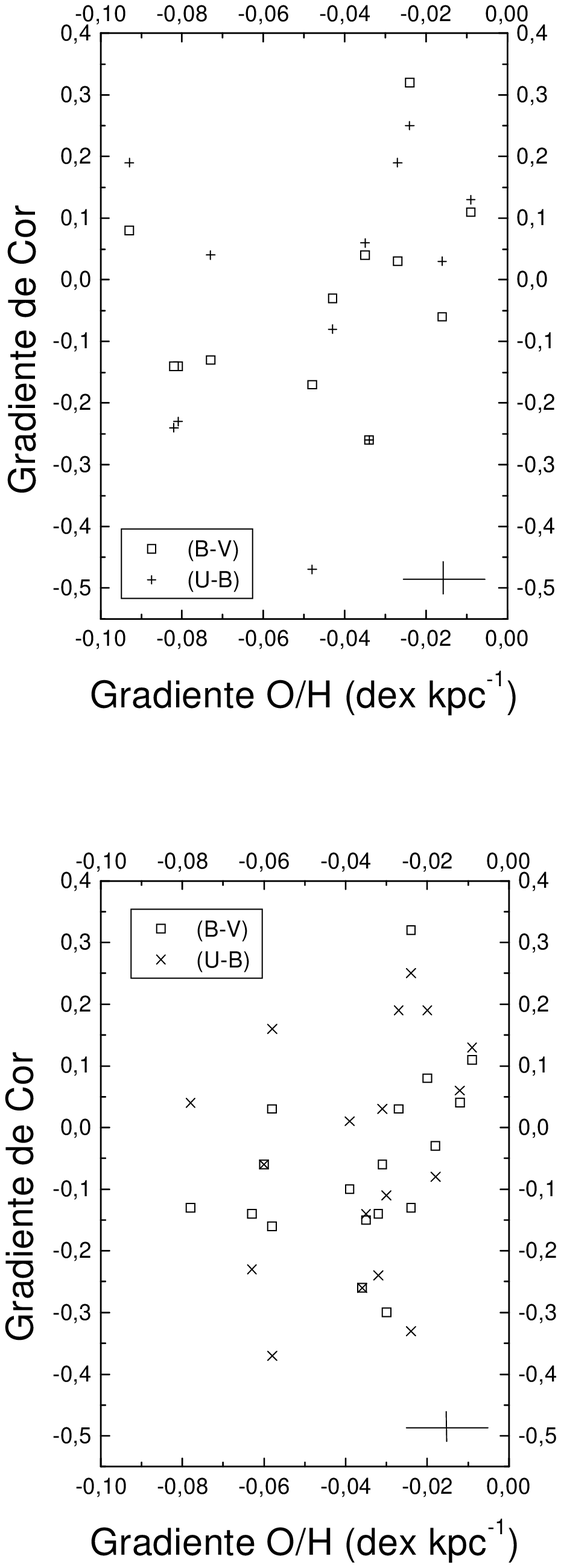}}
\caption{Correla\cao entre os gradientes de cor e os gradientes de abund\^ancia O/H, utilizando os valores
obtidos por MR94 (diagrama superior) e ZKH94 (diagrama inferior). Os quadrados indicam os gradientes em
(B\,-V), enquanto que o sinal + indica aqueles em (U-B). No canto inferior direito de cada gr\'afico, 
barras de erro t\1picas est\~ao desenhadas.} 
\label{martin-2}
\end{figure}

Em nossa amostra, temos 12 gal\'axias em comum com MR94 e 18 com ZKH94. Na Figura 2.14, exibimos nossos
gradientes de cor em fun\cao dos gradientes de abund\^ancia de MR94 (diagrama superior) e de ZKH94
(diagrama inferior).
Os quadrados indicam (B\,-V), enquanto que o sinal + indica (U-B). O primeiro fato a notar nesta figura
\'e a aus\^encia de uma clara correla\cao entre os gradientes fotom\'etricos, tanto $G_{B-V}$ como $G_{U-B}$, 
e o gradiente da abund\^ancia O/H. \'E dif\1cil acreditar que esta aus\^encia de correla\cao seja devido 
a erros na estimativa dos gradientes fotom\'etricos. Estes est\~ao na faixa de 0.02 a 0.05, tipicamente, 
e se encontram indicados na Figura 2.14.
Os erros nas estimativas do gradiente de abund\^ancia s\~ao mais problem\'aticos de serem avaliados, conforme se 
pode constatar pelo gradiente O/H de NGC 2997, bem distinto entre MR94 ($-$0.093 dex kpc$^{-1}$) e ZKH94 ($-$0.020 
dex kpc$^{-1}$). Entretanto, raramente os 
erros no gradiente O/H s\~ao superiores a 0.02 dex kpc$^{-1}$ (ZKH94). Portanto, a alternativa que resta \'e 
a de que esta aus\^encia de correla\cao \'e real. Entretanto, esta alternativa \'e muito interessante, porque, 
sendo o \1ndice de cor sens\1vel potencialmente a dois par\^ametros, quais sejam, a idade e a metalicidade, esta
aus\^encia de correla\cao na Figura 2.14 indica que o gradiente fotom\'etrico n\~ao est\'a correlacionado 
com o gradiente de metalicidade. Portanto, a 
\'unica alternativa que restaria para interpretar tal comportamento seria a de que o excesso de gal\'axias 
barradas com gradiente de cor nulo ou positivo, presente nas Figuras 2.6, 2.7 e 2.8, e Tabelas 2.1 e 2.2, reflete 
uma diferen\c ca no comportamento da {\em idade} m\'edia da popula\cao estelar ao longo de gal\'axias barradas e
ordin\'arias.

\subsection{\'Indices de cor para bojos e discos}

\hskip 30pt A Tabela A.1 exibe os valores dos \1ndices de cor caracter\1sticos de bojos e discos
das gal\'axias em nossa amostra, estimados atrav\'es das equa\coes (2.8) e (2.9), e corrigidos apropriadamente
(ver se\cao 2.3.1). Na Tabela 2.4, apresentamos os valores medianos destes \1ndices.
As gal\'axias est\~ao divididas em categorias, da mesma forma como foram nas 
Tabelas 2.2 e 2.3. Logicamente, para as gal\'axias que apresentam gradientes nulos, n\~ao apresentamos os 
\1ndices de bojos e discos separadamente, j\'a que s\~ao, de fato, apro-ximadamente iguais, mas um \'unico 
valor que se aplica ao longo de toda a gal\'axia. 
Este valor foi tomado como sendo o \1ndice de cor efetivo da gal\'axia, como apresentado no RC3.
Al\'em disso, fizemos uma avalia\cao separada para aquelas gal\'axias vistas de face. Os dados referentes
a estas gal\'axias se encontram \`a direita na tabela. Os erros ($\sigma$) em cada \1ndice de cor est\~ao 
expressos entre par\^enteses, e foram calculados segundo a defini\cao de erro padr\~ao para uma amostra 
de objetos com erros independentes, i.e.,

\eq
{\sigma} = {\delta \over \sqrt{n}},
\eeq

\noindent onde $\delta$ \'e o desvio padr\~ao das medidas e $n$ \'e o n\'umero de medidas.

{\small
\begin{table}
\caption{Valores medianos dos \1ndices de cor caracter\1sticos de bojos e discos para as gal\'axias em nossa 
amostra, separadas
por categoria. Os \1ndices determinados para gal\'axias vistas de face se encontram \`a direita. Os erros 
($\sigma$) est\~ao apresentados entre par\^enteses.}
\begin{center}
\begin{tabular}{||cccc|ccc||}
\hline
\hline
		      & Cor	& Bojo   	  & Disco  	 & Cor   	& Bojo         & Disco \\
	    	      &		& Total	    	  &   	 	 &	 	& Face	       &  	\\
\hline
G $\geq$ 0.1          & (B\,-V)	& 0.34(0.03)  	  & 0.53(0.04) 	 & (B\,-V)      &  0.36(0.03)  & 0.55(0.05) \\
-0.1 $<$ G $<$ 0.1    &	    	& 0.52(0.03)      & 0.52(0.03)   &   		&  0.57(0.02)  & 0.57(0.02) \\ 
-0.05 $<$ G $<$ 0.05  &	    	& 0.51(0.04)      & 0.51(0.04)   &   		&  0.52(0.03)  & 0.52(0.03) \\ 
G $\leq$ -0.1         &		& 0.64(0.01)  	  & 0.43(0.01)   &  	        &  0.74(0.01)  & 0.53(0.01) \\
\hline
G $\geq$ 0.1          & (U-B)	& -0.35(0.06) 	  & -0.08(0.05)	 &  (U-B)       &  -0.31(0.07) & 0.05(0.07)  \\ 
-0.1 $<$ G $<$ 0.1    &	    	& -0.06(0.03)     & -0.06(0.03)  &  		&  -0.05(0.03) & -0.05(0.03) \\ 
-0.05 $<$ G $<$ 0.05  &	   	& -0.05(0.04)     & -0.05(0.04)  &  		&  -0.01(0.03) & -0.01(0.03) \\
G $\leq$ -0.1         & 	& 0.19(0.02)  	  & -0.14(0.01)	 &              &  0.24(0.03)  & -0.05(0.02) \\
\hline
\hline
\end{tabular}
\end{center}
\end{table}
}

\begin{figure}
\epsfysize=20cm
\centerline{\epsfbox{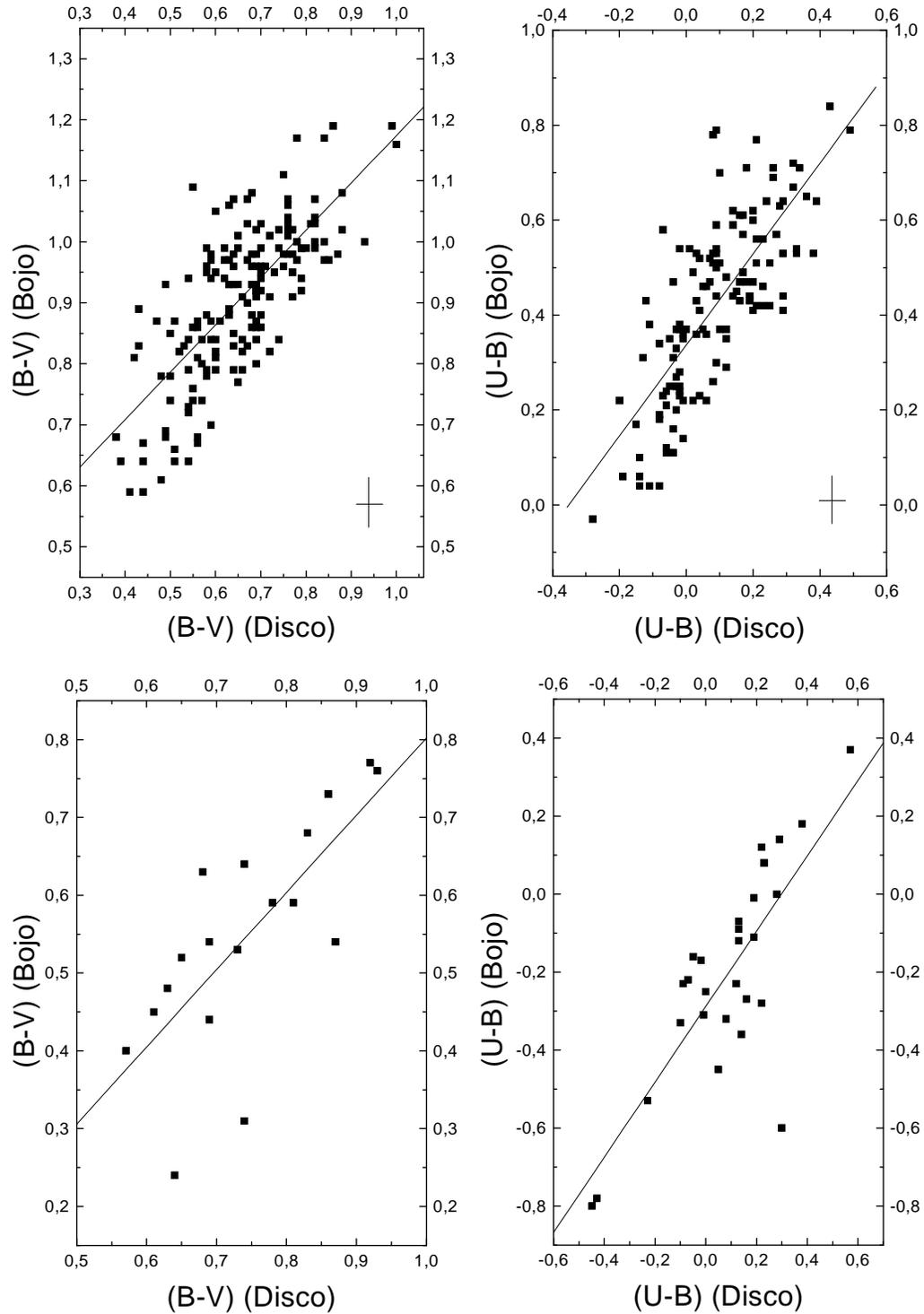}}
\caption{\'Indices de cor do bojo em fun\cao dos \1ndices do disco para as gal\'axias com gradientes 
negativos (pain\'eis superiores) e positivos (pain\'eis inferiores), estimados em primeira
aproxima\c c\~ao (ver se\cao 2.3.1). Percebe-se claramente haver uma 
correla\cao entre ambos os par\^ametros.} 
\label{BulDis1}
\end{figure}

\begin{figure}
\epsfysize=20cm
\centerline{\epsfbox{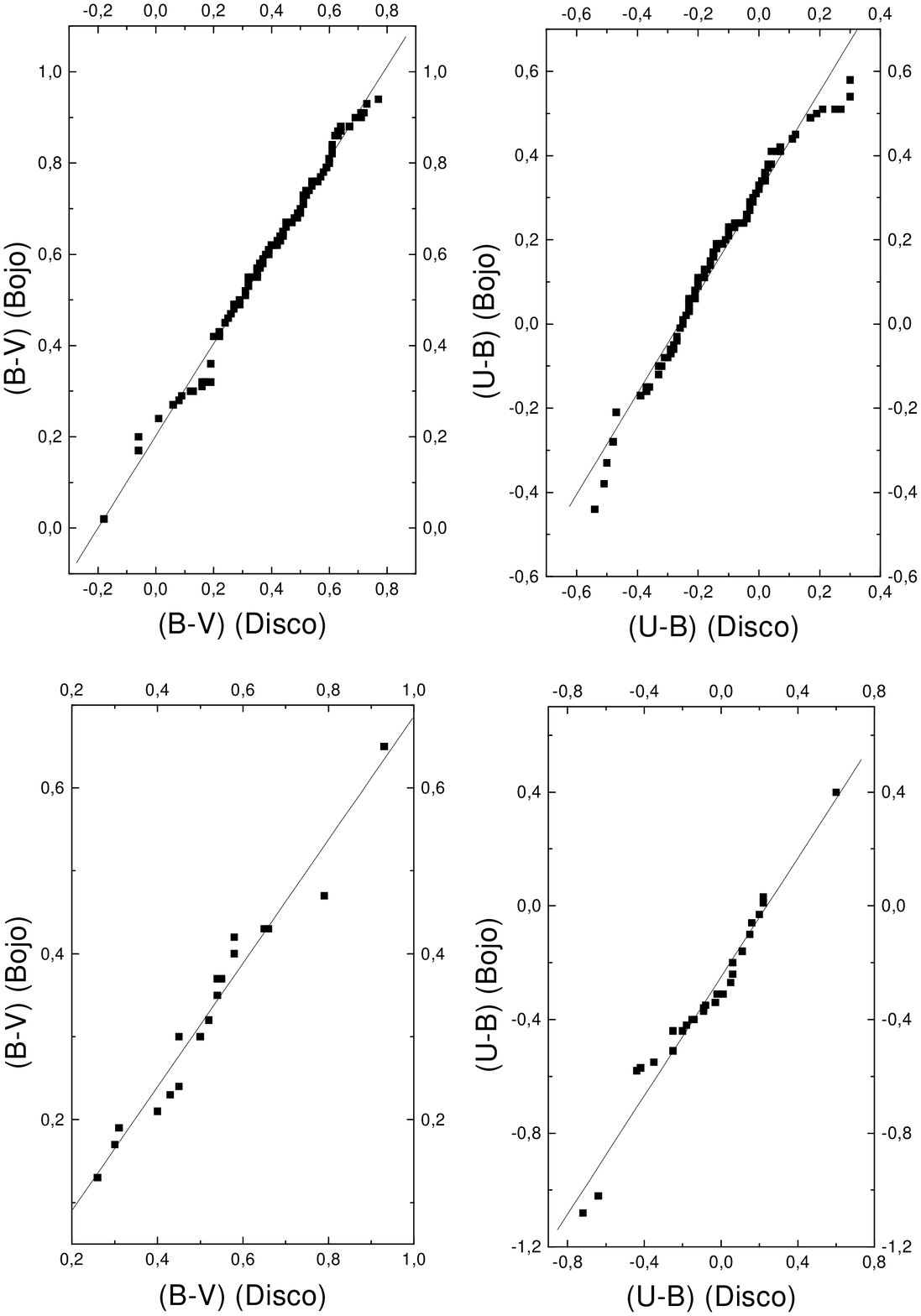}}
\caption{\'Indices de cor do bojo em fun\cao dos \1ndices do disco para as gal\'axias com gradientes 
negativos (pain\'eis superiores) e positivos (pain\'eis inferiores), determinados atrav\'es das 
equa\coes (2.8) e (2.9) (ver se\cao 2.3.1).} 
\label{BulDis2}
\end{figure}

Considerando somente as gal\'axias vistas de face (setor \`a direita na Tabela 2.4), percebe-se que o \1ndice de cor 
das gal\'axias que apresentam
gradientes nulos \'e aproximadamente igual ao \1ndice de cor do disco das gal\'axias com o t\1pico 
gradiente negativo. Mais ainda, o \1ndice do disco permanece aproximadamente constante ao longo das tr\^es
classes de gradientes, enquanto o bojo apresenta \1ndices cada vez menores. Estes resultados se aplicam tanto
a (B\,-V) quanto a (U-B). Isto significa que os discos das gal\'axias que apresentam diferentes 
comportamentos no gradiente, t\^em o mesmo \1ndice de cor em comum.

Passando a analisar o comportamento dos \1ndices de cor para toda a amostra (setor \`a esquerda na Tabela 2.4), 
vemos que o \1ndice de cor 
das gal\'axias que apresentam gradientes nulos continua sendo aproximadamente igual ao \1ndice de cor dos
discos de gal\'axias com o {\em raro} gradiente {\em positivo}, tanto em (B\,-V) quanto em (U-B). Entretanto,
o \1ndice das gal\'axias com gradientes nulos assume valores mais intermedi\'arios entre os valores de bojo 
e disco das gal\'axias com o t\1pico gradiente negativo, embora mais pr\'oximos dos \1ndices dos discos.
Mais uma vez, isto ocorre em ambas as cores. Como esta amostra pode estar mais afetada pela extin\cao intr\1nseca, 
devido \`a maior inclina\cao dos discos, n\~ao se pode descartar que seja esta a causa de tal comportamento.

A Figura 2.15 exibe os \1ndices de cor caracter\1sticos dos bojos em fun\cao dos \1ndices dos discos para as
gal\'axias com gradientes negativos (pain\'eis superiores) e positivos (pain\'eis inferiores). Nesta figura,
os \1ndices s\~ao aqueles determinados atrav\'es da estimativa aproximada descrita na se\cao 2.3.1.
Percebe-se claramente haver uma correla\cao entre ambos os par\^ametros.

A Figura 2.16 \'e an\'aloga \` a anterior. Neste caso, por\'em, os \1ndices s\~ao aqueles determinados
atrav\'es das equa\coes (2.8) e (2.9). Evidentemente, neste caso, a correla\cao entre os \1ndices de
bojos e discos surge naturalmente, uma vez que, subtraindo a equa\cao (2.8) da (2.9), obtemos:

\eq
{{(X-Y)}_{d}} = {{(X-Y)}_{b} + G}.
\eeq

Assim, a estreita correla\cao apresentada nesta \'ultima figura indica, t\~ao somente, que as gal\'axias
apresentam gradientes semelhantes, dentro de cada categoria de gradiente. 
Deste ponto de vista, \'e interessante observar que, se bojo e disco 
estivessem passando por processos de 
evolu\cao n\~ao correlacionados, ent\~ao dever\1amos observar uma maior varia\cao no valor do gradiente. 
Portanto, o fato do gradiente ser aproximadamente o mesmo indica uma clara correla\cao evolutiva entre estas 
duas componentes.

\section{Discuss\~ao e conclus\~oes}

\hskip 30pt Nesta se\c c\~ao, iremos discutir os resultados apresentados na se\cao anterior, separando os
resultados relativos aos gradientes de cor em gal\'axias daqueles que se referem aos \1ndices de cor
de bojos e discos.

\subsection{Gradientes de cor}

\hskip 30pt Um dos primeiros resultados apresentados na se\cao 2.4 foi o de que gal\'axias vistas de 
perfil tendem a ter gradientes de cor mais negativos (ver Tabela 2.1 e Figura 2.6), mas este
comportamento s\'o \'e observado com clareza no \1ndice (U-B). Certamente que este efeito est\'a relacionado ao 
fato de que, para gal\'axias vistas de perfil, o avermelhamento diferencial provocado pela extin\cao interestelar \'e
maior, e tamb\'em mais dif\1cil de ser corrigido. Para que os gradientes em (U-B) sejam mais
acentuados para gal\'axias vistas de perfil,
devido ao avermelhamento interestelar, \'e preciso que as regi\~oes centrais das gal\'axias sofram com maior
intensidade os efeitos do avermelhamento do que as regi\~oes perif\'ericas. Assim, este resultado indica que
a extin\cao interestelar \'e maior no bojo do que no disco das gal\'axias espirais, i.e., o bojo cont\'em
mais poeira do que o disco. Isso est\'a de acordo com os resultados apresentados por de Jong (1996c). Neste
trabalho, de Jong tamb\'em concluiu que os gradientes de cor observados em gal\'axias n\~ao podem ser 
explicados somente pelo excesso de avermelhamento nas regi\~oes centrais, mas que, de fato, o avermelhamento
tem um papel insignificante. Este autor conclui que os gradientes podem ser explicados como um efeito conjunto da
varia\cao da idade e da metalicidade estelar ao longo de gal\'axias espirais.

Outros comportamentos que podem ser observados nas distribui\coes dos gradientes (Figura 2.7 e Tabela 2.1)
s\~ao: {\bf (i)} -- gal\'axias barradas t\^em gradientes menos negativos, mas, novamente, este comportamento
somente ocorre quando se considera o \1ndice (U-B), e, {\bf (ii)} -- gal\'axias barradas t\^em distribui\coes 
mais alargadas, tanto em (U-B) quanto em (B\,-V). Estes dois comportamentos podem ser vistos
como compat\1veis com o cen\'ario de evolu\cao secular. A atenua\cao dos gradientes de cor em gal\'axias
barradas pode decorrer dos efeitos de evolu\cao secular relacionados a barras, j\'a que estes promovem
uma homogeneiza\cao das popula\coes estelares ao longo da gal\'axia, conforme vimos nas se\coes 1.1.3 e
2.1. Ademais, este efeito deve mesmo ser mais vis\1vel na cor (U-B), j\'a que quando ocorre um surto 
de forma\cao estelar, as varia\coes nesta cor s\~ao maiores do que em
(B\,-V), conforme se pode concluir dos modelos de evolu\cao fotom\'etrica de Larson \& Tinsley (1978).
O fato das gal\'axias barradas possu\1rem distribui\coes mais alargadas pode ser interpretado de
algumas maneiras distintas, por\'em ainda dentro do cen\'ario de evolu\cao secular. Por um lado, a presen\ca
de uma barra pode n\~ao ser uma condi\cao suficiente para que haja a homogeneiza\cao das popula\coes 
estelares. A quantidade de g\'as dispon\1vel no disco da gal\'axia pode ter a\1 um papel fundamental.
Uma barra em um disco pobre em g\'as pode n\~ao coletar a quantidade de g\'as necess\'aria para induzir
uma forma\cao estelar central suficientemente poderosa a ponto de tornar similares as popula\coes estelares do
bojo e disco. Por outro lado, sendo as condi\coes favor\'aveis, a presen\ca da barra homogeneiza as
popula\coes estelares ao longo da gal\'axia, atenuando os gradientes de cor, ou at\'e mesmo tornando-os
positivos, no caso em que a forma\cao estelar central, induzida pela barra, \'e ainda recente (o que
torna as cores centrais mais azuladas). De fato, outro resultado que apresentamos, ainda relativo \`a 
distribui\cao dos gradientes de cor, \'e o de que a fra\cao de gal\'axias com gradientes nulos ou positivos
\'e maior entre as gal\'axias barradas (Figura 2.8). Todas essas observa\coes n\~ao somente podem decorrer
do cen\'ario de evolu\cao secular como tamb\'em o favorecem.

Os resultados que apresentamos em rela\cao \`as tr\^es categorias de gradientes (negativos, nulos e positivos),
na se\cao 2.4.2, podem ser discutidos em duas etapas distintas. Primeiro, consideraremos toda a amostra, para,
em seguida, analisar os resultados que se referem somente \`as gal\'axias vistas de face.

Considerando toda a amostra deste estudo, vemos que 70\% das gal\'axias s\~ao barradas 
(SAB+SB). Atrav\'es da Tabela 2.2, 
vimos que, no entanto, essa fra\cao \'e menor entre as gal\'axias com o t\1pico gradiente negativo, sendo
igual a 64\% em (B\,-V) e 59\% em (U-B). Al\'em disso, essa fra\cao aumenta entre as gal\'axias com 
gradientes nulos, indo para 87\% em (B\,-V) e 81\% em (U-B). Portanto, h\'a uma diferen\ca bastante significativa
entre a fra\cao de gal\'axias barradas que t\^em gradientes nulos, e essa mesma fra\cao entre as gal\'axias com
gradientes negativos. Essa diferen\ca \'e ligeiramente mais acentuada quando utilizamos o crit\'erio mais 
restritivo para definir o gradiente nulo. 
Em outras palavras, as gal\'axias barradas est\~ao super-representadas na amostra de objetos com gradientes 
nulos ou positivos.
Esse resultado favorece o cen\'ario de evolu\cao secular, uma vez que
mostra que as gal\'axias barradas tendem a ter gradientes de cor nulos, conforme previsto. A import\^ancia
desse resultado \'e ainda maior na medida em que ocorre de maneira semelhante quer consideremos o \1ndice
(U-B) ou o \1ndice (B\,-V). Se voltarmos os olhos apenas para as gal\'axias vistas de face, 79\% das gal\'axias
s\~ao barradas. Esse aumento \'e provavelmente devido ao fato de ser mais f\'acil identificar uma gal\'axia
barrada quando esta \'e vista de face do que quando \'e vista de perfil. A diferen\ca 
entre a fra\cao de gal\'axias barradas que t\^em gradientes nulos, e essa mesma fra\cao entre as gal\'axias com
gradientes negativos \'e, nesse caso, menos acentuada, mas ainda est\'a presente (ver Tabela 2.3). 
A fra\cao de gal\'axias barradas que
t\^em gradientes negativos \'e de 75\% em (B\,-V) e de 73\% em (U-B), subindo para 91\% em (B\,-V) e 83\% em
(U-B), entre as gal\'axias com gradientes nulos. No entanto, a diferen\ca \'e ainda significativa e, mais
uma vez, o comportamento \'e semelhante nos dois \1ndices de cor estudados. Al\'em disso, 
se considerarmos o crit\'erio mais restritivo para definir o gradiente nulo, a diferen\ca torna-se 
substancialmente mais acentuada. Enfim, o resultado permanece favor\'avel ao cen\'ario de evolu\cao secular, 
mesmo quando se considera somente as gal\'axias vistas de face.

Ainda em rela\cao \`as tr\^es categorias de gradientes, vimos que a maior fra\cao de gal\'axias barradas
ocorre entre as gal\'axias com gradientes positivos, tanto ao considerarmos toda a amostra, quanto ao 
avaliarmos somente as gal\'axias vistas de face. Entretanto, esse comportamento s\'o \'e observado em (U-B), 
j\'a que, como mencionado antes, esta cor \'e mais sens\1vel aos surtos de forma\cao estelar recente.
Por outro lado, a fra\cao de gal\'axias com AGN's aumenta progressivamente das gal\'axias
com gradientes negativos para aquelas com gradientes nulos e, por fim, para as gal\'axias com o raro
gradiente positivo. Esse resultado ocorre nos dois \1ndices de cor, e \'e v\'alido para gal\'axias vistas
de face ou n\~ao. Apesar de uma maior fra\cao de AGN's entre as gal\'axias com gradientes positivos ser
um resultado esperado, o fato de haver um excesso de gal\'axias barradas com gradientes
po-sitivos, pode estar indicando uma conex\~ao entre a presen\ca de barras e de AGN's. Essa conex\~ao
pode favorecer o cen\'ario proposto por Shlosman, Frank \& Begelman (1989) e Shlosman, Begelman \& Frank (1990),
no qual a barra \'e respons\'avel pela manuten\cao de AGN's, coletando g\'as at\'e as regi\~oes centrais 
das gal\'axias. Por outro lado, o fato de haver gal\'axias barradas com gradientes positivos, mas que n\~ao t\^em
AGN's pode indicar que a coleta de g\'as nem sempre chega \`as pequenas escalas de dist\^ancia ao centro, exigidas
para alimentar o AGN, ou ainda que a simples coleta de g\'as para as regi\~oes centrais n\~ao \'e uma condi\cao 
suficiente para originar um AGN. Ho, Filippenko \& Sargent (1997) mostram que, enquanto parece estar bem
estabelecido que as barras de fato aumentam a forma\cao estelar nuclear em gal\'axias, a conex\~ao entre
barras e AGN's \'e muito menos clara, sendo, portanto, uma quest\~ao que ainda precisa de uma 
investiga\cao mais detalhada.

Vamos nos atentar agora \`a compara\cao dos gradientes obtidos por n\'os com aqueles determinados em PH98
(se\cao 2.4.3). Vimos, como era de se esperar, que existe uma boa correla\cao entre os gradientes determinados
em cada trabalho (Figura 2.10 -- pain\'eis inferiores). Al\'em disso, vimos que n\~ao existe nenhuma diferen\ca 
sistem\'atica entre os gradientes calculados em cada trabalho (Figura 2.11) e que as diferen\cas m\'edias s\~ao
bastante pequenas. Vimos tamb\'em que os gradientes em (U-B) e em (B\,-V)
se correlacionam entre si (Figura 2.10 -- painel superior para PH98 e Figura 2.9(c) para este trabalho). No
entanto, a correla\cao entre os gradientes (U-B) e (B\,-V) \'e mais significativa no nosso trabalho. O fato de 
haver tal correla\cao indica que os mesmos fen\^omenos f\1sicos s\~ao respons\'aveis pela origem
dos gradientes em ambas as cores. Um resultado semelhante foi apresentado por Balcells \& Peletier (1994)
para os gradientes (U-R) e (B\,-R), utilizando, entretanto, apenas 18 gal\'axias.
A melhor correla\cao apresentada por n\'os est\'a provavelmente vinculada
ao fato de que utilizamos m\'etodos estat\1sticos mais robustos do que PH98 para determinar os gradientes.
Assim, de certa forma, nossos gradientes s\~ao ligeiramente mais acurados.

Ainda utilizando os resultados em PH98, mostramos que existe uma leve tend\^encia de as gal\'axias com
gradientes nulos ou positivos apresentarem uma maior concentra\cao central de luz, ou seja, um menor
brilho superficial efetivo. No entanto, a dispers\~ao \'e muito grande. Este resultado pode ser interpretado
favoravelmente ao cen\'ario de evolu\cao secular, j\'a que o transporte de g\'as das regi\~oes perif\'ericas 
para as regi\~oes centrais, induzido pela barra, dando origem a surtos de forma\cao estelar central, 
deve tornar a distribui\cao de massa (e, portanto, luminosidade) mais centralmente concentrada. Veremos, no
Cap\1tulo 3, que obtivemos um resultado an\'alogo, que se refere, por\'em, ao brilho superficial efetivo
da componente bojo, separadamente.

Vamos agora explorar os dados apresentados em M95, que se referem a uma morfologia quantitativa de barras.
Nesse trabalho, Martin determina certos par\^ametros morfol\'ogicos em barras de uma amostra de gal\'axias.
Alguns deste par\^ametros, como a raz\~ao axial, comprimento do eixo maior (normalizado pela isofota de 25 
mag arcsec$^{-2}$), e a elipticidade aparente das barras, podem ser considerados como uma medida da for\ca
da barra, ou seja, de sua capacidade em coletar g\'as para as regi\~oes centrais. \'E interessante notar que
MR94 encontram que os gradientes de abund\^ancia O/H em gal\'axias barradas tornam-se menos acentuados
\`a medida que a elipticidade ou o comprimento da barra aumentam. Portanto, isto indica que gal\'axias com
barras fortes t\^em gradientes de abund\^ancia O/H menos acentuados, em completo acordo com o cen\'ario
de evolu\cao secular. Surpreendentemente, entretanto, n\~ao existe nenhuma correla\cao entre os par\^ametros
de for\ca das barras em 45 gal\'axias de nossa amostra, que foram estudadas em M95, com os gradientes de cor
que n\'os determinamos (Figura 2.13). Seria mais natural esperarmos 
um resultado an\'alogo ao encontrado em MR94, i.e., 
que as gal\'axias com barras mais fortes exibissem gradientes menos acentuados. Este resultado mostra que
n\~ao podemos supor que os gradientes de cor (U-B) e (B\,-V) s\~ao completamente equivalentes aos gradientes
de abund\^ancia O/H, e, portanto, os gradientes de cor s\~ao mais sens\1veis \`a idade dos surtos 
de forma\cao estelar. Enquanto que os gradientes de cor s\~ao mais representativos da popula\cao estelar, i.e., 
tipo espectral, temperatura e luminosidade estelares, o gradiente O/H est\'a mais relacionado \`a
abund\^ancia qu\1mica do meio interestelar, e \`a hist\'oria de forma\cao estelar na gal\'axia, j\'a que o 
meio interestelar \'e enriquecido com Oxig\^enio atrav\'es de explos\~oes de supernovas de tipo 2, que s\~ao
produzidas por estrelas de popula\cao I, que s\~ao jovens e t\^em vida relativamente curta.

Ainda em rela\cao aos gradientes de abund\^ancia O/H, utilizando os trabalhos MR94 e ZKH94, elaboramos
a Figura 2.14, procurando correla\coes entre esses gradientes e os gradientes de cor. Corroborando a
n\~ao-equival\^encia entre estas duas classes de gradientes, que encontramos no par\'agrafo anterior, as
correla\coes s\~ao inexistentes, apesar de os erros estimados serem pequenos. Portanto, esta aus\^encia 
de correla\cao indica que os gradientes fotom\'etricos s\~ao mais sens\1veis ao fator idade do que 
\`a metalicidade.

\subsection{\'Indices de cor}

\hskip 30pt Mais uma vez, vamos separar a discuss\~ao dos resultados relativos aos \1ndices de cor, analisando
primeiramente somente os resultados referentes \`as gal\'axias vistas de face, para depois passar aos resultados
obtidos quando toda a amostra \'e considerada.

Temos, em nossa amostra, 124 gal\'axias vistas de face com o gradiente em (B\,-V) determinado, o que corresponde
a 52\% do total. Em rela\cao ao gradiente (U-B), temos 104 gal\'axias, ou 51\% do total de gal\'axias com o 
gradiente (U-B) determinado. Avaliando somente as gal\'axias vistas de face, podemos verificar, atrav\'es da
Tabela 2.4, que a cor das gal\'axias com gradientes nulos \'e bastante similar \`a cor dos discos das gal\'axias
com gradientes negativos, tanto em (B\,-V), quanto em (U-B). Esse comportamento corrobora o cen\'ario de 
evolu\cao secular, j\'a que indica que \'e o bojo que se azula na atenua\cao dos gradientes de cor, e n\~ao
o disco que se avermelha. Portanto, \'e material do disco que \'e coletado para as regi\~oes centrais na 
atenua\cao dos gradientes de cor. De fato, outro resultado importante que pode ser extra\1do da Tabela 2.4
\'e o de que a cor do disco se mant\'em constante nas tr\^es categorias de gradientes, mas o bojo torna-se
cada vez mais azul, quando se parte das gal\'axias com gradientes negativos para as com gradientes nulos e
para as com gradientes positivos. Considerando-se toda a amostra, os resultados n\~ao s\~ao t\~ao significativos, 
j\'a que o \1ndice das gal\'axias com gradientes nulos assume valores mais intermedi\'arios entre os valores de bojo 
e disco das gal\'axias com o t\1pico gradiente negativo, embora mais pr\'oximos dos \1ndices dos discos. Este
comportamento est\'a certamente ligado ao fato de as corre\coes por inclina\cao nas cores de gal\'axias espirais, 
como a que aplicamos neste trabalho (se\cao 2.3.1), serem por demais incertas. Portanto, \'e muito mais confi\'avel
utilizar os resultados referentes \`as gal\'axias vistas de face somente.

Na Figura 2.15, mostramos que as cores de bojos e discos se correlacionam, resultado que j\'a foi obtido por
Peletier \& Balcells (1996) para os \1ndices (U-R), (B\,-R), (R-K) e (J-K), por\'em para uma amostra
de gal\'axias muito menor (somente 30 objetos). Esses autores concluem que, assumindo que as metalicidades
em bojos e discos sejam id\^enticas, ent\~ao as diferen\cas entre as idades estelares nestas componentes
n\~ao pode ser maior do que 30\%, o que \'e compat\1vel com o cen\'ario de evolu\cao secular.

\subsection{Barras recorrentes e a import\^ancia do cen\'ario de evolu\cao secular}

\hskip 30pt Norman, Sellwood \& Hasan (1996) sugerem que a forma\cao da barra, sua dissolu\cao e
conseq\"uente constru\cao do bojo possa ser um fen\^omeno recorrente, ou seja, que ocorra diversas
vezes para uma mesma gal\'axia. J\'a \'e um fato bem estabelecido que uma barra se desenvolve
naturalmente em um disco estelar, atrav\'es de instabilidades din\^amicas (e.g., Binney \& Tremaine 1987).
Tamb\'em j\'a est\'a claro que o aumento na concentra\cao central de massa, induzido pela barra, \'e capaz
de destru\1-la. No entanto, ainda n\~ao est\'a claro se uma gal\'axia que j\'a foi barrada, e teve sua barra
dissolvida, pode desenvolver uma nova barra (Friedli 1999). No processo de dissolu\cao da barra, \'e muito
prov\'avel que a dispers\~ao de velocidades das estrelas no disco aumente, o que torna mais dif\1cil
o desenvolvimento de uma nova barra, j\'a que as barras t\^em maior facilidade de se originar em discos
``frios'', i.e., com baixa dispers\~ao de velocidade, conforme o crit\'erio de Toomre 
(e.g., Binney \& Tremaine 1987)\footnote{Devemos a G. Gilmore o fato de ter-nos chamado a aten\cao para este 
ponto.}.

Em nossa amostra de gal\'axias, temos gal\'axias barradas que n\~ao tem o gradiente nulo, mas o t\1pico gradiente 
negativo. Isso pode ser um ind\1cio de que a presen\ca de uma barra n\~ao \'e uma condi\cao suficiente para
que o gradiente de cor seja atenuado. A quantidade de g\'as dispon\1vel no disco deve ser um outro par\^ametro
importante nesse sentido, como j\'a foi discutido anteriormente. Por outro lado, a barra nesse caso pode ter sido
rec\'em-formada, n\~ao tendo tido tempo ainda para homogeneizar as popula\coes estelares ao longo da gal\'axia.
Inversamente, tamb\'em temos gal\'axias em nossa amostra que, apesar de n\~ao serem barradas, exibem gradientes
de cor nulos. Podemos interpretar este caso como sendo o de uma gal\'axia que teve a sua barra dissolvida
recentemente e, por isso, ainda apresenta o gradiente atenuado. Ap\'os um certo per\1odo de tempo, a evolu\cao
da gera\cao de estrelas formada nas regi\~oes centrais da gal\'axia, que se formou atrav\'es do g\'as 
transportado pela barra, bem como a aus\^encia de novos surtos de forma\cao estelar central, podem tornar o
gradiente mais negativo novamente.

Apesar de nossa amostra n\~ao ser completa, podemos fazer algumas considera\coes frut\1feras.
Considerando que o fen\^omeno de barras seja realmente recorrente, que todas as gal\'axias espirais (pelo
menos as espirais de tipo tardio) sofram desta instabilidade, e assumindo uma idade m\'edia de 10 Giga-anos para essas
gal\'axias, podemos interpretar a fra\cao de gal\'axias em cada categoria de gradiente como se segue.
Vamos imaginar que todas as gal\'axias se formem com gradientes de cor negativos.
Ora, temos 65\% das gal\'axias com gradientes negativos e 35\% com gradientes nulos ou positivos, tornados
assim devido aos processos de evolu\cao secular relacionados \`a barras. Portanto, o per\1odo em que a barra
mant\'em o gradiente atenuado, ou at\'e mesmo positivo, \'e de cerca de 3 a 4 Giga-anos. 
Em verdade, como a presen\ca de uma barra pode n\~ao ser uma condi\cao suficiente para a atenua\cao
dos gradientes, este valor \'e apenas um limite inferior. Podemos adotar
que o tempo decorrido ap\'os a forma\cao da barra para que o gradiente seja atenuado \'e da ordem da escala
de tempo que leva para que ocorra o espessamento da barra, i.e., 1 Giga-ano (ver se\cao 1.1.3). Se o tempo
que leva para que o gradiente volte a tornar-se negativo, ap\'os a dissolu\cao da barra,
for suficientemente pequeno para que possa ser
desprezado, podemos estimar o tempo de vida da barra como sendo de cerca de 5 Giga-anos. Portanto, 
devemos esperar que metade das gal\'axias espirais sejam barradas, o que est\'a em acordo com as observa\c c\~oes, 
pelo menos no que se refere a gal\'axias do Universo local.

Se, ao contr\'ario, barras em discos estelares n\~ao s\~ao recorrentes, podemos interpretar a fra\cao 
de gal\'axias em cada categoria de gradiente, tendo em mente dois distintos cen\'arios para a forma\cao 
de bojos. O cen\'ario monol\1tico prev\^e gradientes de cor negativos, j\'a que a popula\cao estelar 
do bojo \'e mais velha do que a do disco, neste cen\'ario. Portanto, de acordo com essas suposi\coes e os
nossos resultados, 65\% das gal\'axias espirais se formam atrav\'es do 
cen\'ario monol\1tico. No cen\'ario de evolu\cao
secular a popula\cao estelar do bojo deve ter a mesma idade, ou ser mais jovem, que a do disco e, portanto, 
gal\'axias formadas deste modo devem apresentar gradientes de cor nulos ou positivos. Assim, 35\% das 
gal\'axias se formariam, nesta conjectura, atrav\'es do cen\'ario de evolu\cao secular. Em uma conjectura
mista, poder\1amos supor que todas as gal\'axias se formem atrav\'es do cen\'ario monol\1tico, mas que, em
pelo menos 35\% das gal\'axias, os processos de evolu\cao secular relacionados \`a barras contribuam
para a constru\cao dos seus bojos.
   \newpage \chapter{Fotometria Superficial}

\hskip 30pt No cap\1tulo anterior, conclu\1mos, entre outras coisas, que 25\% das gal\'axias avaliadas
neste estudo possuem os mesmos \1ndices de cor em bojos e discos, caracterizando uma homegeneiza\cao de 
popula\coes estelares nestas componentes, compat\1vel com a previs\~ao do cen\'ario de evolu\cao secular.
Neste cap\1tulo, vamos utilizar uma t\'ecnica de decomposi\cao bi-dimensional (2D) para modelar o bojo e o disco,
com o objetivo de estimar os par\^ametros caracter\1sticos dos perfis de luminosidade de bojos e discos, 
em 39 gal\'axias da amostra. 

O interesse na separa\cao das componentes bojo e disco de gal\'axias \'e 
devido ao fato de que a avalia\cao dos par\^ametros fotom\'etricos que descrevem as distribui\coes de brilho 
destas componentes, separadamente, nos fornece informa\coes importantes relativas \`a estrutura das gal\'axias. 
Desta forma, a avalia\cao destes par\^ametros contribui tamb\'em para a avalia\cao de cen\'arios de 
forma\cao e evolu\cao de gal\'axias.
A t\'ecnica de decomposi\cao 2D, aplicada \`a imagem de uma gal\'axia, ajusta um perfil exponencial para o disco, 
segundo a lei
de King, e um perfil que segue a lei de S\'ersic, para o bojo. Desta forma, obtemos um modelo para o bojo
e para o disco de cada
gal\'axia, que corresponde, aproximadamente, devido \`as limita\coes do programa e dos modelos 
escolhidos, \`a estrutura real da gal\'axia.

Atrav\'es dos perfis de luminosidade dos modelos de cada gal\'axia, podemos determinar, entre outros 
par\^ametros, as raz\~oes entre
as luminosidades e os di\^ametros de bojo e disco em cada gal\'axia. 
A partir das estimativas destes par\^ametros, vamos
verificar a exist\^encia de correla\coes que possam indicar a ocorr\^encia dos efeitos de evolu\cao 
secular em gal\'axias barradas. Em particular, pretendemos verificar se existe uma correla\cao entre os
gradientes de cor e as raz\~oes bojo/disco. Se, por exemplo, as gal\'axias com gradientes nulos 
possu\1rem bojos mais proeminentes, ent\~ao esse resultado poderia ser um ind\1cio de que a barra realmente
transfere g\'as e estrelas do disco para o bojo, n\~ao somente homogeneizando as popula\coes estelares
ao longo da gal\'axia, mas tamb\'em contribuindo, dessa forma, para a constru\cao do bojo.

Iniciaremos este cap\1tulo descrevendo os crit\'erios para a sele\cao das 39 gal\'axias de nossa amostra 
escolhidas para a
decomposi\cao bojo/disco. Como as imagens destas gal\'axias, que utilizamos para aplicar o programa, foram
obtidas atrav\'es do DSS, na se\cao 3.2 apresentamos 
informa\coes relevantes a respeito 
destas im\nolinebreak agens. Em seguida, descrevemos resumidamente o programa que realiza a 
decomposi\cao
bojo/disco, e o procedimento que adotamos na obten\cao dos par\^ametros desejados, na se\cao 3.3. 
A se\cao 3.4 exibe nossos resultados. Na se\cao 3.5, apresentamos os resultados da fotometria superficial
aplicada em imagens adquiridas por n\'os, em CCD, no LNA, como
um estudo comparativo. A discuss\~ao acerca destes resultados, 
bem como nossas conclus\~oes, s\~ao apresentadas na se\cao 3.6.

\section{Sele\cao da sub-amostra}

\hskip 30pt O tempo de processamento do programa para a decomposi\cao bojo/disco em uma gal\'axia \'e
da ordem de 30 a 120 minutos, dependendo do tamanho da imagem da gal\'axia, bem como dos valores iniciais
dos par\^ametros de entrada. No entanto, para maximizar a qualidade do ajuste do modelo \`a estrutura
verdadeira da gal\'axia, o programa deve ser aplicado em um modo semi-interativo, i.e., 
o usu\'ario deve aplicar o programa diversas vezes para a mesma gal\'axia, variando os valores dos
par\^ametros de entrada, at\'e obter um ajuste com a qualidade desejada. Os detalhes a respeito deste e de
outros aspectos relativos ao programa ser\~ao apresentados na se\cao 3.3. No entanto, \'e 
importante salientar agora que o tempo de c\'alculo para a determina\cao dos modelos de bojo e disco para
cada gal\'axia \'e demasiado longo.

Al\'em disso, as imagens que utilizamos para modelar as gal\'axias s\~ao imagens do DSS (ver pr\'oxima
se\c c\~ao). Estas imagens
foram obtidas em placas fotogr\'aficas com exposi\coes relativamente profundas e, sendo assim, 
devido \`a pequena
faixa din\^amica destes detectores, algumas imagens encontram-se saturadas na regi\~ao central das 
gal\'axias.

Assim, devido ao longo tempo de processamento do programa e ao problema de satura\cao nas imagens do DSS,
selecionamos uma sub-amostra de gal\'axias, da nossa amostra total, para a qual realizamos a 
decomposi\cao bojo/disco. Os crit\'erios satisfeitos pelas gal\'axias contidas nesta sub-amostra s\~ao:
\begin{enumerate}
\item Gradientes de cor bem determinados.
\item Baixa inclina\cao da normal ao disco da gal\'axia com rela\cao \`a linha de visada, excluindo, assim, 
gal\'axias vistas de perfil.
\item Aus\^encia de perturba\c c\~oes, tais como faixas de poeira e gal\'axias companheiras.
\item \'Area central de satura\c c\~ao pequena em rela\cao ao tamanho aparente do bojo.
\end{enumerate}

Utilizando estes crit\'erios, obtivemos uma sub-amostra inicial de 61 objetos. No entanto, no decorrer do 
trabalho, encontramos gal\'axias cuja \'area central de satura\cao \'e maior do que a que hav\1amos
previamente estimado, o que impossibilitou obtermos modelos confi\'aveis atrav\'es do programa. Portanto,
nossa sub-amostra final de gal\'axias, para as quais obtivemos os modelos de bojos e discos, \'e constitu\1da
por 39 objetos. Estas gal\'axias encontram-se na Tabela A.2, no Ap\^endice A.

Inspecionando as Tabelas A.1 e A.2, pode-se verificar que, em nossa sub-amostra, as fra\coes de gal\'axias
barradas (SAB+SB) e ordin\'arias (SA) s\~ao similares \`as mesmas fra\coes na amostra total. Al\'em disso,
a distribui\cao das gal\'axias na sub-amostra em rela\cao \`as tr\^es categorias de gradientes de cor 
(negativos, nulos e positivos; ver se\cao 2.4.2) tamb\'em \'e similar \`a mesma distribui\cao na amostra
total. Portanto, a nossa sub-amostra \'e representativa da nossa amostra total de gal\'axias.

\section{As imagens do DSS}

\hskip 30pt Como j\'a foi dito, as imagens que utilizamos para a decomposi\cao bojo/disco das gal\'axias em
nossa sub-amostra s\~ao parte do ``Digitized Sky Survey''. O DSS \'e um conjunto de placas
fotogr\'aficas digitalizadas, com imagens obtidas atrav\'es de c\^ameras Schmidt, que cobrem todo o c\'eu. 
O DSS \'e o resultado de um esfor\c co do
``Space Telescope Science Institute'' (STScI) de elaborar um cat\'alogo de estrelas--guia (o ``Guide Star 
Catalog'' -- GSC) para as opera\coes de apontamento do telesc\'opio espacial Hubble (ver Lasker \et 1990).

As placas fotogr\'aficas que constituem o DSS foram obtidas em tr\^es levantamentos. Para o hemisf\'erio
Norte (+90$^{\circ} \geq \delta \geq +6^{\circ}$), as imagens foram obtidas no observat\'orio Palomar, 
na Calif\'ornia, Estados Unidos, no telesc\'opio Oschin, que \'e uma c\^amera Schmidt com espelho de
1.2 metros de di\^ametro, no monte Palomar. A banda fotom\'etrica utilizada \'e semelhante \`a banda V de
Johnson--Morgan. Em verdade, foram utilizadas duas bandas bastante similares, cujos comprimentos de onda
centrais e larguras (entre par\^enteses) s\~ao, respectivamente, 576 (114) e 565 (140), em nanometros.
Exposi\coes t\1picas foram da ordem de 20 minutos. 
Para o hemisf\'erio Sul ($-20^{\circ} \geq \delta \geq -90^{\circ}$), foi utilizada a c\^amera Schmidt
de 1.2 metros de di\^ametro do observat\'orio de Siding Spring, Austr\'alia. A banda fotom\'etrica utilizada
\'e similar \`a banda J (ou $B_{J}$) de Couch--Newell (ver Fukugita, Shimasaku \& Ichikawa 1995), com
comprimento de onda caracter\1stico de 450 nanometros e largura \`a meia altura de 150 nanometros. 
Exposi\coes t\1picas foram da ordem de 60 a 70 minutos. Para a
regi\~ao equatorial do hemisf\'erio celeste, a mesma c\^amera Schmidt foi utilizada.

Para viabilizar a distribui\cao do DSS, a equipe do STScI procedeu \`a compress\~ao de suas imagens. O
algoritmo de compress\~ao utilizado \'e baseado na transforma\cao H (ver White, Postman \& Lattanzi 1992, e
refer\^encias a\1 contidas). Neste trabalho, White, Postman e Lattanzi realizaram a fotometria superficial
de gal\'axias na imagem original e em diversas outras imagens com distintos fatores de compress\~ao. A
conclus\~ao \'e a de que, para fatores de compress\~ao at\'e da ordem de 50, os res\1duos das magnitudes 
entre a imagem original e as imagens comprimidas s\~ao da mesma ordem de grandeza da incerteza intr\1nseca
na fotometria da imagem original, i.e., cerca de 10\%. A fotometria superficial de gal\'axias {\em n\~ao}
\'e afetada pela compress\~ao at\'e fatores da ordem de 50. As imagens do DSS sofreram compress\~ao com
fatores da ordem de 10, o que permitiu que fossem armazenadas em um conjunto de 100 CD's (``Compact Discs'').

Para obter as imagens DSS das gal\'axias em nossa sub-amostra, utilizamos o ``Skyview Virtual 
Observatory'', que \'e um servi\c co prestado pelo Laborat\'orio de Astrof\1sica de Altas Energias (LHEA)
da NASA (``National Aeronautics \& Space Administration''). As im\nolinebreak agens podem ser obtidas atrav\'es do 
seguinte endere\c co na ``World Wide Web'' (WWW): http://skyview.gsfc.nasa.gov/. Tamb\'em obtivemos uma 
imagem diretamente dos CD's do DSS, verificando que as imagens obtidas na WWW s\~ao id\^enticas \`as dos 
CD's.

As placas fotogr\'aficas utilizadas na confec\cao do DSS s\~ao, ao contr\'ario dos CCD's, detectores 
n\~ao-lineares. Quando um feixe
de luz incide sobre uma regi\~ao de uma placa fotogr\'afica, originam-se sais de prata nesta regi\~ao com
uma determinada densidade $\rho$, denominada densidade fotogr\'afica. Entretanto, $\rho$ n\~ao \'e sempre 
linearmente proporcional ao n\'umero de
f\'otons que incidiram nesta regi\~ao da placa fotogr\'afica durante a exposi\cao desta ao feixe de luz.
Para interpretar corretamente as imagens do DSS, precisamos conhecer a intensidade da luz incidente $I$ 
em cada ``pixel'' da imagem. A curva caracter\1stica que relaciona $I$ e $\rho$ nas imagens do DSS pode
ser encontrada em Lasker \et (1990), e \'e representada pela seguinte equa\c c\~ao:

\eq
{I(\rho)} = {\Phi_{1}\rho + \Phi_{2}\ln({e^{(B\rho)}}^{C_{1}} - 1) + \Phi_{3}{e^{(B\rho)}}^{C_{2}} + \Phi_{4}},
\eeq

\noindent onde $B$, $C_{1}$ e $C_{2}$ valem, respectivamente, 0.1, 0.7 e 1.0. Os valores dos coeficientes $\Phi$
tamb\'em s\~ao fornecidos por Lasker \et e dependem de a imagem ter sido obtida em Palomar ou em Siding Spring.

No entanto, os valores em cada ``pixel'' das imagens originais do DSS {\em n\~ao} s\~ao iguais \`a densidade 
fotogr\'afica, mas sim diretamente proporcionais, sendo o fator de convers\~ao igual a 6553.4, conforme pode ser 
visto nos manuais dos CD's do DSS.

Assim, para interpretar corretamente as imagens do DSS, realizamos o seguinte procedimento. Inicialmente,
dividimos o valor de cada ``pixel'' em todas as imagens pelo fator de convers\~ao 6553.4, tornando o valor em cada 
``pixel'' igual \`a densidade fotogr\'afica. Em seguida, para linearizar as imagens, aplicamos a equa\cao 
(3.1) em cada ``pixel'' de cada imagem. Assim, como produto final, obtivemos imagens linearizadas, com o valor
em cada ``pixel'' igual \`a intensidade de luz incidente. \'E importante salientar que as intensidades obtidas
atrav\'es da equa\cao (3.1) s\~ao normalizadas pela intensidade do fundo de c\'eu local.

\section {Decomposi\cao bojo/disco}

\hskip 30pt A distribui\cao de luz em gal\'axias pode ser estudada atrav\'es de perfis radiais de
brilho superficial, em geral, em unidades de magnitudes por segundo de arco ao quadrado. 
Estes podem ser decompostos, para gal\'axias espirais, em 2 componentes: uma que se refere ao bojo, 
e outra correspondente ao disco. A fun\cao que mais se utiliza para descrever a componente disco do perfil de
brilho superficial de uma gal\'axia espiral \'e uma fun\cao exponencial, tamb\'em conhecida como perfil de King 
(ver Mihalas \& Binney 1981). Esta fun\cao pode ser descrita como:

\eq
{\Sigma_{d}(r)} = {\Sigma_{0} e^{-r/h}},
\eeq

\noindent em luminosidade (erg $s^{-1}$). Ou como:

\eq
{\mu_{d}(r)} = {\mu_{0} + 1.086r/h},
\eeq

\noindent em magnitudes. Nas equa\coes acima, $\Sigma_{d}(r)$ e $\mu_{d}(r)$ representam o brilho superficial do
disco em fun\cao da dist\^ancia ao centro $r$. $\Sigma_{0}$ e $\mu_{0}$ representam o brilho superficial central do
disco, e $h$ \'e o raio caracter\1stico do disco.

Para a componente bojo, a fun\cao mais utilizada \'e aquela sugerida por de Vaucouleurs (ver Caon, Capaccioli \&
D'Onofrio 1993 e refer\^encias a\1 contidas), descrita como:

\eq
{\Sigma_{b}(r)} = {\Sigma_{e} 10^{-3.33[{(r/r_{e})}^{1/4}-1]}},
\eeq

\noindent em luminosidade, ou:

\eq
{\mu_{b}(r)} = {\mu_{e} + 8.325 [{(r/r_{e})}^{1/4} - 1]},
\eeq

\noindent em magnitudes. Enquanto $\Sigma_{b}(r)$ e $\mu_{b}(r)$ indicam o brilho superficial do bojo em fun\cao da 
dist\^ancia galactoc\^entrica $r$, $r_{e}$ \'e o raio efetivo do bojo, i.e., aquele que cont\'em metade da
luminosidade total da gal\'axia. $\Sigma_{e}$ e $\mu_{e}$ representam o brilho superficial efetivo, i.e., aquele
na dist\^ancia galactoc\^entrica $r_{e}$.

Embora a lei de de Vaucouleurs represente bem o perfil de brilho
para gal\'axias el\1pticas, e tamb\'em para alguns bojos, v\'arios trabalhos mostraram que alguns
bojos s\~ao melhor representados por um perfil
puramente exponencial (e.g., de Jong 1996a). Nesse caso, torna-se interessante utilizar o perfil generalizado
de de Vaucouleurs, proposto por S\'ersic (Caon, Capaccioli \& D'Onofrio 1993 e refer\^encias a\1 contidas).
O perfil de S\'ersic \'e descrito como:

\eq
{\Sigma_{b}(r)} = {\Sigma_{e} 10^{-b_{n}[{(r/r_{e})}^{1/n}-1]}},
\eeq

\noindent em luminosidade, ou:

\eq
{\mu_{b}(r)} = {\mu_{e} + c_{n} [{(r/r_{e})}^{1/n} - 1]},
\eeq

\noindent em magnitudes, onde $b_{n} = 0.868 n + 0.142$, e $c_{n} = 2.5 b_{n}$, com $n$ denominado por \1ndice
de S\'ersic. No caso em que $n$ \'e igual a 4, temos a conhecida lei de de Vaucouleurs, enquanto que $n = 1$
nos fornece uma lei exponencial semelhante \`aquela utilizada para descrever discos.

A inser\cao de mais um par\^ametro livre para representar os perfis radiais de brilho superficial de gal\'axias
espirais certamente produz melhores ajustes. Por\'em, uma motiva\cao maior para se utilizar o perfil de S\'ersic
vem do fato de que o \1ndice $n$ parece se correlacionar com alguns par\^ametros fundamentais de gal\'axias, tais
como a raz\~ao bojo/disco (Andredakis, Peletier \& Balcells 1995).

Como j\'a foi mencionado, utilizamos um programa (de Souza 1997) para realizar a
decomposi\cao bojo/disco bi-dimensional nas 39 gal\'axias de nossa sub-amostra, adotando a 
equa\cao 3.2 para a componente disco,
e a equa\cao 3.6 para a componente bojo. As imagens utilizadas por este programa s\~ao as imagens linearizadas do
DSS, transformadas em uma matriz de dados, na qual cada posi\cao (linha/coluna) representa um ``pixel'' da
imagem, com o valor da intensidade no ``pixel'' correspondente.

Entretanto, para que o programa possa executar corretamente a minimiza\cao dos erros durante a decomposi\c c\~ao, 
\'e necess\'ario multiplicar a imagem do DSS em intensidade (linearizada e normalizada pelo fundo de c\'eu local)
por uma constante, de modo a converter o ru\1do estat\1stico na imagem em um ru\1do de car\'ater Poiss\^onico, 
tal como em uma observa\cao direta.
Assim, definindo $I_{s}$ como a intensidade do c\'eu ($\approx 1$), e $\sigma_{s}$ o seu desvio padr\~ao absoluto,
precisamos multiplicar a imagem com valores em intensidade $I$ por uma constante $k$, obtendo ent\~ao valores em 
contagens: $C = k I$. As contagens do c\'eu seguir\~ao a mesma express\~ao: $C_{s} = k I_{s}$, e o seu desvio
padr\~ao absoluto (i.e., o ru\1do na imagem), em contagens, \'e: ${\sigma _{cs}}^{2} = k^{2} {\sigma_{s}}^{2}$. 
Para que $\sigma _{cs}$
obede\c ca a estat\1stica de Poisson, deve seguir a express\~ao: ${\sigma _{cs}}^{2} = C_{s}$. Portanto, temos:

\eq
{{\sigma _{cs}}^{2}} = {k^{2} {\sigma_{s}}^{2}} = {C_{s}} = {k I_{s}}.
\eeq

Assim, podemos determinar $k$:

\eq
{k} = {{I_{s}} \over {{\sigma_{s}}^{2}}}.
\eeq

O valor da intensidade do c\'eu, $I_{s}$, o desvio padr\~ao absoluto do c\'eu, em intensidade, $\sigma_{s}$, 
bem como a constante $k$, s\~ao determinados separadamente para cada gal\'axia, atrav\'es da tarefa 
{\sc imstatistics} do pacote {\sc iraf} (``Image Reduction \& Analysis Facility''). Esta tarefa foi aplicada
em 5 regi\~oes distintas de cada imagem, nas quais se observa, essencialmente, somente o fundo de c\'eu.
Os valores medianos de cada par\^ametro foram adotados. Antes de executar o programa,
ainda \'e necess\'ario subtrair a contribui\cao do fundo de c\'eu. Para tanto, simplesmente subtra\1mos da
imagem a constante $C_{s} = k I_{s}$. Al\'em disso, objetos que n\~ao fazem parte de nossa an\'alise, tais como,
estrelas e gal\'axias de fundo (bem como defeitos na imagem), s\~ao retirados da imagem, atrav\'es da tarefa 
{\sc imedit} do {\sc iraf}.

O programa cont\'em v\'arios par\^ametros de entrada a serem fornecidos pelo usu\'ario, entre eles:

\begin{itemize}
\item o tamanho do ``pixel'' em segundos de arco (igual a 1.7 para as imagens do DSS);
\item os valores iniciais das coordenadas do centro da gal\'axia;
\item os valores iniciais do brilho superficial central do disco e do brilho superficial efetivo do bojo;
\item os valores iniciais do raio caracter\1stico do disco e do raio efetivo do bojo;
\item os valores iniciais do \^angulo de posi\cao e da elipticidade de bojo e disco;
\item o valor inicial do \1ndice de S\'ersic;
\item e o valor do ``seeing'' (adotado como sendo igual a 1 ``pixel'').
\end{itemize}

Com estes valores e a matriz de dados correspondente \`a imagem da gal\'axia, o programa procura ajustar os 
perfis de disco e bojo (i.e., as equa\coes 3.2 e 3.6, respectivamente), variando n\~ao somente os par\^ametros dos
perfis, mas tamb\'em os par\^ametros geom\'etricos (elipticidade e \^angulo de posi\c c\~ao). A varia\cao 
destes par\^ametros \'e controlada de modo a maximizar a qualidade do ajuste. No ajuste, considera-se que
as isofotas dos bojos e discos s\~ao elipses perfeitas, e que o disco se extende at\'e o centro da gal\'axia. Assumimos
que n\~ao existem erros residuais relativos \`a subtra\cao do c\'eu, mas apenas um ru\1do
de car\'ater Poiss\^onico na matriz de dados. Quando a varia\cao na qualidade do ajuste cai abaixo de um certo
limite de converg\^encia, o programa \'e interrompido, gerando os par\^ametros a seguir:

\begin{itemize}
\item os valores finais do brilho superficial central do disco e do brilho superficial efetivo do bojo;
\item os valores finais do raio caracter\1stico do disco e do raio efetivo do bojo;
\item o valor final do \1ndice de S\'ersic;
\item e as raz\~oes entre as luminosidades e os di\^ametros de bojo e disco.
\end{itemize}

Desta forma, o programa \'e aplicado para cada gal\'axia cerca de 3 a 5 vezes, tipicamente. Isso porque \'e 
necess\'ario variar os valores iniciais dos par\^ametros de entrada, at\'e encontrar o ajuste mais satisfat\'orio.
Isto decorre do car\'ater altamente n\~ao linear dos perfis de brilho, que, em alguns casos, gera solu\coes esp\'urias.
Por exemplo, se existem barras ou bra\c cos espirais, a orienta\cao do eixo maior da gal\'axia tende a sofrer 
varia\coes muito bruscas, particularmente se esta est\'a sendo vista de face.

\'E comum encontrar na literatura decomposi\coes bojo/disco que utilizam somente uma dimens\~ao ao longo da
imagem da gal\'axia, em geral ao longo do seu eixo-maior. De toda a informa\cao contida na imagem completa
da gal\'axia, extrai-se um valor caracter\1stico para se construir um perfil que se ajuste em uma \'unica
dire\c c\~ao. Este m\'etodo tem a vantagem de aumentar a raz\~ao sinal/ru\1do, mas perde toda a informa\cao
presente na imagem da gal\'axia, relativa a componentes n\~ao-axissim\'etricas, como barras, por
e-xemplo. O ajuste bi-dimensional de modelos a imagens de gal\'axias tem poucos exemplos na literatura 
(e.g., de Jong 1996a e refer\^encias a\1 contidas). Apesar de o modelo utilizado neste trabalho n\~ao incluir uma barra, 
a t\'ecnica utilizando um ajuste bi-dimensional \'e certamente mais confi\'avel do que um m\'etodo uni-dimensional, 
n\~ao somente para gal\'axias barradas, que correspondem a 85\% das gal\'axias em nossa sub-amostra, mas 
para gal\'axias em geral.

Com os par\^ametros dos perfis ajustados, o programa constr\'oi imagens artificiais da gal\'axia modelo, 
bem como do bojo e do disco separadamente. Em seguida, utilizando a tarefa {\sc ellipse} do {\sc iraf}, 
constru\1mos os perfis de brilho da gal\'axia original, da gal\'axia modelo, do bojo e do disco, al\'em
dos perfis de elipticidade, \^angulo de posi\cao e do coeficiente de Fourier b4, para a gal\'axia
original e para a gal\'axia modelo. O comportamento do coeficiente b4 indica o quanto as isofotas da
gal\'axia s\~ao distintas de uma elipse perfeita. Um valor positivo para b4 indica isofotas com uma
sub-componente ``disky'', enquanto que um valor negativo indica isofotas ``boxy''. Se b4 \'e igual a 0, 
ent\~ao a isofota \'e uma elipse perfeita. Todo esse procedimento nos permite avaliar a qualidade do
ajuste determinado pelo programa e ainda outros par\^ametros estruturais da gal\'axia. Al\'em disso, 
subtra\1mos da imagem original as imagens sint\'eticas, obtendo imagens residuais, com o objetivo de
avaliar eventuais sub-estruturas.

A Figura 3.1 exibe os resultados da decomposi\cao bojo/disco para a gal\'axia NGC 488, como um exemplo.
Nesta figura, temos, da esquerda para a direita e de cima para baixo:

\begin{itemize}
\item imagem original (passada por um filtro mediano e transformada para brilho superficial) com o mapa
isofotal sobreposto;
\item imagem residual total (imagem original $-$ modelo completo);
\item imagem residual do disco (imagem original $-$ modelo do bojo);
\item imagem residual do bojo (imagem original $-$ modelo do disco);

\item perfis de brilho na banda V (em magnitudes por segundo de arco ao quadrado em fun\cao do raio em
``pixels''):

\begin{itemize}
\item pontos com barras de erro: imagem original;
\item linha com tra\c cos curtos: modelo do disco;
\item linha com  tra\c cos longos: modelo do bojo;
\item linha cheia: modelo completo;
\item pontos com linha cheia: res\1duo total em magnitudes por segundo de arco ao quadrado
(imagem original - modelo + 28);
\end{itemize}

\item perfis de elipticidade:

\begin{itemize}
\item pontos com barras de erro: imagem original;
\item linha cheia: modelo completo;
\end{itemize}

\item perfis do \^angulo de posi\c c\~ao:

\begin{itemize}
\item pontos com barras de erro: imagem original;
\item linha cheia: modelo completo;
\end{itemize}

\item perfis do coeficiente de Fourier b4:

\begin{itemize}
\item pontos com barras de erro: imagem original;
\item linha cheia: modelo completo;
\end{itemize}

\end{itemize}

Para calibrar nossos dados, de modo a deixar nossos perfis compat\1veis com as magnitudes observadas
na literatura para as gal\'axias em nossa sub-amostra, utilizamos a tarefa {\sc imexamine} do {\sc iraf}.
Com a tarefa {\sc imexamine}, calculamos a magnitude da gal\'axia, na imagem em contagens, em sete valores
distintos de abertura, comparando os resultados com a magnitude V aparente publicada em LdV83,85 para as
mesmas aberturas. A constante de calibra\cao $C_{C}$ para cada gal\'axia \'e tomada como sendo a mediana
das diferen\c cas entre essas magnitudes, somada \`a constante de transforma\cao para brilho superficial
e subtraindo o termo de corre\cao para a extin\cao Gal\'actica (ver se\cao 2.3.1), utilizando os mapas
de Schlegel, Finkbeiner \& Davis (1998). Assim, 

\eq
{C_{C}} = {mag(LdV83,85) - mag({\sc imexamine}) + 5 \log_{10} (1.7) - 3.1 \times E(B-V)}.
\eeq

\begin{figure}
\epsfysize=22cm
\centerline{\epsfbox{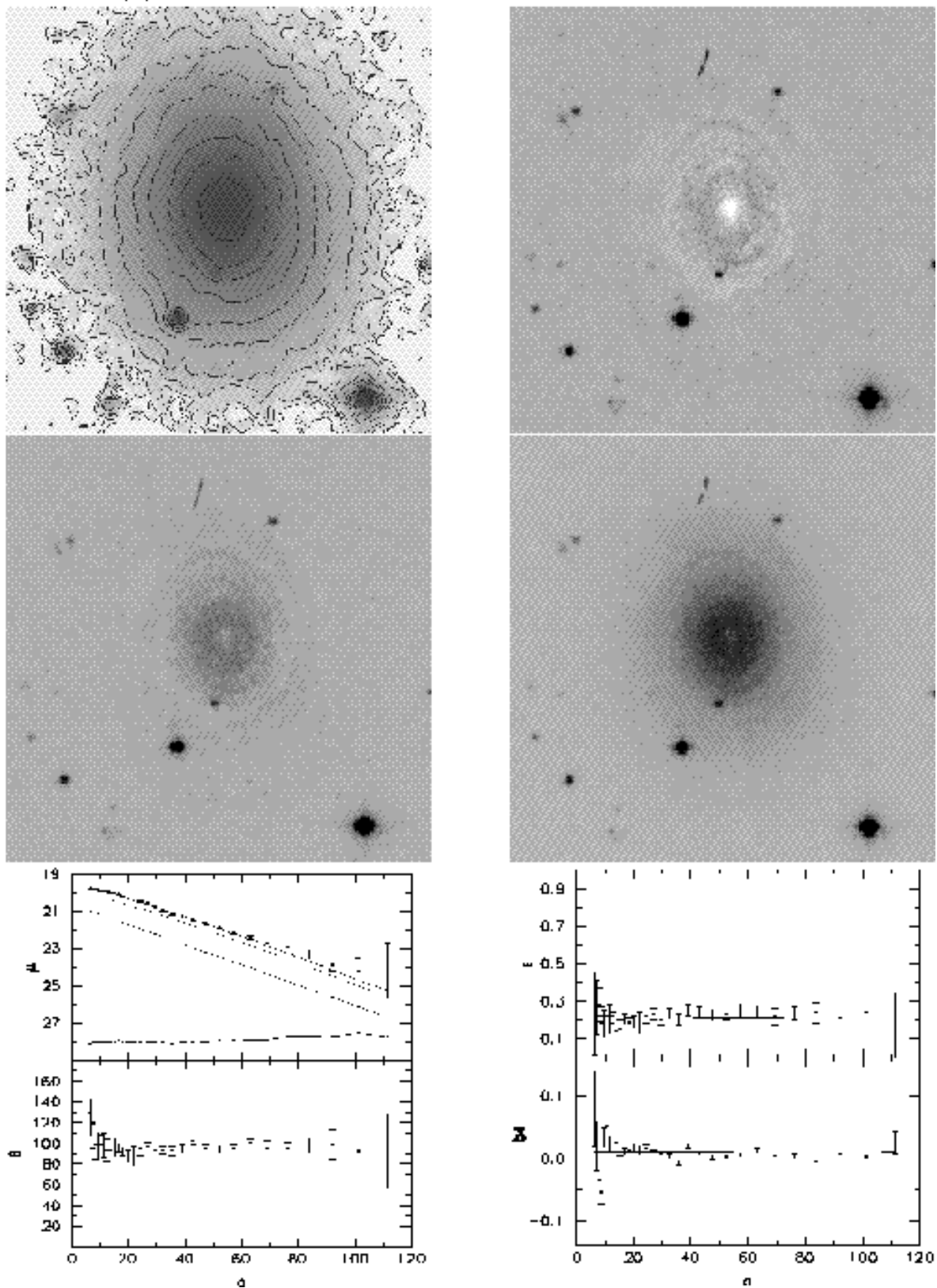}}
\caption{Resultados da decomposi\cao bojo/disco para a gal\'axia Sb(r) NGC 488. No alto est\'a especificada a
dimens\~ao das imagens em segundos de arco. Para legendas, ver texto.} 
\label{N488}
\end{figure}

Muita informa\cao pode ser extra\1da da Fig. 3.1. Em rela\cao \`as imagens, v\^e-se prontamente que o bojo \'e
dominante nessa gal\'axia. De fato, NGC 488 apresenta a maior raz\~ao entre as luminosidades de bojo e disco 
(igual a 3.1) entre as gal\'axias de nossa sub-amostra. Tamb\'em se verifica que os bra\c cos espirais 
s\~ao mais aparentes nas imagens residuais, o que
\'e esperado, j\'a que os bra\c cos n\~ao s\~ao ajustados no modelo. Os perfis mostram que o ajuste 
alcan\c cado \'e muito bom, e que o bojo segue um perfil puramente exponencial, semelhante ao disco.

\section{Resultados}

\hskip 30pt A Tabela A.2 exibe os resultados da decomposi\cao bojo/disco para as 39 gal\'axias de nossa sub-amostra.
Como j\'a foi dito, um dos objetivos deste estudo \'e o de verificar se existe uma correla\cao entre os
gradientes de cor e as raz\~oes bojo/disco.

Verificamos que n\~ao h\'a qualquer correla\cao entre as raz\~oes bojo/disco e os gradientes de cor.
Uma poss\1vel explica\cao para a aus\^encia dessa correla\cao em nossos dados pode vir do problema relacionado
\`a satura\cao das imagens do DSS na regi\~ao central da gal\'axias. O estudo que realizamos na pr\'oxima se\c c\~ao, 
comparando os resultados da decomposi\cao bojo/disco em imagens do DSS e em imagens CCD, mostra que a satura\cao
central nas imagens DSS afeta substancialmente os perfis de brilho das gal\'axias e, evidentemente, os 
perfis de bojos e discos.
Veremos que o \1ndice de S\'ersic e os par\^ametros caracter\1sticos dos perfis dos discos s\~ao muito perturbados
por esse problema. No entanto, veremos tamb\'em que o raio efetivo do bojo \'e um par\^ametro que 
pode ser determinado com certa confiabilidade atrav\'es das imagens do DSS, e resulta em uma an\'alise mais acurada.

\begin{figure}
\epsfysize=20cm
\centerline{\epsfbox{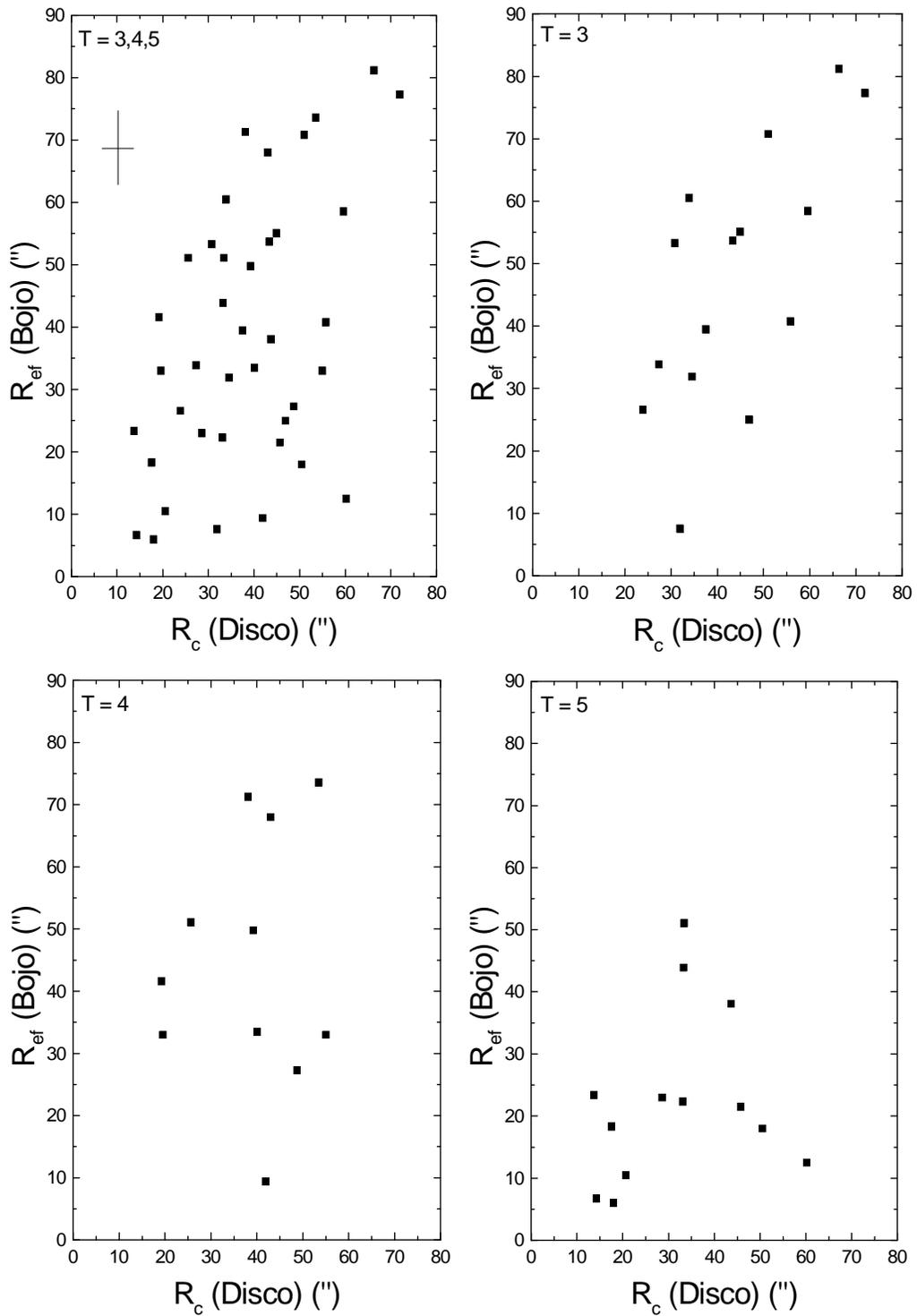}}
\caption{A correla\cao entre os raios efetivos de bojos e os raios caracter\1sticos de discos.} 
\label{bd7}
\end{figure}

\begin{figure}
\epsfysize=20cm
\centerline{\epsfbox{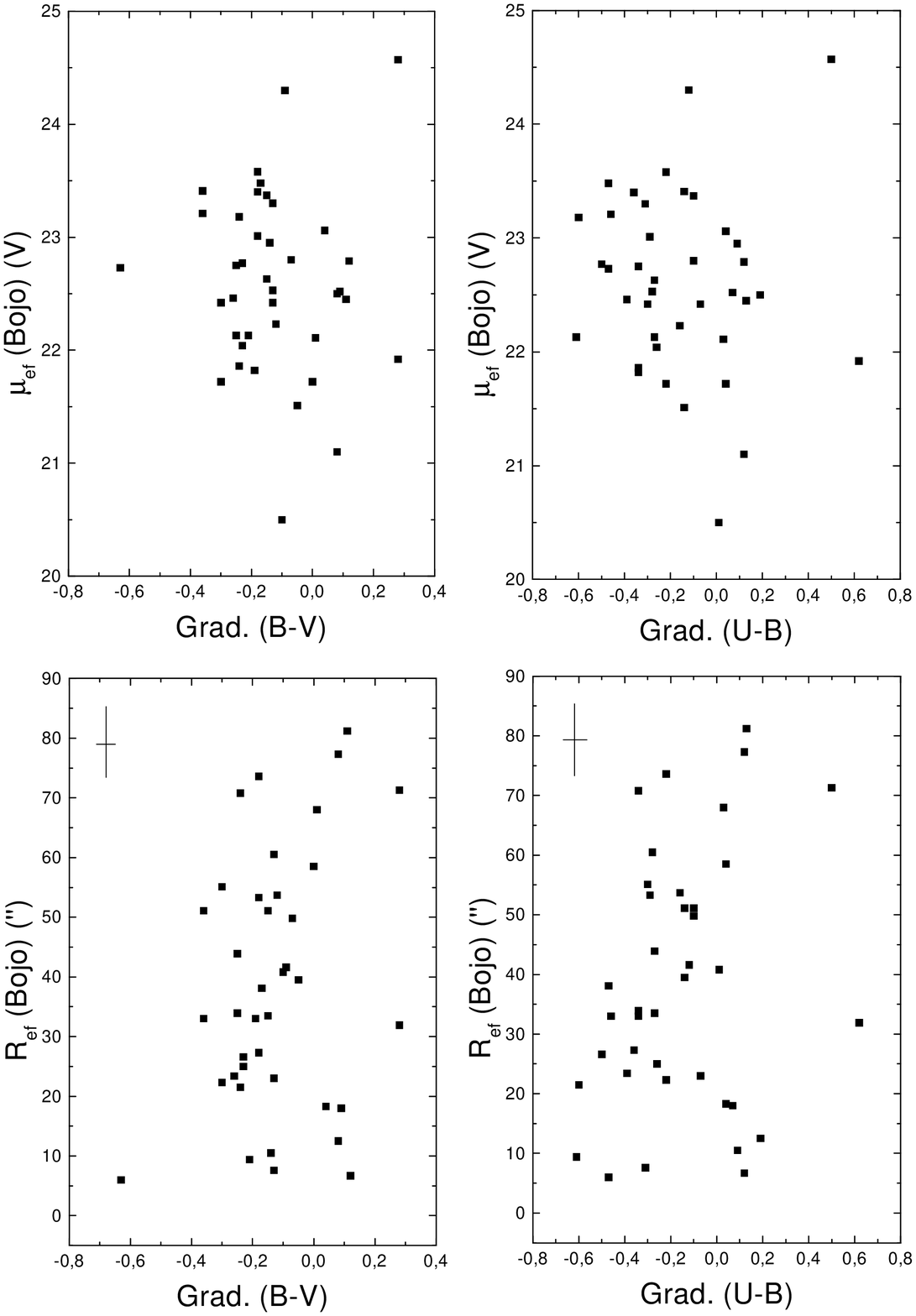}}
\caption{Par\^ametros caracter\1sticos dos perfis dos bojos em fun\cao dos gradientes de cor.} 
\label{bd1}
\end{figure}

Com a inten\cao de verificar o comportamento apontado por de Jong (1996b) e por Courteau, de Jong \& Broeils (1996), 
de que existe uma correla\cao entre as escalas de comprimento de bojos e discos em gal\'axias espirais, comparamos
os raios efetivos dos bojos e os raios caracter\1sticos dos discos que determinamos para as gal\'axias de nossa 
sub-amostra (Figura 3.2). Nesta figura, podemos ver que existe uma ligeira correla\c c\~ao, quando se considera toda a
sub-amostra, i.e., T = 3, 4 e 5. Vemos que a correla\cao me-lhora substancialmente ao considerarmos somente as
gal\'axias Sb's (T = 3), e que n\~ao h\'a correla\cao para o restante das gal\'axias (T = 4 e T = 5). A correla\cao
apresentada nos trabalhos citados acima tamb\'em apresenta uma elevada dispers\~ao, de modo que a correla\cao que
apresentamos aqui pode ser considerada como uma confirma\cao desse resultado.

Na Figura 3.3, mostramos os par\^ametros caracter\1sticos dos perfis dos bojos em fun\cao dos gradientes de
cor das gal\'axias. Os pain\'eis superiores se referem ao brilho superficial efetivo na banda V em magnitudes
por segundo de arco ao quadrado, e os pain\'eis inferiores dizem respeito ao raio efetivo em segundos de arco.
Vemos que existe uma fraca correla\c c\~ao, com elevada dispers\~ao, ou seja, uma leve tend\^encia, de as gal\'axias com
gradientes nulos ou positivos apresentarem bojos com uma concentra\cao central de luz elevada (i.e., um valor
reduzido para o brilho superficial efetivo) tanto em (B\,-V) quanto em (U-B). Apesar dos problemas j\'a apontados, 
relativos \`a determina\cao do brilho superficial efetivo do bojo em imagens do DSS, essa leve tend\^encia pode
estar indicando que os efeitos de evolu\cao secular, ao atenuar os gradientes de cor, contribuem tamb\'em para
que a distribui\cao de massa (e, portanto, luminosidade) no bojo seja mais concentrada. Al\'em disso, vemos, 
nos pain\'eis inferiores da Figura 3.3, que tamb\'em h\'a uma leve tend\^encia de as gal\'axias com gradientes
nulos ou positivos apresentarem bojos maiores.

Tamb\'em procuramos outras correla\coes poss\1veis entre os par\^ametros adquiridos na decomposi\cao bojo/disco
e os gradientes de cor, bem como, com os \1ndices de cor de bojos e discos. No entanto, nenhum resultado 
significativo pareceu-nos digno de nota.

\section{Observa\coes no LNA}

\hskip 30pt Iremos apresentar, nesta se\c c\~ao, resultados obtidos atrav\'es do imageamento em CCD
de 14 gal\'axias de nossa amostra,
realizado no Observat\'orio do Pico dos Dias (OPD -- Bras\'opolis), junto ao Laborat\'orio Nacional de
Astrof\1sica (LNA -- Itajub\'a).

Esse estudo nos permitir\'a fazer compara\coes com os resultados do Cap\1tulos 2 e da se\cao anterior. No Cap\1tulo 2,
mostramos um estudo de gradientes de cor em gal\'axias, realizado atrav\'es de 
medidas que se utilizaram de t\'ecnicas de fotometria
fotoel\'etrica de abertura. J\'a nas se\coes anteriores, o estudo apresentado baseia-se em placas 
fotogr\'aficas digitalizadas. Assim, com os dados coletados no LNA, podemos comparar gradientes de cor obtidos
atrav\'es de fotometria fotoel\'etrica de abertura com aqueles obtidos no imageamento em CCD. Da mesma
forma, vamos comparar os resultados da decomposi\cao bojo/disco em imagens do DSS com os resultados
da decomposi\cao nas imagens em CCD por n\'os obtidas.

O restante desta se\cao assim segue. Na pr\'oxima subse\c c\~ao, descreveremos sucintamente a obten\cao e o 
tratamento das imagens coletadas no LNA. A subse\cao 3.5.2 apresenta o estudo comparativo dos gradientes de cor  
e dos resultados na decomposi\cao bojo/disco. A discuss\~ao e nossas 
conclus\~oes acerca dos resultados desta se\cao s\~ao apresentadas na subse\cao 3.5.3.

\subsection{Aquisi\cao e tratamento das imagens}

\hskip 30pt As observa\coes fotom\'etricas foram realizadas em Outubro de 1997, e em Julho e Agosto de 1998,
no telesc\'opio com espelho de 60 cent\1metros de di\^ametro localizado no OPD--LNA. A \'optica deste
telesc\'opio \'e do tipo Ritchey--Chr\'etien, e a raz\~ao focal no foco Cassegrain \'e f/13.5. Foi utilizada
uma c\^amera direta do LNA, acoplada a um detector CCD SITe SI003AB, fino e retro-iluminado
(o CCD n\'umero 101 do LNA).
Este CCD possui 1024 por 1024 ``pixels'' com dimens\~ao de 24 micrometros cada. Como a escala de placa \'e 
de 0.57 segundos de arco por ``pixel'', o campo total observado \'e de aproximadamente 10 por 10 minutos
de arco. O ganho neste CCD \'e de 5 el\'etrons por ADU, com um ru\1do de leitura de 5.5 el\'etrons.

A Tabela 3.1 exibe um resumo das observa\c c\~oes. Nesta tabela, mostramos as gal\'axias observadas em 
cada noite, bem como os filtros utilizados, o ``seeing'' (em segundos de arco)\footnote{O ``seeing''
\'e uma deforma\cao
na imagem, provocada pela turbul\^encia atmosf\'erica. Seu valor pode ser estimado como sendo a largura
a meia altura no perfil Gaussiano de uma estrela.} e uma estimativa aproximada da qualidade 
fotom\'etrica da noite.

\begin{table}[bh]
\caption{Resumo das observa\c c\~oes.}
\begin{center}
\begin{tabular}{||ccccc||}
\hline
\hline
Noite	& Gal\'axias & Filtros & ``Seeing'' ('') & Qualidade Fotom\'etrica? \\
\omit \\
\hline
27/10/97 & N1637;N7479 & B;V;R   & 1.0 		& sim \\
26/07/98 & N151	       & B;V     & 1.8		& sim \\
27/07/98 & N7496;N7531 & B;V     & 1.8		& n\~ao \\
17/08/98 & N289        & B;V     & 1.8		& n\~ao \\
18/08/98 & N782        & B;V;R;I & 1.5 		& sim \\
	 & N7314       & B;V     &		& \\
19/08/98 & N613        & B;V     & 1.5		& n\~ao \\
20/08/98 & N488;N7755  & B;V     & 1.5 		& n\~ao \\
	 & N6769;N6923 & B;V;R;I & 		& \\
\hline
22/07/98 & N6890       & B;V;R;I & --		& -- \\
\hline
\hline
\end{tabular}
\end{center}
{\footnotesize{Nota: a gal\'axia NGC 6890 foi observada no telesc\'opio com espelho de 
1.6 metros do OPD--LNA, com o mesmo detector por n\'os utilizado.}}
\end{table}

A fotometria foi realizada em bandas largas, com os filtros B, V, R e I do sistema de Cousins. Foram
realizadas 6 exposi\coes na banda B, 5 na banda V e 3 nas bandas R e I, tipicamente. A realiza\cao de
exposi\coes m\'ultiplas tem o objetivo de facilitar a remo\cao de raios c\'osmicos, bem como de evitar
imagens com artif\1cios provocados pelo mal acompanhamento do telesc\'opio. O tempo de
integra\cao em cada exposi\cao \'e de 300 segundos. Com o intuito de realizar a fotometria absoluta 
de cada gal\'axia, utilizamos as estrelas--padr\~ao publicadas em Graham (1982). Na Tabela 3.2, exibimos
as estrelas--padr\~ao observadas em cada noite, bem como os coeficientes de calibra\cao em cada filtro, 
para cada noite, calculados atrav\'es da tarefa {\sc imexamine} do {\sc iraf}.

\begin{table}[hbt]
\caption{Resumo das calibra\coes fotom\'etricas.}
\begin{center}
\begin{tabular}{||cccccc||}
\hline
\hline
Noite	& Estrelas & $C_{C}$ (B) & $C_{C}$ (V) &  $C_{C}$ (R) &  $C_{C}$ (I) \\
\omit \\
\hline
27/10/97 & 18-N-E2 &        --        &        --        & 20.30 $\pm$ 0.03 &        --          \\
  	 & 34-S-E3	& & & & \\
26/07/98 & 39-S-E8 & 19.56 $\pm$ 0.06 & 19.79 $\pm$ 0.03 &        --        &        --          \\
  	 & 47-V-E8	& & & & \\
  	 & 48-W-E8	& & & & \\
  	 & 18-P-E8	& & & & \\
27/07/98 & 20-Q-E1 & 19.58 $\pm$ 0.03 & 19.86 $\pm$ 0.04 &        --        &        --          \\
  	 & 44-S-E1	& & & & \\
17/08/98 & 39-S-E8 & 19.63 $\pm$ 0.05 & 19.82 $\pm$ 0.09 &        --        &        --          \\
  	 & 47-V-E8	& & & & \\
  	 & 48-W-E8	& & & & \\
  	 & 18-P-E8	& & & & \\
18/08/98 & 39-S-E8 & 19.75 $\pm$ 0.10 & 19.96 $\pm$ 0.08 & 20.06 $\pm$ 0.03 & 19.59 $\pm$ 0.03   \\
  	 & 47-V-E8	& & & & \\
  	 & 48-W-E8	& & & & \\
  	 & 18-P-E8	& & & & \\
19/08/98 & 20-Q-E1 & 19.41 $\pm$ 0.13 & 19.67 $\pm$ 0.14 &        --        &        --          \\
  	 & 35-R-E1	& & & & \\
  	 & 44-S-E1	& & & & \\
20/08/98 & 39-S-E8 & 19.76 $\pm$ 0.03 & 19.89 $\pm$ 0.05 & 20.02 $\pm$ 0.03 & 19.55 $\pm$ 0.03   \\
  	 & 47-V-E8	& & & & \\
  	 & 48-W-E8	& & & & \\
  	 & 18-P-E8	& & & & \\
\hline
22/07/98 & 98-E6   & 22.60 $\pm$ 0.25 & 22.61 $\pm$ 0.27 & 22.41 $\pm$ 0.37 & 22.30 $\pm$ 0.42   \\
  	 & 27-R-E9	& & & & \\
\hline
\hline
\end{tabular}
\end{center}
{\footnotesize{Notas: infelizmente, na noite de 27/10/97 n\~ao foi poss\1vel realizar a observa\cao
de estrelas--padr\~ao. Dessa forma, foi utilizada a constante de calibra\cao da noite anterior. O erro 
apresentado para cada constante de calibra\cao da noite corresponde ao desvio padr\~ao nas constantes 
determinadas para cada estrela.}}
\end{table}

Os coeficientes de calibra\cao para cada estrela--padr\~ao, $C_{C}$, foram calculados atrav\'es da seguinte
rela\c c\~ao:

\eq
{C_{C}} = {mag(Graham) - mag(instrum.) + k x},
\eeq

\noindent onde $mag(Graham)$ \'e a magnitude da estrela fornecida por Graham (1982), $k$ \'e o coeficiente
de extin\cao atmosf\'erica e $x$ \'e a massa de ar (ver se\cao 2.1). $mag(instrum.)$ \'e definido por:

\eq
{mag(instrum.)} = {-2.5 \log_{10} {{f} \over {t}}},
\eeq

\noindent com $f$ sendo igual ao fluxo de energia irradiada (em ADU's) pela estrela em uma abertura igual
\`a largura a meia altura de seu perfil Gaussiano, e $t$ sendo o tempo de exposi\cao (ver se\cao 2.1). Os 
coeficientes $k$ para os filtros B, V, R e I, s\~ao, respectivamente, 0.28, 0.16, 0.12 e 0.08, 
correspondendo aos valores de inverno determinados pela equipe t\'ecnica do LNA. Para cada noite, foi
determinado um coeficiente de calibra\cao m\'edio, atrav\'es daqueles calculados para cada estrela--padr\~ao 
individualmente. O erro no coeficiente de calibra\cao foi tomado como sendo o desvio padr\~ao nos 
coeficientes determinados para cada estrela.

Para determinar o coeficiente de calibra\cao para cada gal\'axia, j\'a transformando os valores em 
magnitudes para brilho superficial, utilizamos a seguinte rela\c c\~ao:

\eq
{C_{\mu}} = {2.5 \log_{10} (t) - k x + C_{C} + 5 \log_{10} (l) - R \times E(B-V)},
\eeq

\noindent onde $l$ \'e a dimens\~ao do ``pixel'' em segundos de arco, e $R \times E(B-V)$ \'e a
absor\cao em magnitudes provocada pela extin\cao Gal\'actica (ver se\cao 2.3.1). $E(B-V)$ foi extra\1do
dos mapas de Schlegel, Finkbeiner \& Davis (1998), e o valor de $R$ nos filtros B, V, R e I, \'e, 
respectivamente, 4.1, 3.1, 2.3 e 1.5 (ver Binney \& Merrifield 1998).

Os CCD's s\~ao detectores que se caracterizam pela alta efici\^encia qu\^antica, baixo ru\1do e alta
sensibilidade. No entanto, certos procedimentos devem ser realizados no tratamento das imagens obtidas
em CCD's. Esses procedimentos podem ser resumidos como se segue.

\paragraph{Elimina\cao do ``bias''.} O ``bias'' \'e um mecanismo artificial, de origem eletr\^onica, 
que aumenta a efici\^encia da transfer\^encia de carga nos CCD's. Como o seu valor varia ligeiramente
de ``pixel'' para ``pixel'', se obt\'em uma imagem de ``bias'' em uma exposi\cao com tempo de
integra\cao igual a 1 segundo, e com o obturador fechado. Para melhorar a determina\cao da imagem de
``bias'', obtivemos 25 imagens em cada noite. Essas imagens foram combinadas, utilizando o valor
m\'edio em cada ``pixel'', atrav\'es da tarefa {\sc zerocombine} do {\sc iraf}. A imagem final de
``bias'' deve ser subtra\1da de cada imagem a ser analisada.

\paragraph{Subtra\cao do ``dark current''.} Este \'e um sinal causado por el\'etrons t\'ermicos e, 
portanto, proporcional ao tempo de integra\c c\~ao. No nosso caso, este sinal pode ser desprezado.

\paragraph{Normaliza\cao pelo ``flatfield''.} Trata-se de um mapeamento corretivo da varia\cao da
sensibilidade ao longo do CCD em exposi\coes com fontes de luz uniformes. Normalmente, utiliza-se,
como fonte de luz uniforme, regi\~oes claras do c\'eu ao nascer ou p\^or do Sol, ou ainda, exposi\coes
tomadas na c\'upula, iluminada com luz difusa. Obtivemos 25 imagens de ``flatfield'' por noite para
cada filtro e, utilizando a tarefa {\sc flatcombine} do {\sc iraf}, essas imagens foram combinadas, 
utilizando o valor m\'edio em cada ``pixel'', normalizado pelo valor modal de cada imagem, 
para obter imagens finais de ``flatfield''. \'E 
necess\'ario obter imagens de ``flatfield'' para cada filtro, pois a varia\cao na sensibilidade do
CCD \'e dependente da distribui\cao espectral da luz incidente. As imagens a serem analisadas devem
ser divididas pela imagem de ``flatfield'' correspondente.

\paragraph{Cosm\'etica.} CCD's podem possuir alguns ``pixels'' ``quentes'', i.e., que saturam
rapidamente. Estes podem ser eliminados por processos de interpola\cao linear, ou por mascaramento.

As imagens analisadas foram corrigidas do ``bias'' e do ``flatfield'', bem como cortadas das
regi\~oes n\~ao \'uteis do CCD, e corrigidas pelo ``overscan''\footnote{O ``overscan'' \'e uma 
corre\cao adicional ao ``bias''. Trata-se de uma regi\~ao do CCD que n\~ao \'e exposta \`a luz
incidente, mesmo durante uma exposi\cao normal. Desse modo, \'e an\'alogo ao ``bias'', mas com um tempo
de exposi\cao longo, e determinado no momento da aquisi\cao da imagem a ser analisada.}, 
atrav\'es da tarefa {\sc ccdproc} do {\sc iraf}.

Para eliminar raios c\'osmicos e aumentar a raz\~ao sinal/ru\1do das imagens a serem analisadas, 
utilizamos a tarefa {\sc imcombine} do {\sc iraf}, combinando as exposi\coes obtidas para cada gal\'axia, 
em cada filtro separadamente, utilizando o valor mediano em cada ``pixel''. Para determinar 
pequenos deslocamentos entre exposi\coes sucessivas, cada i-magem individual foi inspecionada, para
determinar a posi\cao de 3 estrelas de refer\^encia. Esses pequenos deslocamentos foram calculados
pela tarefa {\sc imcentroid} do {\sc iraf}, e utilizados pela tarefa {\sc imcombine} na confec\c c\~ao
da imagem combinada.

O passo seguinte consiste na subtra\cao do fundo de c\'eu. Para determinar a imagem do fundo de c\'eu, 
inicialmente editamos a imagem combinada de cada gal\'axia, em cada filtro, retirando a gal\'axia e
estrelas presentes, atrav\'es da tarefa {\sc imedit} do {\sc iraf}. Em seguida, atrav\'es da tarefa
{\sc imstatistics}, determinamos o valor m\'edio da imagem editada, bem como o seu desvio padr\~ao.
Todos os valores que estavam fora dos limites valor m\'edio $\pm$ 3 $\times$ o desvio padr\~ao, foram
substitu\1dos pelo valor m\'edio, atrav\'es da tarefa {\sc imreplace}. Utilizando a tarefa
{\sc imsurfit}, ajustamos um polin\^omio de Legendre de grau 2 (que corresponde a um ajuste
linear) \`a imagem assim obtida, o que nos fornece uma imagem sint\'etica do fundo de c\'eu. Da imagem
combinada de cada gal\'axia subtra\1mos a imagem sint\'etica do fundo de c\'eu. A imagem com o c\'eu
subtra\1do foi ent\~ao editada, eliminando-se os objetos que n\~ao pertencem \`a nossa an\'alise, tais
como estrelas e regi\~oes H{\sc ii} presentes na gal\'axia. Obtivemos, assim, a imagem final de cada
gal\'axia, em cada filtro, pronta para ser estudada.

\subsection{O estudo comparativo}

\hskip 30pt Para determinar os gradientes de cor das gal\'axias por n\'os observadas \'e necess\'ario, 
inicialmente, calcular o perfil radial de brilho superficial de cada gal\'axia, em cada filtro utilizado.
Para tanto, utilizamos a tarefa {\sc ellipse} do {\sc iraf}. Essa tarefa ajusta isofotas (n\1veis de
mesma intensidade) \`a imagem da gal\'axia em estudo, gerando uma tabela com 26 informa\c c\~oes, entre essas,
a intensidade m\'edia, o \^angulo de posi\c c\~ao, a elipticidade, e o coeficiente de Fourier b4 (ver se\cao 
3.3). Inicialmente, a {\sc ellipse} foi executada com os centros das isofotas livres, para uma melhor 
determina\cao do centro da gal\'axia. Este foi ent\~ao fixo para a execu\cao final da tarefa. Adicionando 
o valor de $C_{\mu}$ \`a intensidade m\'edia de cada isofota, obtivemos os perfis de brilho superficial
desejados.

No entanto, os perfis obtidos com o {\sc iraf} consistem do brilho superficial em fun\cao do raio em
``pixels''. Esses perfis foram ent\~ao transformados para as unidades adotadas em LdV83,85 e 
tamb\'em no Cap\1tulo 2, i.e., o brilho superficial (em magnitudes por segundo de arco ao quadrado) em 
fun\cao do logaritmo do raio em unidades de 0.1 minutos de arco. Posteriormente, subtra\1mos os perfis
em cada banda, para obtermos perfis dos \1ndices de cor. Obtivemos, portanto, uma tabela semelhante
\`aquela que elaboramos com os dados do LdV83,85, para cada gal\'axia. 
Por fim, utilizamos o programa {\sc progress} da mesma
forma que no Cap\1tulo 2, determinando os gradientes de cor para cada gal\'axia. Estes s\~ao apresentados
na Tabela 3.3.

\begin{table}
\caption{Gradientes de cor.}
\begin{center}
\begin{tabular}{||ccccc||}
\hline
\hline
Gal\'axia	& $G(B-V)$ (LdV) & $G(B-V)$ & $G(B-R)$ & $G(V-I)$ \\
\omit \\
\hline
N151		& -0.36		& -0.28     & --       & --       \\
N289		& -0.19		& -0.20     & --       & --       \\
N488		& -0.13		& -0.03     & --       & --       \\
N613		& 0.01		& -0.03     & --       & --       \\
N782		& -0.31		& -0.36     & -0.47    & -0.10    \\
N1637		& -0.22		& -0.11     & --       & --       \\
N6769		& -0.20		& 0.16      & -0.17    & -0.18    \\
N6890		& -0.21		& -0.19     & -0.17    & -0.06    \\
N6923		& -0.38		& -0.37     & -0.42    & -0.28    \\
N7314		& -0.21		& -0.13     & --       & --       \\
N7479		& -0.25		& -0.10     & --       & --       \\
N7496		& 0.23		& 0.31      & --       & --       \\
N7531		& -0.23		& -0.21     & --       & --       \\
N7755		& -0.13		& -0.18     & --       & --       \\
\hline
\hline
\end{tabular}
\end{center}
{\footnotesize{Notas: a primeira coluna identifica a gal\'axia; na segunda, apresentamos o gradiente de cor
(B\,-V) obtido no Cap\1tulo 2, e, na terceira, aquele determinado com as imagens adquiridas no LNA. As demais
colunas exibem os gradientes em outros \1ndices de cor, tamb\'em determinados com as imagens adquiridas 
no LNA. O erro t\1pico \'e de 0.02--0.03 magnitudes.}}
\end{table}

\begin{figure}
\epsfysize=21.8cm
\centerline{\epsfbox{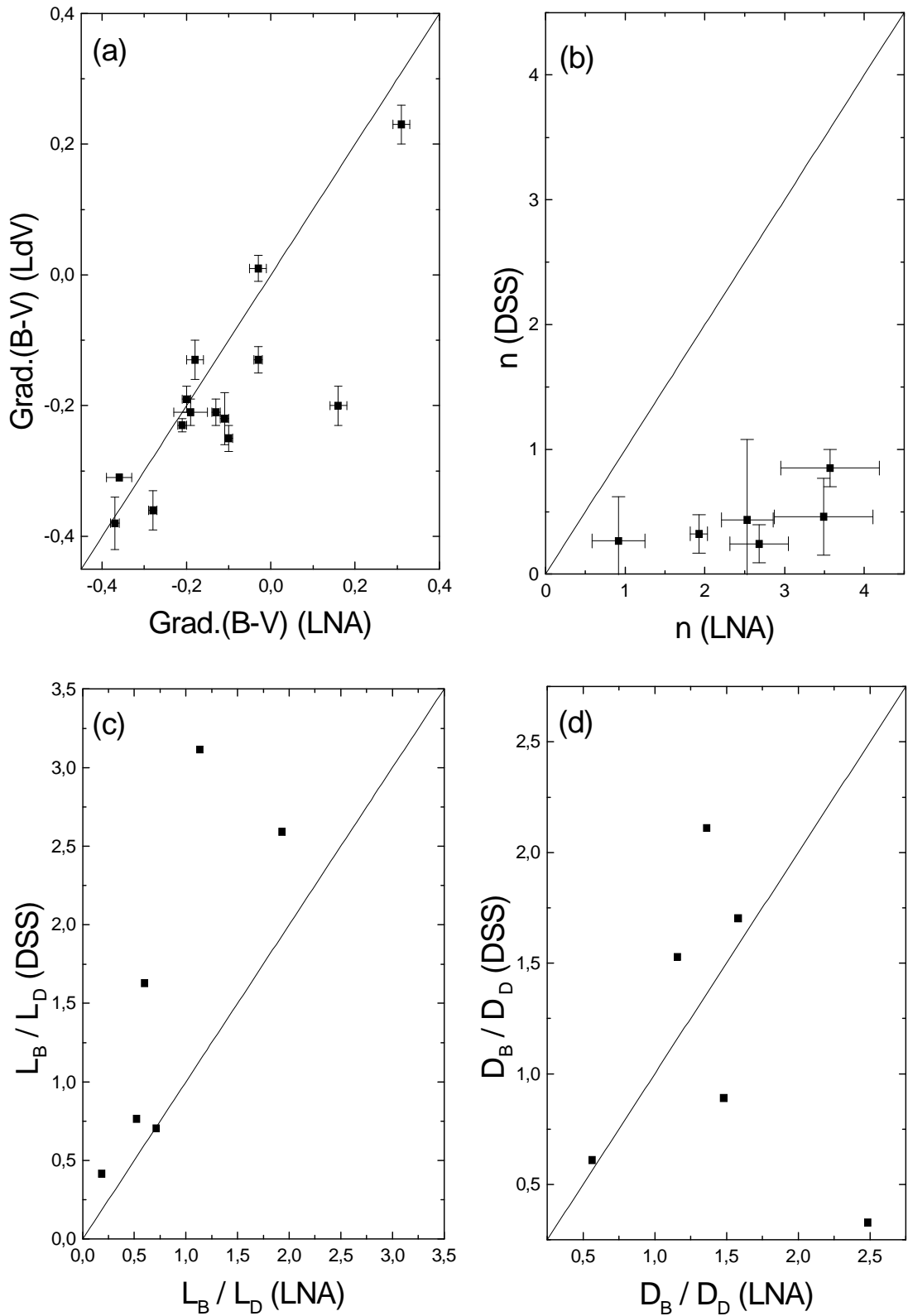}}
\caption{(a): compara\cao entre os gradientes obtidos dos dados do LdV83,85 e aqueles determinados atrav\'es do
imageamento em CCD. (b), (c) e (d): compara\cao entre os \1ndices de S\'ersic e as raz\~oes bojo/disco 
determinados nas imagens DSS e no imageamento CCD. A linha cheia representa perfeita equival\^encia.} 
\label{LNA1}
\end{figure}

Com os dados apresentados na Tabela 3.3, nota-se que, para 7 das 14 gal\'axias, os gradientes determinados
atrav\'es das imagens em CCD s\~ao os mesmos (dentro dos erros) aos determinados atrav\'es dos dados de
fotometria fotoel\'etrica de abertura do LdV83,85. Considerando as 3 categorias de gradientes que 
definimos na se\cao 2.4.2, os gradientes de 12 das 14 gal\'axias continuam na mesma categoria na fotometria
em CCD. Duas gal\'axias, NGC 488 e NGC 6769, apresentam grandes discrep\^ancias. No entanto, na Tabela 3.1 se
v\^e que ambas foram observadas em uma mesma noite, de qualidade n\~ao-fotom\'etrica. A Figura 3.4(a)
mostra os gradientes (B\,-V) obtidos atrav\'es dos dados do LdV83,85 em fun\cao daqueles obtidos atrav\'es
do imageamento CCD. A linha cheia representa perfeita equival\^encia. Nota-se uma boa correla\c c\~ao, 
o que d\'a suporte aos resultados da an\'alise dos gradientes, discutidos no Cap\1tulo 2.

Para 6 das 14 gal\'axias observadas por n\'os no LNA realizamos a decomposi\cao bojo/disco, utilizando 
imagens do DSS (ver Cap\1tulo 3). S\~ao elas: NGC 151, NGC 289, NGC 488, NGC 613, NGC 7479 e NGC 7755.
Utilizamos os mesmos procedimentos aplicados \`as imagens do DSS na decomposi\cao bojo/disco dessas 6
gal\'axias, ao utilizar as imagens em CCD. As Figuras 3.4 e 3.5 apresentam as compara\coes entre os
resultados da decomposi\cao bojo/disco em imagens DSS e em imagens CCD. Mais uma vez, a linha cheia 
representa perfeita equival\^encia.

\begin{figure}
\epsfysize=22cm
\centerline{\epsfbox{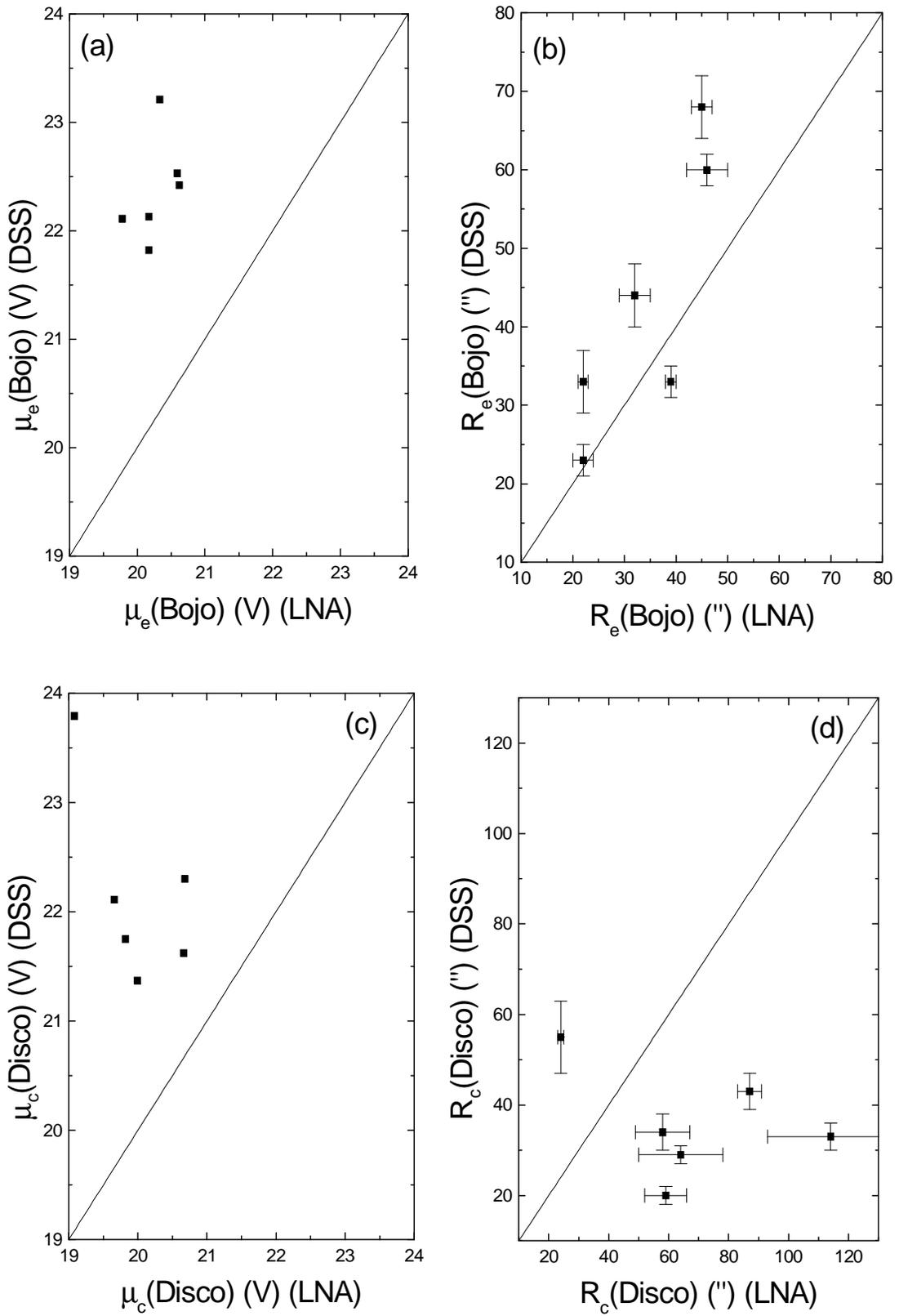}}
\caption{Compara\coes entre os par\^ametros caracter\1sticos dos perfis de bojos e discos obtidos atrav\'es da
decomposi\cao em imagens DSS e no imageamento CCD. A linha cheia representa perfeita equival\^encia.} 
\label{LNA2}
\end{figure}

A Figura 3.4(b) apresenta a compara\cao entre os \1ndices de S\'ersic. Verifica-se prontamente que
os \1ndices obtidos em imagens DSS s\~ao sempre menores que aqueles nas imagens CCD, al\'em de apresentarem
uma faixa de valores expressivamente mais estreita. 
Este comportamento reflete muito provavelmente o fato de que a satura\cao das placas do DSS afetam o perfil 
de luminosidade do bojo, tornando-o muito mais achatado do que de fato o \'e.
Na Figura 3.4(c), comparamos as raz\~oes entre as
luminosidades de bojo e disco, e na Figura 3.4 (d) as raz\~oes entre os di\^ametros de bojo e disco. N\~ao h\'a
barras de erro, pois esses valores s\~ao calculados atrav\'es das imagens--modelo. Apesar de ser uma 
amostra estatisticamente pequena, nota-se, em ambas as compara\c c\~oes, que a distribui\cao dos
pontos exibe uma clara tend\^encia de que o comportamento das raz\~oes bojo/disco \'e aproximadamente o mesmo, tanto
nas imagens DSS, quanto nas imagens CCD. Ou seja, valores elevados para as raz\~oes bojo/disco nas imagens DSS tamb\'em
resultam ser mais elevados nas imagens CCD, o mesmo ocorrendo com os valores baixos. No entanto, h\'a um ponto que se 
faz excess\~ao, correspondente \`a gal\'axia NGC 289. Verifica-se, al\'em disso, que as raz\~oes entre as 
luminosidades de bojo e disco obtidas atrav\'es das imagens do DSS tendem a ser maiores que aquelas obtidas 
atrav\'es do imageamento em CCD.

Nas Figuras 3.5(a) e (b), apresentamos as compara\coes entre os par\^ametros ca-racter\1sticos dos perfis 
dos bojos, enquanto que as Figuras 3.5(c) e (d) se referem aos par\^ametros caracter\1sticos dos perfis 
dos discos. A Figura 3.5(a) mostra que o brilho superficial efetivo do bojo, na banda V, em magnitudes
por segundo de arco ao quadrado, determinado nas imagens DSS, \'e sempre maior que aquele determinado nas
imagens CCD. Os pontos est\~ao substancialmente afastados da linha de equival\^encia. Entretanto, 
bojos com alta concentra\cao central de luz (i.e., valores baixos para o brilho superficial efetivo) nas
imagens DSS, tamb\'em t\^em essa propriedade nas imagens CCD. O mesmo ocorre em bojos com pequena
concentra\cao central de luz, ou seja, valores elevados para o brilho superficial efetivo. A Figura 3.5(b)
exibe a compara\cao dos raios efetivos (em segundos de arco) dos bojos. Nota-se, mais uma vez, que o 
comportamento deste par\^ametro \'e o mesmo, tanto nas imagens DSS, quanto nas imagens em CCD. Tamb\'em
pode-se notar que o raio caracter\1stico dos bojos tende a ser maior nas imagens do DSS.
A compara\cao dos raios efetivos dos bojos mostra que este par\^ametro \'e mais confi\'avel de ser determinado
nas imagens do DSS do que o brilho superficial efetivo do bojo, apesar de ser sistematicamente maior que 
o valor mais acurado determinado a partir da fotometria CCD.  

A Figura 3.5(c) mostra que o brilho superficial central dos discos (na banda V, em magnitudes por segundo de 
arco ao quadrado) determinado em imagens do DSS \'e tamb\'em sempre maior que aquele determinado no imageamento CCD.
Mais uma vez, os pontos est\~ao substancialmente afastados da linha de equival\^encia. Com excess\~ao de
um \'unico ponto, que corresponde, mais uma vez, \`a gal\'axia NGC 289, verifica-se que a amplitude da
faixa de valores \'e bastante semelhante tanto nas imagens do DSS quanto naquelas em CCD, sendo pouco
maior do que 1 magnitude por segundo de arco ao quadrado. A Figura 3.5(d) mostra que os raios 
caracter\1sticos dos discos (em segundos de arco) nas imagens DSS tendem a ser expressivamente menores do
que aqueles obtidos com as imagens CCD. Esta compara\cao mostra que a dimens\~ao do disco em imagens DSS deve 
estar afetada pelo processo de subtra\cao do fundo de c\'eu.

\subsection{Discuss\~ao e conclus\~oes acerca do estudo comparativo}

\hskip 30pt A princ\1pio, dever\1amos esperar que os gradientes de cor em gal\'axias, determinados
atrav\'es de fotometria fotoel\'etrica de abertura (como s\~ao aqueles obtidos no Cap\1tulo 2, 
atrav\'es dos dados do LdV83,85) tivessem valores menos pronunciados do que aqueles
calculados atrav\'es de fotometria em CCD. 

Imaginemos o caso de uma gal\'axia com o t\1pico gradiente 
negativo, ou seja, com um \1ndice de cor central maior (mais avermelhado) do que o seu \1ndice de cor
perif\'erico, mais azulado. Em se tratando de dados obtidos atrav\'es de fotometria de abertura,
a maior abertura pela qual se realiza a fotometria da gal\'axia n\~ao cont\'em
somente a regi\~ao perif\'erica desta (mais azulada), mas tamb\'em cont\'em a regi\~ao central, mais 
avermelhada. De fato, a maior abertura utilizada permite, em geral, que a luz de toda a gal\'axia a-tinja o
detector. Portanto, em fotometria de abertura, o \1ndice de cor atribu\1do \`a regi\~ao perif\'erica da 
gal\'axia em quest\~ao n\~ao \'e t\~ao distinto do \1ndice da regi\~ao central. No caso do imageamento
em CCD, o \1ndice de cor medido na regi\~ao perif\'erica n\~ao sofre essa perturba\cao vinda da luz emitida
pela regi\~ao central da gal\'axia. Dessa forma, os gradientes de cor calculados no imageamento em CCD
s\~ao, a princ\1pio, mais acentuados do que aqueles obtidos atrav\'es de dados em fotometria de abertura.
Essa conclus\~ao permanece inalterada para o caso do mais incomum gradiente positivo. No entanto, 
para gal\'axias com um \1ndice de cor constante desde a regi\~ao central at\'e a regi\~ao perif\'erica, 
o gradiente de cor obtido ser\'a nulo tanto no caso da fotometria de abertura, quanto no imageamento 
em CCD.

Entretanto, os resultados apresentados na Tabela 3.3 e na Figura 3.4(a) mostram que os gradientes de cor
em gal\'axias, calculados atrav\'es da fotometria em CCD, s\~ao essencialmente equivalentes \`aqueles
determinados por fotometria fotoel\'etrica de abertura. 
Portanto, o efeito da perturba\cao do \1ndice de cor perif\'erico da gal\'axia, em fotometria de abertura,
parece poder ser desprezado.
Esse resultado \'e coerente com aquele apresentado
nos dois diagramas inferiores da Figura 2.10, nos quais se compara os gradientes por n\'os determinados 
com aqueles obtidos em PH98. Como, nesse \'ultimo trabalho, tamb\'em foram utilizados dados obtidos no 
imageamento em CCD, esses diagramas s\~ao, ainda que de maneira mais indireta, tamb\'em uma compara\cao
entre os gradientes obtidos por fotometria de abertura e aqueles determinados no imageamento em CCD.

Al\'em disso, n\~ao se nota, na Figura 3.4(a), qualquer diferen\c ca sistem\'atica entre os gradientes
obtidos na Cap\1tulo 2 (i.e., por fotometria de abertura) e os gradientes obtidos no presente cap\1tulo 
(por imageamento em CCD).

Esses resultados indicam que o estudo apresentado aqui permaneceria essencialmente o mesmo, fossem os 
gradientes de cor determinados atrav\'es do imageamento em CCD no lugar dos dados extra\1dos do LdV83,85.

Vejamos agora os resultados referentes \`a decomposi\cao bojo/disco. A Figura 3.4(b) mostra que o 
\1ndice de S\'ersic para o bojo, calculado pelas imagens do DSS, \'e sempre menor do que aquele calculado
pelas imagens em CCD. Em particular, as imagens do DSS sugerem \1ndices sempre menores do que 1, o que
corresponde a um perfil inicialmente relativamente plano, com uma queda posterior. Por outro lado, as
imagens em CCD fornecem \1ndices que variam de 1 (perfil puramente exponencial) a 4 (lei de de Vaucouleurs).
Esse resultado \'e esperado, j\'a que sabemos que as imagens do DSS, por serem obtidas em exposi\coes
profundas, sofrem de satura\cao na regi\~ao central (mais brilhante) das gal\'axias. De fato, os perfis
determinados para os bojos na decomposi\cao bojo/disco nas imagens do DSS s\~ao bastante distintos nas 
imagens em CCD. Portanto, as imagens do DSS n\~ao nos fornecem valores confi\'aveis para o \1ndice de
S\'ersic do bojo. Evidentemente, esse efeito perturba tamb\'em os outros par\^ametros caracter\1sticos do
bojo, bem como os do disco.

A Figura 3.4(c) mostra que existe uma correla\cao relativamente boa entre a raz\~ao bojo/disco (luminosidade)
determinada em imagens DSS e aquela determinada em imagens CCD. No entanto, h\'a uma tend\^encia de os bojos nas
imagens DSS serem mais luminosos. Esse efeito est\'a relacionado ao problema mencionado no par\'agrafo acima, 
\linebreak referente ao \1ndice de S\'ersic, e tamb\'em pode ser avaliado em conjunto com as Figuras 3.4(d) 
e 3.5(a) e (b).
O fato de os bojos nas imagens DSS possu\1rem \1ndices de S\'ersic menores do que 1, significa que seus perfis
n\~ao possuem o caracter\1stico aumento abrupto nas regi\~oes centrais, como no caso de um perfil mais pr\'oximo ao
de de Vaucouleurs. Assim, os bojos nas imagens DSS t\^em uma menor concentra\cao central de luz, o que \'e o que
de fato mostra a Figura 3.5(a). Assim sendo, \'e de se esperar que os raios efetivos dos bojos sejam maiores nas
imagens DSS, como se pode verificar na Figura 3.5(b). Por fim, a raz\~ao bojo/disco (di\^ametro) determinada em
imagens do DSS tamb\'em deve ser afetada, no sentido de os bojos serem maiores nas imagens do DSS. \'E o que mostra
a Figura 3.4(d).

Entretanto, apesar de os par\^ametros caracter\1sticos de bojo e disco serem afetados substancialmente pela
satura\cao central das imagens do DSS, as Figuras 3.4(c) e (d) mostram que existe uma correla\cao relativamente
boa entre as raz\~oes bojo/disco determinadas em imagens DSS e aquelas determinadas em CCD. De fato, os bojos
mais luminosos, ou maiores, nas imagens do DSS, tamb\'em o s\~ao nas imagens em CCD. O mesmo ocorre com o
brilho supeficial efetivo do bojo (Figura 3.5(a)) e o raio efetivo do bojo (Figura 3.5(b)). Portanto, embora
os efeitos da satura\cao central nas imagens do DSS n\~ao possam ser deprezados, os resultados da decomposi\cao
bojo/disco em imagens do DSS s\~ao coerentes com aqueles em que o imageamento CCD \'e utilizado.

O efeito da sub-estimativa para o \1ndice de S\'ersic dos bojos nas imagens do DSS nos par\^ametros caracter\1sticos
dos discos pode ser avaliado atrav\'es das Figuras 3.5(c) e (d). Observa-se que o brilho superficial central
dos discos, bem como seus raios caracter\1sticos, s\~ao menores, i.e., os discos s\~ao menos luminosos nas
imagens DSS, em concord\^ancia com as Figuras 3.4(c) e (d) e com o discutido nos par\'agrafos acima. Por
outro lado, o espalhamento dos pontos nos par\^ametros caracter\1sticos dos discos \'e maior do que nos dos
bojos (Figuras 3.5(a) e (b)) e, com o pequeno n\'umero de pontos (seis), torna-se dif\1cil afirmar se existe
ou n\~ao uma correla\cao entre os par\^ametros determinados com as imagens do DSS e aqueles determinados 
atrav\'es do imageamento CCD.

A conclus\~ao geral que extra\1mos do estudo comparativo realizado nesta se\cao pode ser resumida no
seguinte. Os resultados e conclus\~oes apresentados no estudo dos gradientes de cor (Cap\1tulo 2) permaneceriam
essencialmente os mesmos se, ao inv\'es de utilizarmos dados obtidos atrav\'es de fotometria de abertura, 
tiv\'essemos utilizado imagens em CCD para a determina\cao dos gradientes de cor das 257 gal\'axias de
nossa amostra. Mais ainda, os resultados e conclus\~oes (de car\'ater qualitativo) 
obtidos na decomposi\cao bojo/disco das 39 
gal\'axias de nossa sub-amostra (Cap\1tulo 3) n\~ao sofreriam mudan\c cas fundamentais, fossem as imagens que
utilizamos obtidas com CCD's no lugar de placas fotogr\'aficas digitalizadas. Enfim, os resultados e conclus\~oes
apresentados nesta Disserta\cao permanecem essencialmente os mesmos se os dados aqui utilizados forem 
substitu\1dos por dados obtidos 
atrav\'es da mais moderna e acurada t\'ecnica de imageamento em CCD.

\section{Discuss\~ao e conclus\~oes}

\hskip 30pt Do estudo comparativo que apresentamos na se\cao anterior, podemos concluir que, apesar da satura\cao que as 
imagens do DSS apresentam na regi\~ao central das gal\'axias, que influencia de maneira fundamental os par\^ametros
que obtivemos para os perfis de brilho de bojos e discos, podemos ressaltar certas conclus\~oes com os dados que
obtivemos.

Al\'em do estudo comparativo, apresentamos, neste cap\1tulo, 3 principais resultados referentes aos par\^ametros
obtidos na decomposi\cao bojo/disco. O primeiro deles (Figura 3.2) diz respeito \`a correla\cao entre as escalas
de comprimento de bojos e discos. Esta correla\cao foi identificada por de Jong (1996b) e por Courteau, 
de Jong \& Broeils (1996). Estes trabalhos utilizaram amostras de gal\'axias uma ordem de grandeza
maiores do que a nossa, por\'em constitu\1das basicamente por gal\'axias espirais de tipo tardio. 
O m\'etodo de decomposi\cao utilizado nestes trabalhos foi variado. de Jong utilizou um m\'etodo bi-dimensional, 
enquanto Courteau e Broeils utilizaram um m\'etodo uni-dimensional. Quanto ao \1ndice de S\'ersic do bojo, 
de Jong utilizou um valor fixo (n = 1, 2 ou 4) que melhor se ajustava, mais um disco exponencial, enquanto
Courteau e Broeils utilizaram n = 1 ou 4 (fixo) mais um disco exponencial. De qualquer forma, \'e interessante
notar que conseguimos reproduzir a correla\cao observada, para as gal\'axias em nossa amostra com T = 3. Nos trabalhos
de de Jong e de Courteau, de Jong \& Broeils n\~ao se faz men\cao a respeito de a correla\cao ser diferente 
entre tipos morfol\'ogicos distintos. A dispers\~ao apresentada, tanto no nosso trabalho quanto nos acima citados, 
\'e elevada. Uma poss\1vel raz\~ao para essa elevada dispers\~ao vem dos erros envolvidos nos m\'etodos de
decomposi\c c\~ao, bem como nas hip\'oteses assumidas. Baggett, Baggett \& Anderson (1998) mostram que a diferen\ca t\1pica
nos ajustes realizados por diferentes autores \'e da ordem de 100\%.
A import\^ancia dessa correla\cao reside no fato de que ela mostra que a forma\cao das componentes bojo e disco em 
gal\'axias espirais n\~ao ocorre de forma totalmente distinta e separada, como no cen\'ario monol\1tico. Portanto, 
esta correla\cao tem sido invocada para corroborar o cen\'ario de evolu\cao secular, no qual as forma\coes de bojo
e disco est\~ao vinculadas uma \`a outra.

O segundo resultado est\'a apresentado na Figura 3.3 (pain\'eis superiores). Vimos que existe uma leve tend\^encia
de as gal\'axias com gradientes nulos ou positivos apresentarem bojos com uma maior concentra\cao central
de luz, apesar de ser estatisticamente marginal. Esse resultado pode indicar que 
os efeitos de evolu\cao secular, ao transportar material do disco para o bojo, 
n\~ao somente homogeneizam as popula\coes estelares ao longo da gal\'axia, o que se reflete em gradientes de cor 
nulos ou positivos, mas tamb\'em aumentam a concentra\cao central de massa nos bojos, o que se reflete em uma maior
concentra\cao central de luz. Entretanto, esse resultado deve ser considerado com extremo cuidado, j\'a que a
dispers\~ao \'e bastante elevada. Por outro lado, a Figura 2.12 apresenta um resultado an\'alogo, com dados de
PH98. Nesta figura, vemos que as gal\'axias que apresentam gradientes de cor nulos ou positivos tamb\'em tendem a
apresentar uma maior concentra\cao central de luz. Nesse caso, por\'em, a concentra\cao central de luz se refere ao perfil
de brilho da gal\'axia como um todo, e n\~ao somente ao perfil do bojo separadamente.

Por fim, mostramos que as gal\'axias com gradientes de cor nulos ou positivos t\^em uma leve tend\^encia a 
apresentar bojos maiores (Figura 3.3 -- pain\'eis inferiores). Embora esse resultado
tamb\'em deva ser considerado com extremo cuidado, j\'a que a 
correla\cao \'e bastante fraca, vimos que o raio efetivo do bojo \'e uma par\^ametro que pode ser obtido com
certa confian\ca das imagens do DSS. Esse resultado tamb\'em \'e compat\1vel com o cen\'ario de evolu\cao secular, 
j\'a que nas gal\'axias em que os processos de evolu\cao secular teriam ocorrido, ou seja, nas gal\'axias em que os
gradientes de cor s\~ao atenuados, os bojos devem ser mais proeminentes dentro deste cen\'ario.
   \newpage \chapter{Conclus\~oes e perspectivas}

\section{Conclus\~oes}

\hskip 30pt Realizamos um estudo estat\1stico da distribui\cao radial dos \1ndices de cor (B\,-V) e (U-B) em
uma amostra de 257 gal\'axias de tipos Sb, Sbc e Sc, ordin\'arias e barradas, utilizando dados da literatura, obtidos
atrav\'es de t\'ecnicas de fotometria fotoel\'etrica de abertura. Estes dados foram tratados utilizando
m\'etodos estat\1sticos robustos para determinar os gradientes de cor nas  gal\'axias, bem como os \1ndices
de cor caracter\1sticos de bojos e discos, separadamente. Podemos destacar, entre os resultados principais deste estudo:
\begin{enumerate}

\item  Encontramos na amostra analisada 3 categorias de gradiente, com 
os respectivos percentuais:  65\% das gal\'axias possuem gradientes de cor negativos, i.e., mais vermelhos na regi\~ao 
central do que na regi\~ao perif\'erica, um comportamento que pode ser considerado normal. Cerca de 25\%  possuem 
gradientes de cor nulos e 10\% possuem gradientes de cor 
positivos. Do total de $25\%$ das gal\'axias que apresentam gradientes nulos, 
a fam\1lia de gal\'axias barradas (SAB+SB) contribui com o percentual de  aproximadamente $87\%$ e $81\%$ na cores 
(B\,-V) e (U-B), respectivamente. 
Portanto, as gal\'axias barradas est\~ao super-representadas 
nesta categoria de objetos, indicando que a barra atua no sentido de homogeneizar a popula\cao estelar ao longo de 
gal\'axias espirais de tipo tardio. 

\item Considerando o \1ndice (B\,-V), 37\% das gal\'axias barradas possuem gradientes de cor nulos ou positivos, 
enquanto que 24\% das gal\'axias ordin\'arias apresentam gradientes com essas propriedades. Por outro lado, 
considerando o \1ndice (U-B), 47\% das gal\'axias barradas possuem gradientes de cor nulos ou positivos, 
enquanto que apenas 19\% das gal\'axias ordin\'arias t\^em essas categorias de gradientes.
Esse resultado \'e an\'alogo ao de que gal\'axias barradas parecem ter tamb\'em tend\^encia a apresentar 
gradientes da abund\^ancia O/H menos 
acentuados (MR94; ZKH94). Como os \1ndices de cor representam a popula\cao estelar que domina a emiss\~ao de 
radia\c c\~ao, este resultado indica que
as popula\coes estelares de bojos e discos, em grande parte das gal\'axias barradas, s\~ao similares, ou ainda que
a popula\cao do bojo \'e mais jovem do que a popula\cao do disco. Este resultado est\'a
em desacordo com modelos de forma\cao de gal\'axias espirais que se baseiam exclusivamente no cen\'ario monol\1tico de 
forma\c c\~ao. Por outro lado, esse resultado est\'a em acordo com o cen\'ario de
evolu\cao secular para a forma\cao e/ou constru\cao de bojos, j\'a que este cen\'ario prev\^e que barras estelares
atuam induzindo forma\cao estelar central, e homogeneizando a po-pula\cao estelar ao longo de gal\'axias.
\item A cor ao longo das gal\'axias com gradientes de cor nulos \'e similar \`a cor dos discos das gal\'axias
com gradientes de cor negativos, indicando portanto que as 
modifica\coes ocorrem no bojo e n\~ao no disco. Este resultado tamb\'em est\'a de acordo com as previs\~oes do cen\'ario
de evolu\cao secular, e indica que  deve haver um fluxo de g\'as das regi\~oes perif\'ericas para as 
regi\~oes centrais de gal\'axias, que provoca surtos de forma\cao estelar central, homogeneizando a 
popula\cao estelar ao longo de gal\'axias.
\item Confirmando os resultados obtidos por Peletier \& Balcells (1996), que indicam haver uma correla\cao
entre os \1ndices de cor (U-R), (B\,-R), (R-K) e (J-K) de bojos e discos em uma amostra de 30 gal\'axias espirais, 
encontramos, de maneira an\'aloga, uma correla\cao entre os \1ndices de cor (B\,-V) e (U-B) de bojos
e discos para as gal\'axias em nossa amostra, que \'e substancialmente maior. Essa correla\cao tem
sido utilizada para argumentar que as forma\coes das componentes bojo e disco em gal\'axias espirais
s\~ao eventos conexos, e que a diferen\ca na idade das popula\coes estelares de bojo e disco \'e 
pequena, favorecendo o cen\'ario de evolu\cao secular.
\end{enumerate}

Os resultados acima citados favorecem  o cen\'ario de evolu\cao secular, indicando que, mesmo ap\'os formados, os bojos 
s\~ao estruturas que evoluem.  Estes resultados podem ser interpretados  \`a luz da hip\'otese de
recorr\^encia de barras, sugerida por Norman, Sellwood \& Hasan (1996), e analisados no sentido de
estimar a import\^ancia do cen\'ario de evolu\cao secular na forma\cao e/ou constru\cao de bojos e, 
portanto, na forma\cao de gal\'axias. Admitindo que a fra\cao de objetos em uma determinada fase de evolu\cao 
\'e proporcional ao tempo de perman\^encia nesta 
fase, mostramos que o tempo de vida  de uma barra deve ser da ordem de 5 Giga-anos. Por outro lado, nossos 
resultados mostram
que os processos de evolu\cao secular, relacionados a barras, t\^em um papel fundamental na evolu\cao de
pelo menos 35\% das gal\'axias espirais brilhantes de tipo tardio.

Al\'em disso, realizamos a decomposi\cao dos perfis de brilho de 39 gal\'axias de nossa amostra, em perfis
de brilho separados, para bojos e discos, atrav\'es de um algoritmo bi-dimensional (de Souza 1997) 
aplicado a imagens do DSS,
obtendo par\^ametros estruturais caracter\1sticos dos bojos e discos destas gal\'axias. Estes
par\^ametros foram explorados em conjunto com os gradientes de cor das gal\'axias e os \1ndices de
cor de bojos e discos, determinados na primeira etapa do trabalho. Os principais resultados obtidos
nesta etapa podem ser assim resumidos:
\begin{enumerate}
\item Gal\'axias com gradientes de cor nulos ou positivos t\^em uma leve tend\^encia a apresentarem bojos
maiores e com maior concentra\cao central de luz. Estes resultados podem estar indicando que os processos de evolu\cao
secular, ao transportar material do disco para o bojo, n\~ao somente homogeneizam as popula\coes estelares
destas componentes, como tamb\'em contribuem para a forma\cao e/ou constru\cao de bojos mais proeminentes e
com uma distribui\cao de massa mais concentrada. Entretanto, estes resultados t\^em de ser avaliados com
extremo cuidado, j\'a que a dispers\~ao a-presentada \'e bastante 
significativa, e requerem testes mais conclusivos. 
\item Confirmando os resultados apresentados por de Jong (1996b) e Courteau, de Jong \& Broeils (1996), que 
indicam haver uma correla\cao entre as escalas de comprimento de bojos e discos em gal\'axias espirais de tipo tardio, 
encontramos uma correla\cao entre o raio efetivo dos bojos e o raio caracter\1stico dos discos das 39 gal\'axias
de nossa sub-amostra. Essa correla\cao indica que as forma\coes das componentes bojo e disco de gal\'axias 
espirais s\~ao eventos conexos, favorecendo, mais uma vez, o cen\'ario de evolu\cao secular.
\end{enumerate}

Utilizando imagens em CCD, adquiridas no OPD/LNA--CNPq para 14 gal\'axias de nossa amostra, 
realizamos um estudo comparativo, a partir do qual pode-se concluir que:
\begin{enumerate}
\item Os gradientes de cor determinados atrav\'es de dados obtidos com uso de t\'ecnicas de fotometria fotoel\'etrica
de abertura se correlacionam bem com os gradientes determinados atrav\'es de imagens obtidas em CCD. Este resultado indica 
que aqueles resultados obtidos na an\'alise do Cap\1tulo 2 s\~ao robustos. 
\item A satura\cao da qual sofrem as imagens do DSS na regi\~ao central de gal\'axias brilhantes afeta de
maneira fundamental os perfis de brilho destas gal\'axias. No entanto, estas imagens podem ser utilizadas para
avaliar perfis de brilho em estudos estat\1sticos, i.e., que envolvem um n\'umero relativamente grande de objetos.
Em particular, o raio efetivo do bojo \'e um par\^ametro que pode-se obter atrav\'es destas imagens com um bom grau 
de confiabilidade.
\end{enumerate}

\section{Perspectivas}

\hskip 30pt Pretendemos dar continuidade ao estudo da import\^ancia dos processos de evolu\cao secular
na evolu\cao de gal\'axias, bem como na forma\cao e/ou constru\cao de bojos, e na rela\cao dos processos 
din\^amicos e cinem\'aticos relacionados a barras e a atividade nuclear observada em gal\'axias com AGN's. 
Um estudo importante neste 
sentido ser\'a aplicar o que foi realizado neste trabalho para uma amostra de gal\'axias espirais de tipo
jovem, i.e., S0's e Sa's. N\~ao h\'a at\'e o momento estudos que indiquem que os processos de evolu\cao secular ocorram
nesta classe de gal\'axias. Por outro lado, outros estudos (e.g., Friedli \& Benz 1995) mostram que as barras
em gal\'axias de tipo jovem t\^em propriedades distintas daquelas apresentadas por barras em gal\'axias de tipo tardio.

Em meu  projeto de Doutoramento, pretendo  realizar um estudo direto das principais fam\1lias de \'orbitas peri\'odicas
em um potencial que modele a distribui\c c\~ao de massa em uma gal\'axia fortemente barrada,
incluindo as \'orbitas que se afastam do plano do disco da gal\'axia, isto \'e, aquelas que
d\~ao \`a barra uma importante estrutura vertical. A estrutura vertical de barras estelares
n\~ao tem sido estudada com freq\"u\^encia na literatura. Com o conhecimento adquirido assim da cinem\'atica
de estrelas em barras, em \'orbitas que fujam ao plano do disco, poderemos criar um diagn\'ostico
para identificar a estrutura vertical de barras em gal\'axias vistas de face.
Esse diagn\'ostico ser\'a \'util na solu\c c\~ao de v\'arios problemas, entre eles na compreens\~ao 
das diferen\c cas entre barras em gal\'axias de tipo jovem e aquelas em gal\'axias de tipo tardio. 
Ao obter espectros para uma amostra de gal\'axias fortemente barradas vistas de face, poderemos usar
esse diagn\'ostico para, por exemplo, identificar barras rec\'em-formadas. 
Realizando o imageamento das gal\'axias em nossa amostra nas bandas B, V, R, I e K, poderemos estudar
a import\^ancia das barras na forma\c c\~ao e/ou constru\c c\~ao de bojos, buscando, portanto,
v\1nculos para os modelos de forma\c c\~ao e evolu\c c\~ao de gal\'axias.

   \newpage
\addcontentsline{toc}{chapter}{REFER\^ENCIAS}

\Large

REFER\^ENCIAS

\normalsize

Abraham, R.G., Valdes, F., Yee, H.K.C. \& van den Bergh, S.: 1994, ApJ, 432, 75 

Abraham, R.G., Tanvir, N.R., Santiago, B.X., Ellis, R.S., Glazebrook, K. \& 

van den Bergh, S.: 1996, MNRAS, 279, L47 

Alladin, S.M. \& Narasimhan, K.S.V.S.: 1982, Physics Reports, 92(6), 339

Andredakis, Y.C., Peletier, R.F. \& Balcells, M.: 1995, MNRAS, 275, 874

Athanassoula, E. \& Bureau, M.: 1999, accepted for publication in ApJ, astro-ph/9904206

Avila-Reese, V. \& Firmani, C.: 1998, in Star Formation in Early-Type 
Galaxies, ASP Conf. Ser., ed. by P. Carral \& J. Cepa, astro-ph/9808165

Baggett, W.E., Baggett, S.M. \& Anderson, K.S.J.: 1998, ApJ, 116, 1626

Balcells, M. \& Peletier, R.F.: 1994, AJ, 107(1), 135

Baugh, C.M., Cole, S. \& Frenk, C.S.: 1996, MNRAS, 283, 1361

Berentzen, I., Heller, C.H., Shlosman, I. \& Fricke, K.J.: 1998, MNRAS, 300, 49 

Binney, J. \& Merrifield, M.: 1998, in Galactic Astronomy, Princeton Series in Astrophysics

Binney, J. \& Tremaine, S.: 1987, in Galactic Dynamics, Princeton Series in Astrophysics 

Bouwens, R., Cay\'on, L. \& Silk, J.: 1998, astro-ph/9812193, accepted for publication in ApJ

Bruzual, G.A., Magris, G.C. \& Calvet, N.: 1988, ApJ, 333, 673

Bureau, M. \& Athanassoula, E.: 1999, accepted for publication in ApJ, astro-ph/9903061

Bureau, M. \& Freeman, K.C.: 1999, accepted for publication in AJ, astro-ph/9904015

Bureau, M., Freeman, K.C. \& Athanassoula, E.: 1999, in When and How do Bulges Form and
Evolve?, ed. by C.M. Carollo, H.C. Ferguson \& R.F.G. Wyse, Cambridge: CUP, astro-ph/9901246

Carollo, C.M., Stiavelli, M., de Zeeuw, P.T. \& Mack, J.: 1997, AJ, 114(6), 2366

Caon, N., Capaccioli, M. \& D'Onofrio, M.: 1993, MNRAS, 265, 1013

Combes, F.: 1999, in Building Galaxies: from the Primordial Universe to the Present, 
Proceedings of Rencontres de Moriond, 13-20 March 1999, ed. by F. Hammer, T.X. Thuan, V. Cayatte, 
B. Guiderdoni and J. Tran Thanh Van (Ed. Frontieres), astro-ph/9904031

Combes, F. \& Sanders, R.H.: 1981, AA, 96, 164 

Courteau, S., de Jong, R.S. \& Broeils, A.H.: 1996, ApJ, 457, L73

de Jong, R.S.: 1996a, AASS, 118, 557

de Jong, R.S.: 1996b, AA, 313, 45

de Jong, R.S.: 1996c, AA, 313, 377

de Jong, R.S. \& van der Kruit, P.C.: 1994, AASS, 106, 451

de Souza, R.E.: 1997, Tese de Livre Doc\^encia, Departamento de Astronomia, Instituto 
Astron\^omico e Geof\1sico, Universidade de S\~ao Paulo

de Souza, R.E. \& dos Anjos, S.: 1987, AASS, 70, 465

de Vaucouleurs, G.: 1963, ApJS, 8, 31 

de Vaucouleurs, G., de Vaucouleurs, A., Corwin, H.G., Buta, R.J., Paturel, G. \&
Fouque, P.: 1991, in Third Reference Catalog of Bright Galaxies, Springer-Verlag, New York {\bf (RC3)}  

Eggen, O.J., Lynden-Bell, D. \& Sandage, A.R.: 1962, ApJ, 136, 748

Elmegreen, D.M.: 1998, in Galaxies and Galactic Structure, Prentice Hall

Elmegreen, D.M. \& Elmegreen, B.G.: 1982a, MNRAS, 201, 1021 

Elmegreen, D.M. \& Elmegreen, B.G.: 1982b, MNRAS, 201, 1035 

Erwin, P. \& Sparke, L.S.: 1998, in Galaxy Dynamics, proceedings of a conference 
held at Rutgers University, 8-12 Aug 1998. To appear in ASP Conference Series,
edited by D.R. Merritt, M. Valluri, and J.A. Sellwood, astro-ph/9811345

Erwin, P. \& Sparke, L.S.: 1999, astro-ph/9906262, accepted by ApJL

Evans, R.: 1994, MNRAS, 266, 511

Fran\c cois, P., Vangioni-Flam, E. \& Audouze, J.: 1990, ApJ, 361, 487

Friedli, D.: 1999, in The Evolution of Galaxies on Cosmological Timescales, 
ed. by J.E. Beckman \& T.J. Mahoney, ASP Conf. Ser., astro-ph/9903143

Friedli, D. \& Benz, W.: 1993, AA, 268, 65 

Friedli, D. \& Benz, W.: 1995, AA, 301, 649 

Friedli, D. \& Martinet, L.: 1993, AA, 277, 27 

Frogel, J.A.: 1985, ApJ, 298, 528

Fukugita, M., Shimasaku, K. \& Ichikawa, T.: 1995, PASP, 107, 945

Giovanelli, R., Haynes, M.P., Salzer, J.J., Wegner, G., da Costa, L.N., \& Freudling, W.: 
1994, AJ, 107(6), 2036

Giovanelli, R., Haynes, M.P., Salzer, J.J., Wegner, G., da Costa, L.N., \& Freudling, W.: 
1995, AJ, 110(3), 1059

Graham, J.A.: 1982, PASP, 94, 244

Ho, L.C., Filippenko, A.V. \& Sargent, W.L.W.: 1997, ApJ, 487, 591 

Hubble, E.P.: 1926, ApJ, 64, 321 

Hubble, E.P.: 1936, in The Realm of the Nebulae, New Haven: Yale University Press 

Humason, M.L.: 1936, ApJ, 83, 10 

Ibata, R.A. \& Gilmore, G.F.: 1995, MNRAS, 275, 605

Kauffmann, G. \& White, S.D.M.: 1993, MNRAS, 261, 921

Kitchin, C.R.: 1998, in Astrophysical Techniques, Institute of Physics Publishing, 
Bristol and Philadelphia

Knapen, J.H.: 1998, in The Central Regions of the Galaxy and Galaxies, IAU Symp.184, p.93

Koopmann, R.A. \& Kenney, J.D.P.: 1998, ApJ, 497, L75 

Kormendy, J.: 1982, ApJ, 257, 75

Kormendy, J.: 1993, in Galactic Bulges, IAU Symp.153, p.209

Kormendy, J. \& Bender, R.: 1996, ApJ, 464, L119

Kormendy, J. \& Illingworth, G.: 1983, ApJ, 265, 632

Kuijken, K. \& Merrifield, M.R.: 1995, ApJ, 443, L13

Lahav, O., Naim, A., Buta, R.J., Corwin, H.G., de Vaucouleurs, G., Dressler, A., Huchra, J.P., 
van den Bergh, S., Raychaudhury, S., Sodr\'e Jr., L. \& Storrie-Lombardi, M.C.: 1995, Science, 267, 859

Larson, R.B.: 1990, PASP, 102(653), 709 

Larson, R.B. \& Tinsley, B.M.: 1978, ApJ, 219, 46

Lasker, B.M., Sturch, C.R., McLean, B.J., Russell, J.L., Jenkner, H. \&
Shara, M.M.: 1990, AJ, 99, 2019

Longo, G. \& de Vaucouleurs, A.: 1983, Univ. Texas Monograph in Astronomy No. 3 {\bf (LdV83)}  

Longo, G. \& de Vaucouleurs, A.: 1985, Univ. Texas Monograph in Astronomy No. 3A {\bf (LdV85)}  

Marleau, F.R. \& Simard, L.: 1998, ApJ, 507, 585

Martin, P.: 1995, AJ, 109(6), 2428 {\bf (M95)}

Martin, P. \& Roy, J.R.: 1994, ApJ, 424, 599 {\bf (MR94)}

McWilliam, A. \& Rich, R.M.: 1994, ApJS, 91, 749

Merrifield, M.R. \& Kuijken, K.: 1999, AA, 345, L47

Mihalas, D. \& Binney, J.: 1981, in Galactic Astronomy: Structure and Kinematics, 
W.H. Freeman and Co.

Morgan, W.W.: 1958, PASP, 70, 364 

Morgan, W.W. \& Mayall, N.U.: 1957, PASP, 69, 409 

Norman, C.A., Sellwood, J.A. \& Hasan, H.: 1996, ApJ, 462, 114

Peletier, R.F.: 1989, PhD Thesis, University of Groningen, The Netherlands

Peletier, R.F. \& Balcells, M.: 1996, AJ, 111, 2238

Peletier, R.F., Valentjin, E.A., Moorwood, A.F.M. \& Freudling, W.: 1994, AASS, 108, 621

Pfenniger, D.: 1993, in Galactic Bulges, IAU Symp.153, p.387

Prugniel, Ph. \& H\'eraudeau, Ph.: 1998, AASS, 128, 299 {\bf (PH98)}

Renzini, A.: 1999, in When and How do Bulges Form and Evolve?, ed. by C.M. Carollo, 
H.C. Ferguson \& R.F.G. Wyse (Cambridge University Press), astro-ph/9902108

Rich, R.M. \& Terndrup, D.M.: 1997, PASP, 109, 571

Roberts, M.S. \& Haynes, M.P.: 1994, ARAA, 32, 115

Rousseeuw, P.J.: 1984, Journal of The American Statistical Association, 79(388), 871

Rousseeuw, P.J. \& Leroy, A.M.: 1987, in Robust Regression and Outlier Detection, 
Wiley-Interscience, New York

Sakamoto, K., Okumura, S.K., Ishizuki, S. \& Scoville, N.Z.: 1999, 
in When and How do Bulges Form and Evolve?, ed. by C.M. Carollo, H.C. Ferguson \&

R.F.G. Wyse, Cambridge University Press, astro-ph/9902005

Sandage, A.: 1961, in The Hubble Atlas of Galaxies, Carnegie Institution of Washington

Sandage, A.: 1975, in Galaxies and the Universe, Stars and Stellar Systems, Vol. 9, Chap. 1, 
ed. by A. Sandage, M. Sandage \& J. Kristian, The University of Chicago Press 

Schlegel, D.J., Finkbeiner, D.P. \& Davis, M.: 1998, ApJ, 500, 525

Schweizer, F.: 1982, ApJ, 252, 455

Searle, L., Sargent, W.L.W. \& Bagnuolo, W.G.: 1973, ApJ, 179, 427

Sellwood, J.A.: 1993, in Galactic Bulges, IAU Symp.153, p.391

Shlosman, I., Begelman, M.C. \& Frank, J.: 1990, Nature, 345, 679

Shlosman, I., Frank, J. \& Begelman, M.C.: 1989, Nature, 338, 45

Silk, J. \& Bouwens, R.: 1999, in Galaxy Evolution, proceedings of a conference held 
in Meudon, September, 21 - 25, 1998, astro-ph/9812057

Silva, D.R. \& Elston, R.: 1994, ApJ, 428, 511

Sodr\'e Jr, L. \& Cuevas, H.: 1994, Vistas in Astronomy, 38, 287 

Sodr\'e Jr, L. \& Cuevas, H.: 1997, MNRAS, 287, 137 

Stetson, P.B., VandenBergh, D.A. \& Bolte, M.: 1996, PASP, 108, 560

Tinsley, B.M.: 1980, Fund. of Cos. Phys., 5, 287

Toomre, A. \& Toomre, J.: 1972, ApJ, 178, 623

Tully, R.B., Pierce, M.J., Huang, J.S., Saunders, W., Verheijen, M.A.W. \& 
Witchalls, P.L.: 1998, astro-ph/9802247

van den Bergh, S.: 1960a, ApJ, 131, 215 

van den Bergh, S.: 1960b, ApJ, 131, 558 

van den Bergh, S.: 1997, AJ, 113, 2054 

van den Bergh, S., Abraham, R.G., Ellis, R.S., Tanvir, N.R., Santiago, B.X. \&
Glazebrook, K.G.: 1996, AJ, 112, 359

van Houten, C.J.: 1961, BAN, 16, 1

V\'eron-Cetty, M.P. \& V\'eron, P.: 1998, ESO Sci. Rep., 18, 1

White, R.L., Postman, M. \& Lattanzi, M.G.: 1992, in Digitised Optical Sky Surveys, 
ed. by H.T. MacGillivray \& E.B. Thomson, Kluwer, p.167

Wyse, R.F.G.: 1998, MNRAS, 293, 429

Wyse, R.F.G., Gilmore, G. \& Franx, M.: 1997, ARAA, 35, 637

Zaritsky, D., Kennicutt, R.C. \& Huchra, J.P.: 1994, ApJ, 420, 87 {\bf (ZKH94)}

Zhang, X.: 1996, ApJ, 457, 125

Zhang, X.: 1998, in The Central Regions of the Galaxy and Galaxies, IAU Symp.184, p.29



\end{document}